\documentclass[]{report}

\usepackage{longtable}
\usepackage[dvips]{graphicx}
\title{ \textbf{Domain Specific Software Architecture for Design Center Automation}}
\date{ Jan 2000}
\author{ Anshuman Sinha, Haritha Nandela, Vijaya Balakrishna}

\begin{document}

\maketitle

\abstract{ \textit{ Domain specific software architecture aims at software reuse through construction of domain architecture reference model.  The constructed reference model presents a set of individual components and their interaction points.  When starting on a new large software project, the design engineer starts with pre-constructed model, which can be easily browsed and picks up opportunities of use in the new solution design.  This report discusses application of domain reference design methods by deriving domain specific reference architecture for a product ordering system in a design center.  The product in this case is instock and special order blinds from different manufacturers in a large supply store.  The development of mature domain specific reference software architecture for this domain is not the objective of this report.  However, this report would like to capture the method used in one such process and that is the primary concern of this report.  This report lists subjective details of such a process applied to the domain of ordering custom and instock blinds from a large home construction and goods supply store.  This report also describes the detailed process of derivation of knowledge models, unified knowledge models and the reference architecture for this domain.  However, this domain model is only partially complete which may not be used for any real applications.  This report is a result of a course project undertaken while studying this methodology.}}

%\tableofcontents
%\listoffigures
%\listoftables
\addtolength{\parskip}{0.5\baselineskip}
% Anshuman Sinha - To reduce para spacing include line \addtolength{\parskip}{-0.5\baselineskip}

\chapter{ Problem Domain}

\section{ Domain Description}

The design center is a one-stop shop for customers to choose, buy and install window-treatments, wallpaper, carpet, and tiles. Typical customers of the Design Center are homeowners, business community, apartment complexes etc. The focus of this project is on the different activities of the Design Center at Home Depot. The goal is to identify these activities of the Design Center, such as product selection, Order Maintenance and Tracking, Billing, Scheduling of appointments for measurement, installation and delivery, etc.

Choosing different products like wallpaper, blinds, etc. can be confusing, The customer is provided with multiple choices of products to choose from. This selection process emphasizes adhering to issues like: customer's budget, time constraints, specifications and expectations. The selection process depends on many different criteria like the dimensions, type of product, part of the house i.e. kitchen, family room etc., color, style the room portrays, location of the room east, west etc., humidity, colors surrounding the product. There are different products being manufactured by many manufacturers. The categorization of these products into meaningful groups can hasten the selection process. Employees of the design-center issue multiple quotes for the same category of products in order to provide the customer with different options to choose from.

Automation hastens this selection process and decision making by the customer to 
by hastening the decision. On the other hand, automation can help the design center 
employees in issuing correct quotes for different selections of the customer.

The design center stocks many varieties of products from different manufacturers.  
Usually there is two-weeks worth of materials in stock. As and when Home Depot introduces a new product into the market they do this on based on districts. Here they test the product sales and decide on an appropriate amount of inventory to maintain which is known as Suggested Order Quantities (SOQ). Product sales are forecasted appropriately to match the sales during the weeks to follow. As the products are sold, the inventory is updated as well. Whenever the inventory dips below the forecasted amount, the product is reordered. Home Depot also needs to track the sales in periodically so that over-stocking does not occur. Another type of report that influences the inventory is the velocity reports - which gives an idea about the products sold well. The design center employee needs to have all this information on hand, which improves their efficacy.

One of the activities that takes place at Home Depot or any other retail store is 
the return of goods. Custom ordered materials that are returned will not be refunded unless there is a really valid reason. Other products are refunded if returned within the thirty days.  When an item is returned to the store it is marked down to zero and the product is treated as a loss. Also the materials on the floor which are damaged, are also marked down i.e. treated as a loss, and all this information is recorded. After the items are marked down, they are sent back to the vendor (RTV). Automating this inventory management will ensure that the required amount of inventory for each product is maintained. Currently the marked downs that are tracked using paper. This also helps in maintaining accurate vendor information, products offered by that particular vendor, maintain the delivery times and quality of service provided by the vendor. Automation helps increase the efficiency and the speed of each transaction.

When a customer places an order with Home Depot, if the product is in stock, the 
customer buys the product. On the other hand if this is a custom order, the order 
details are entered into the system and these orders are automatically faxed to the 
vendors (supplier). The vendors process these orders and supply the required material. Automation will help eliminate the errors in the purchase order by ensuring that the employee enters all the information needed to place the order. It also prompts the employee for any required accessories or additional fittings that enhance the selected product. Depending on the inventory, the order may need to be placed to the manufacturer of the product. Any possible changes requested by the customer are incorporated into the order.

One of the activities that take place at Home Depot is the Customer Service. Most 
of the products sold at the Design center have some sort of warranty -- either a 
two years or a lifetime. When customer reports a problem, employees are present to 
repair and reinstall the product. Home Depot also replaces the product with no additional charge.

In many circumstances a third party contractor is sent from the design center to 
measure the dimensions for the blinds, wallpaper, carpet etc. This involves interaction with the customer to effectively determine the client's needs. After the products are selected they will be ordered and Home Depot will let the contractor know when the supplies arrive. Appointments need to be made for delivery and installation.  Automating this task ensures that the scheduling process is more organized and efficient.

Major issues requests identified:\footnote{ These issues have been identified in an informal discussion with the expert.}

\begin{enumerate}
\item · ·One of the issues is the availability of different types of products and 
their classification.

\item · ·Custom orders or Inventory replacement orders normally take two weeks to 
deliver. But sometimes there is delay and this causes the missing of the installation 
date.

\item · ·One of the requests was that if all the computers in all Home Depots were 
linked there would be easy access to information.

\item · ·Also if there could an easy way of accessing the vendor information with 
product availability.
\end{enumerate}

\subsection{Current Level of Automation}

Currently Home Depot has different systems for different processes. They all use 
a mainframe system, which handles the order placement, billing etc. They have a mobile cart, using which orders for inventory replacement are placed.  A process like updating the inventory with the marked down products i.e. damaged products, is still tracked on paper and handed to the computer room for data entry. The design center employees need to know all the business rules to complete a sales transaction. And to complicate the situation, most of the systems that exist are not user friendly. 

The selection process is not automated at present and this forces the customer to 
look through all of the different catalogs in trying to decide which one to choose. 
If this process is automated, then the design center employee can assist the customer in a more efficient way. When a customer selects a product, it takes valuable time in order to generate a quote. With the proposed automation, the quote generation process can be hastened. Automation can make order placement faster. As the order size increases this feature becomes important. Currently the process dealing with the return of goods is not automated.  

\begin{center}
\begin{table}
\begin{tabular}{|l|c|c|c|} \hline 
Business Process & Automated & Semi-Automated &  Non Automated \\ \hline 
Selection &  &  & * \\ \hline 
Order Management &  & * &  \\ \hline 
Inventory  Management &  & * &  \\ \hline 
Billing & * &  &  \\ \hline 
\end{tabular}
\caption{ This table shows classification of different processes and their present level of automation.}
\end{table}
\end{center}

An example\footnote{ This information was gained when we had an informal discussion about the domain description with the expert.} given by the expert was that the order management process is semi-automated i.e. quotes still need to entered one by one for each window instead of doing a bunch of them together. So currently there is automation in generating a quote but not to the extent required.

\section{ Domain Mental Model and Viewpoints}

\begin{figure}
\centering
\fbox{\includegraphics[height=4in,width=4in]{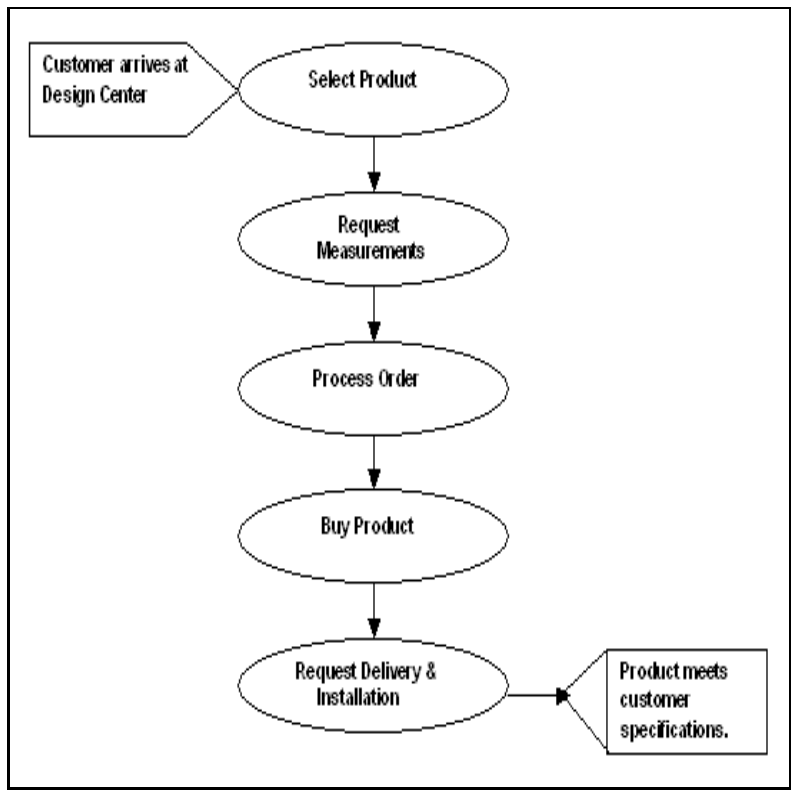}}
\caption{ Domain Mental Model}
\end{figure}

The Domain Mental Model gives a graphical description and the flow of events that occur at the Home Depot.  The fig illustrates the different processes in the mental model. The following details described below highlights the potential activities through the mental model.

\begin{enumerate}
\item · ·The customer arrives at Design Center with requirements for window treatments and/or wallpaper. 

\item · ·If the product is wallpaper, the customers have an idea of their needs. 
If not they consult the employee for advice and make a selection after looking at 
different catalogs.

\item · For window treatments the customer what type of blinds does he need the design center employee assists the customer in deciding on the type depending on the customer requirements.

\item · ·For window treatments after the customer decides the type an appointment 
is made for measuring the dimensions and number of window treatments desired.

\item · ·With the dimensions measured, the customer then consults with a design center employee to decide on different brand choices available for them. The employee generates price quotes for different choices the customer has made. The customer finally selects the product after considering different aspects. The vendor is contacted to find if the product is still in the market for the product.

\item · ·At this moment the customer can still withdraw from the transaction if he 
does not like the choices offered to him

\item · ·If the particular brand is in stock, the customer purchases the product. 
The inventory is then updated.

\item · ·If the product is not in stock an order is placed with the vendor and the 
customer collects after its arrival. Usually the vendors have a two-week delivery 
date after the product is requested

\item · ·If the product is in stock then an order is placed, the customer is billed 
for this purchase and after the payment is made the inventory is updated to reflect 
the change.

\item · ·If the product is not in stock then an order is placed with the vendor, 
the customer is billed for the purchase and when the order arrives at the warehouse 
the customer is notified.

\item · ·There are a couple of choices available the customer can request the product to be delivered and installed or the customer can pick the product and install it.

\item · ·In any case if needed an appointment is scheduled so that a person can deliver the purchased goods and install the same.

\item · ·The terminating event for the mental model is when the product meets the 
customer specifications.
\end{enumerate}

The different viewpoints we need to consider are: the customer's, the design center employee's, the installer's and the vendor's. From the customer's viewpoint, the experience of  buying the product should be as easy and painless as possible. The Design center employee's role is to effectively assist the customer in the making choices, and fast service. The installer's role is to perfectly measure the dimensions and install the product with minimum amount error. The vendor's role is to provide the requested product delivered on an agreed date.  The different organizational units involved in this process are the design center, 
the vendor with whom the order has been placed, the customer and the sub-contractor 
who installs the products. The design center fits right in the middle of this hierarchy, and needs to organize so that the customer's needs are being met. The employees need to coordinate with vendor to place the order and see that it is delivered. They also need to coordinate with the installer so that the measurements and installations are done according the customer's convenience and satisfaction.

\begin{figure}
\centering
\fbox{\includegraphics[height=3in,width=4in]{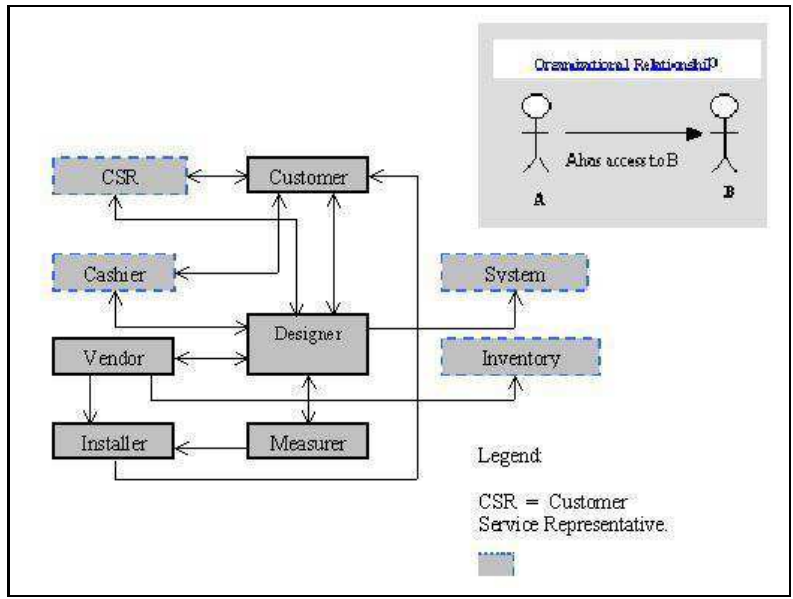}}
\caption{ Organization Chart}
\end{figure}

\subsection{ Organizational Chart}

The above figure shows the organizational chart of the all the persons involved 
at the design center.  Each arrow shows a relation of ``approaches to'' from one 
person to the other.  An ``approach to'' relation would mean either information exchange or a need to consult.  A unidirectional arrow means one way access and a bi-directional arrow implies both way approaches.  For e.g., designer approaches the vendor for placing an order in case of an ex-stock custom ordered product, with the information of type and quantity.  On the other hand vendor gets back to the designer with the information on availability (the product might have been discontinued) and the delivery date of the product. 

The diagram does not illustrate the ``kind of'' data or relationship between the 
persons.  It merely depicts the need for exchange of information between different 
persons involved at the design center.  The organization chart differentiates between people who are in outside the domain by representing them with a dotted line.

\subsection{ Performer Roles}

\begin{enumerate}
\item · ·Designer:  The designer is the primary user of the automation system at 
the design center.  Designer is responsible for helping customer in product selection, contacting measurer for taking the measurements, contacting vendor in case of custom order, maintenance of instock products at the design and contacting system analyst. 

\item · ·Customer:  Customer is the person who buys or has intent to buy a product from the design center.

\item · ·Measurer:  Measurer is the person who takes measurements of the site and reports it back to the designer.  He schedules his visit to the site with customer upon request from the designer.

\item · ·Installer:  Installer is the person who installs the product at the site.  
He schedules his visit to the site with customer and is responsible for the installation of the product.

\item · ·System Analyst: System Analyst is responsible for maintaining the automated system at Home Depot and handling all technical problems within the store for the system.

\item · ·Inventory Manager: Inventory manager deals with the inventory related issues like ordering the products when they reach the critical level, maintaining correct level of stock for instock products based on previous year's sales and receiving the goods from the vendor for instock product.\footnote{ Inventory management and related issues are out of the scope of the domain.  However, the interface to inventory is part of the automation system.}

\item · ·Vendor:  Vendor is a contact in a company that supplies product for the 
design center.  The designer contacts vendor for a custom ordered product.

\item · ·Cashier:  Cashier is the person who accepts payments from the customer against the product or services ordered by the customer.

\item · ·Customer Service Representative: Customer service representative (CSR) is 
responsible for interfacing with customer for return of products and any other special requests by the customer.  After the product is sold CSR is most preferred contact for the customer regarding their previous orders.

\end{enumerate}

Current Level of Requirements Documentation
Many of the activities that occur at the Design Center are processes are generic 
to any retail store. The existing systems do have documentation, which are listed 
below:

\begin{enumerate}
\item · ·The training guide which not only helps new users get acquainted with system but gives a list of the functionality provided by the existing system. The users do have knowledge of the major requirements that are needed by the application. 

\item · ·There is another way of gaining the requirements by reverse engineering the existing application. 

\item · ·There are also a number of manufacturer's catalogs for example Levlor, Decorama, Luxaflex etc which give information about their products regarding price, type, dimensions, quality etc. These catalogs give some selection process requirements for the application.

\item · ·There also exist forms and templates that are currently being used at Home Depot, which also form a part of our requirement documentation. One of the form or template helps the design center employees in collecting all the details from the customer, which helps them in placing the order without any mistakes. There exists another form known as the marked down form that is used for recording the damaged goods. 
\end{enumerate}

The requirements thus consist of some existing documentation, reverse engineering 
the current application and paper based forms and templates.

\section{ Motivations for formal Systems Engineering approach and architecture 
development}

The motivation behind requirements engineering is to get a clearer view of systems 
behavior with environmental changes.  Knowledge engineering is needed to formally 
understand each task performed and its `why' in the system.  Each phase of the formal system engineering process has its own motivation factor, which has been presented, in the following sections.

\subsection{ Motivations for Knowledge Acquisition:}

\begin{enumerate}
\item At the design center the people involved in executing the system manually 
best understand detailed working of the design center. Sessions of knowledge acquisition are necessary to understand the details of present day working of the domain.  Such an interaction is needed to have better understanding of the domain and its interaction between different subsystem such as order tracking and billing.   For example, the different processes followed in custom order and in-stock order of an item.

\item Knowledge Acquisition sessions are needed to understand the present level 
of automation at Design Center.  Parts of the domain are already automated like the 
inventory management and customer information tracking system. There is also a need 
to understand what kind of improvements and changes can be made for a better-integrated design center automation system.

\item Knowledge experts of this domain shall be one of the end users of the system. Therefore, the requirements of the system will be better understood in Knowledge Acquisition sessions.  This will help in setting boundaries around the problem and the solution space and gather the end user viewpoint of the system.

\item To understand the interface and interrelation between each subsystem (inventory management, customer service) such as ordering the items from vendor and markdowns of the items lost, stolen or damaged.  
\end{enumerate}

\subsection{ Motivation for Domain Modeling:}

\begin{enumerate}
\item · ·It assesses crucial timing and resource requirements for subsystems such 
as delivery.  The timing requirements being important for store's profitability and 
customer satisfaction. 

\item · ·To formally model the relationship between the subsystems and entities such as customer, vendor, inventory, measurements call and delivery of goods for this domain.
\end{enumerate}

\subsection{ Motivation for Reference Architecture:}

\begin{enumerate}
\item · ·Reference Architecture serves as a blueprint to the developers of the automation system of design center.  It defines the domain space and helps in deriving the implementation specific solution for Design Center at Home Depot.

\item · ·The reference Architecture is crucial for system maintenance because it 
serves as a map to the implementation and is easier to refer for a system developer 
who is joining new or modifying the system after it has been implemented.  It can 
be referred to, for new product additions at the design center of Home Depot.   Such addition could be the part of maintenance and/or future system upgrade.

\item · ·Reference Architecture for design center at Home Depot can be used for other domain implementations, say design centers at JCPenny, Sears and Lowe's.
\end{enumerate}

\newpage

%% Knowledge Acq.
%% Section Change
\chapter{ Knowledge Acquisition}

\section{  Overview}

Our team uses SEPA to distinguish the functional requirements and implementation 
specific requirements in order to develop the application for the designer at Home 
Depot with continuous evaluation and tractability.

This is obtained by:

\begin{enumerate}
\item · ·Identifying the major activities within the Design center through interactive observation.

\item · ·Understanding the current level of automation and Technology used at the 
Design center

\item · ·Creating an observation table for tractability.

\item · ·Filtering the collected data to generate specific scenarios to develop a 
data driven system

\item · ·Planning KA sessions on the generated scenarios using analysis techniques 
such as

\item · ·Work process Analysis - To conduct observation session to identify the high level tasks accomplished by the performers in the scope of a given scenario.

\item · ·Task Analysis - To investigate and classify tasks involved in carrying out a particular job function identified in a specific scenario and evaluate it against the mental 
model

\item · ·Temporal Analysis - To identify the duration, frequency and critical factors to perform a particular task.    

\item · ·Conceptual Analysis - To understand the definition and relationship of one object to another. It is also used to characterize attributes of resources and relationships between data elements. 

\item · ·Scheduling KA session based on domain expert availability, adequate time 
for the session and the type of KA technique used.

\item · ·Confirming session (date, time and place) with the domain expert(Designer) 
to conduct formal Interview.

\item · ·Motivate the domain expert to provide input for the KA session, in order 
to design a system that helps the expert to do job better.     

\item · ·Providing a sample memo to the domain expert before the KA session, so that expert has thought about the scenario prior to the meeting.

\item · ·Analyzing and documenting the process using flows and templates.

\item · ·Reviewing the domain mental model and the requirements outlined.

\item · ·Constant evaluation in each phase for completeness and accuracy.
\end{enumerate}

\subsection{ Observation worksheet to document activity}

Performer name: Julie, Senior Designer

\begin{longtable}{|p{0.9in}|p{0.7in}|p{1.0in}|p{0.7in}|p{1.4in}|} \hline 
Activity & Location & Information & Resources & Key Problems \\ \hline 
Customer likes to buy a product from design center & Design center at Home Depot, 
Round Rock & Customer can access the information from web or from the catalogs & Web and catalogs & Not enough information provided  \\ \hline 
Customer selects wallpaper in stock or custom orders & Design center at Home Depot, 
Round Rock & Information to the customer is provided by the catalogs and the designer & Catalogs and Designer & The product selected is out of stock or the design center does not deal with the vendor anymore \\ \hline 
Designer needs to co-ordinate with the vendor to get the details(features) on the 
selected item  & Design center at Home Depot, Round Rock & Information to designer 
is provided by the vendor & vendor & Designer may not be able to reach the vendor 
to provide the details on the product useful to the customer  \\ \hline 
Customer orders 5 rolls of wallpaper  & Design center at Home Depot, Round 
Rock & Customer provides information to designer & 5 rolls of wallpaper & 5 rolls 
of wallpaper do not exist. System is not updated to display the actual stock or the product was not ordered or delay in product delivery or unexpected 
need for the product \\ \hline 
Designer orders the product using the mobile cart & Design center at Home Depot, 
Round Rock & Designer enters the information in the mobile cart & Mobile cart &  \\ \hline 
The system displays 5 rolls of wallpaper but only 4 exists, the 5th roll is torn  & Design center at Home Depot, Round Rock & Designer marks down the product to 0 and enters the information on a mark up and marked down sheet & The mark up/mark down sheet & Can lose the sheet and hence the information \\ \hline 
Customer walks in and decides to buy the instock blinds at the design center & Design center at Home Depot, Round Rock & Designer gives an option of different types of blinds and the price range & Catalog and the Designer &  \\ \hline 
Customer selects the Instock blind and decides to buy it  & Design center 
at Home Depot, Round Rock & Designer enter the customer information and generates 
an invoice with the selected products and related SKU\# & Dos program  to enter customer 
information and generate an invoice &  \\ \hline 
Customer pays the bill and is satisfied with the product & Design center at Home 
Depot, Round Rock & Customer pays the bill  & Customer , designer and the front desk 
employee who is in charge of the bill &  \\ \hline 
The mobile cart program updates the entry for the product sold in 24hours & Design 
center at Home Depot, Round Rock & How Information gets updated in the system and 
mobile cart & Mobile cart &  \\ \hline 
Customer decides on custom --order blinds & Design center at Home Depot, Round Rock & Designer provides information on different types of blinds available and price range on the various kinds of blinds & Catalog, customer and designer &  \\ \hline 
Designer takes the information regarding the measurements of the windows to generate an exact price range & Design center at Home Depot, Round Rock & Price range varies on different type of blind and on different vendors and every type of blind has separate feature for which pricing is different & Vendors catalog and Designer and front end DOS application & The vendors information is not available on the system, the designer has to go through all the catalogs and note down the  different features manually, the designer may forget to add a option or may enter certain data incorrectly \\ \hline 
The 
designer generates a invoice for customer to pay the bill & Design center at Home 
Depot, Round Rock & Customer pays the bill  & Customer, designer and the front desk 
employee &  \\ \hline 
Customer requests for measurement and installation  & Design center at Home Depot, 
Round Rock & The designer co-ordinates with the person incharge of measurement/installation and  schedules an appointment with the customer & Customer, designer, Person incharge of measurement installation & No proper co-ordination \\ \hline 
The designer sends the custom order requirements to the vendor and notes the delivery date & Design center at Home Depot, Round Rock & Customer requirements send to the vendor & Designer,vendor &  \\ \hline 
\caption{Observation Worksheet}
\end{longtable}

The above table was used to document the activities observed at the design center.  This was obtained by our frequent visits to the design center and also observing the designer performing these activities. The table also contains the resources used and the problems encountered at the center. Based on these activities we generated the scenarios for this domain.

\section{ KA Team Organization}

The KA team comprises of :
\begin{center}
\begin{enumerate}
\item Anshuman Sinha
\item Haritha Nandela
\item Vijaya Balakrishna
\end{enumerate}
\end{center}

Each member of the team has equal responsibility and has similar roles as the other 
team member.   The responsibilities for the Knowledge Acquisition were shared among 
the team member on the basis of scenarios.  Each team member performed the knowledge acquisition and the documented the report for the specific scenario.

\section{ Scenarios driving KA effort}

The scenarios for the domain are generated based on the Interactive Observation conducted at the design center. Refer to table-1 for Observation worksheet. There are many activities that takes place in the design center, we are mainly focussed on the customer interaction with the design center. They are

\begin{enumerate}
\item · ·Product selection by the customer that is instock or custom-order

\item · ·Customer requesting quote to make a decision on the product

\item · ·Customer buying the product

\item · ·When customer returns the products

\item · ·The process used by the design center if the customer requests for measurement 
and installation
\end{enumerate}

Based on the above topics our scenarios are generated and they traverse through the 
domain mental model. These scenarios are also generated based on the complexity of 
each task.

 The scenarios driving KA effort are

\begin{enumerate}
\item Customer buys In stock product (Blind or wallpaper)

\item Customer selects product from catalog and requests for a quote

\item Customer buys a custom order product

\item Customer requests for measurement, installation and delivery.

\item Return of products by customer
\end{enumerate}

\subsection{ Scenario Descriptions}
Scenarios described below represents the activities the performers at Home Depot. 
Each scenario focuses on certain aspects of the domain thus generating requirements 
for the project. Scenarios described below represents the activities of the performers at Home Depot. Each scenario focuses on certain aspects of the domain thus generating requirements for the project.

\subsection{ Scenario \#1 Customer buys In stock product}
Scenario Description: Jane has just bought a new house recently and she would like 
to have window treatments. Jane new home needs blinds for 18 windows. She needs to 
fix the windows as soon as possible because she is having a tenant come in next week. Jane feels that white color aluminum blinds would suit her home but she would like to confirm her idea. She takes the measurements for the all the windows and walks into Home Depot design center and approaches the designer. Jane gives details regarding her home and her ideas to Julie the designer. The designers takes Jane to the display where all of blinds that are in stock are on display and asks her look at them and decide on the type she likes. Jane ponders over the details and decides that maybe aluminum blinds would be right for her needs and that cream color would match the wallpaper and would look better. They look at a couple of choices that are in stock taking into account the price and other features like color, style etc. Jane agrees that white color 1-inch blinds would suit her purpose better. Julie looks up in the inventory and sees that they have Bali blinds in standard size in stock. The blinds are not available for some window sizes and Julie offers to cut the blinds for Jane. After Julie cuts the blinds, Jane loads up her cart with all the blinds walks to the check out desk and the clerk scans in all the items. Jane pays for the items bought and loads the items into her vehicle and takes them home.

Scenario Objective: To understand the process involved in customer selecting an in 
stock product and also buying the product.

Justification: In this particular scenario we are be considering the customer buying a product be it either wallpaper or Blinds that is currently in the warehouse. The process of dealing with customers who buy in stock product is different from a custom order product. Therefore there is a possibility of identifying additional tasks not identified in other scenarios. In this scenario we will cover the following work processes of the mental model namely the Select product, Process Order, Buy Product. Refer to figure 2.1

Figure-2.1-Depicts the mental model for scenario\#1

\subsection{ Scenario \#2 Customer requests quotes on the product selected from catalog}
Scenario Description: Cynthia walks in to design center, Home Depot with measurements for all her windows in her house to buy blinds. She has no prior selection on the type of product. The designer presents her the various choices available at the design center, like the Instock products or the catalog products. She explains to the customer the advantages of each type of product, the various options the vendor can provide, the different type of vendors available. Based on this information she selects the product from catalog and requests a quote from the designer. Based on the quote obtained she makes her decision whether to buy or not to buy the product.

Scenario Objectives: To understand the process involved in customer approaching the 
designer after selecting the product from catalog, designer generating quote based 
on customer measurements for the type of product selected.

Justification: This scenario traverse through one of the main tasks ``Generate Quote'' of the domain mental model where the initiating event is customer selecting a product from catalog and approaching the designer for quotes. The designer understands the customer requirements and generates quotes as requested. Based on these quotes the customer makes a decision to buy the product. The text represented in bold in the domain mental model below, figure-2.2 represents this part of the scenario.

Figure-2.2-Depicts the domain mental model for scenario\#2

\subsection{ Scenario \#3 Customer buying the custom order product}
Scenario Description: Cynthia gets a quote from the designer based on the products 
she selected from the catalog.  She can now make a decision based on this quote. 
She takes about two days, and finally makes her decision to buy the product and returns to the design center at Home Depot to buy it. The designer confirms with her that she has the right measurements and asks her where she wants the product to be delivered. If Cynthia has not taken correct measurements, the designer asks her to re-measure and gives enough information how to do it.  Cynthia wants it to be delivered at Home depot and she would like to collect it from home depot. Designer generates an invoice based on her selection and tells her about the delivery date, and Cynthia pays for the product at the cash registry. Once the product arrives at the home depot, the designer calls Cynthia and asks her to collect her product.

Scenario Objectives: To understand the process involved in customer approaching the 
designer to buy the product and the designer generating invoice for the customer 
and the customer paying the bill.

Justification: This scenario traverse through one of the main tasks ``Buy product'' 
of the domain mental model where the initiating event is customer receiving a quote 
from the designer and making a decision to buy the product. The designer generates 
the invoice for the customer and the customer pays for the bill. The text represented in bold in the domain mental model below, figure-2.3 below represents this part of the scenario.

Figure-2.3-Depicts the Domain mental model for scenario\#3

\subsection{ Scenario \#4 Customer selects an instock product and requests for measurements, installation and delivery.}

Scenario Description:  Itsy comes to Home Depot Design Center to buy regular blinds 
for her new home in Town Lake.  With the help of the designer she chooses one of 
the blinds that is in stock for living room of the house.  She is not sure of the 
measurements of windows in living room and wants the measurements to be taken by 
the measurer of the Design Center.  She chooses another type of blinds from the catalog, for all other rooms on ground floor of her duplex house.  The measurements of all the windows in those rooms are also to be taken by the measurer.  Itsy will be out of town for the next couple of days and wants the measurer to come only after she is back in town.  She would prefer the following Wednesday after 6:00 p.m. after she gets back from work.  When the measurements are taken she will come back to Home Depot and pay for the blinds. She also wants the blinds to be installed by the installer.  The installer should contact her to make an appointment and agree on suitable time and date. She would like the installation to be done in the weekend.  The blinds are too big to fit in the car and therefore she needs it to be delivered by Home Depot.  She is ready to pay extra for the delivery and installation.

Scenario Objective:  To understand the process involved in taking measurements, installation and delivery of items bought by the customer.

Justification:  The design center offers the service of measuring the site by a suitable person for its customers.  To achieve customer satisfaction and to deliver the right quantity of product in correct size, taking measurements becomes an important activity for the design center and needs to be detailed.  This scenario focuses on this activity to gather all requirements for this process.  The scenario also covers the installation and delivery of an in stock product.  Taking measurements, delivery and Installations are not dependent on the variables like the make of the product and its stock availability.  This scenario pictures a position that happens quite regularly at the design center and the automation system will need to service to such requests. Refer to figure 
2.4 below.

Figure-2.4-Depicts the Domain mental model for scenario\#4

\subsection{ Scenario \#5 Customer returns a product.}

Scenario Description :  Teene comes to Home Depot in Round Rock customer service 
desk with a set of wallpapers that was purchased by her husband for their new house 
in Canyon Oaks.  Wean, her husband, bought the wallpaper at the Home Depot on Research Blvd last month.  She wants to return a part of the purchase made by Wean and buy another set of wallpapers for their living room.  If she doesn't find something that she likes at Home Depot design center, she will try other places in town.  The customer service desk accepts the returned items and forwards it to the designer at the design center.

Scenario Objective:  To understand the process involved in customer returning an 
instock product.

Justification:  The most important goal at Design Center is to sell the product that is suitable to customer's needs and the performers constantly try to achieve this by improving the system.  Home Depot not only tries to achieve this but also has the flexibility of accepting the return of items when the product does not meet customer's specification and satisfaction. This scenario tries to trace how the product return is handled with respect to design center.  Speaking in terms of the effect of return of goods on billing and inventory, the subject is out of the scope of domain.  However, the designer at the design center is responsible for accounting the returned goods and making a decision on returned product in case of in stock product. This scenario tries to trace the process, which happens at the design center after the item has been handed over to the designer by the front desk. Refer to figure 2.5 below

Figure-2.5-Depicts the Domain mental model for scenario\#5

\section{ KA Plan}

\begin{longtable}{|p{0.2in}|p{1.2in}|p{0.7in}|p{0.9in}|p{1.0in}|p{1.0in}|p{0.6in}|} \hline 
S. 
No. & KA Session Title & Domain Expert & Date/\newline Duration & KA Technique & Scenario\\ \hline 
1. & Customer buys Instock product & Jimmy & 2/28/00, 5.00p.m\newline \newline 1.5 
hrs. & Work process Analysis, Task Analysis & \#1 \\ \hline 
2. & Customer buys In stock product & Jimmy & 3/04/00,\newline 11.00a.m\newline \newline 1.5 
hrs.\newline  & Temporal Analysis, Conceptual Analysis & \#1 \\ \hline 
3. & Customer 
selects a custom-order product & Julie & 3/06/00\newline 6.00p.m\newline 1.5 
hrs. & Work process Analysis, Task analysis, Temporal Analysis, Conceptual Analysis & \#2 \\ \hline 
4. & Customer requests quote on the product selected from catalog & Julie & 2/29/00\newline 5.00p.m\newline \newline 1.0 
hrs. & Work process Analysis  & \#2 \\ \hline 
5. & Customer requests quote on the product selected from catalog & Julie & 3/2/00\newline 10.00a.m\newline \newline 1.0 
hrs. & Task Analysis & \#2 \\ \hline 
6. & Customer requests quote on the product selected from catalog & Julie &3/6/00\newline 5.00p.m\newline \newline 1.0 
hrs. & Temporal Analysis, Conceptual Analysis & \#2 \\ \hline 
7. & Customer buys the Custom order product & Julie & 3/14/00\newline 10.00a.m\newline \newline 1.15hrs & Work 
process analysis, Task Analysis & \#3 \\ \hline 
8. & Customer buys the Custom order product & Julie & 3/20/2000\newline 6.00p.m\newline \newline 1.0 
hr. & Temporal Analysis, Conceptual Analysis & \#3 \\ \hline 
9 & Customer requests for measurements. & Lisa & 3/04/00\newline 7.00 
p.m\newline \newline 1.0 hr & Work process analysis, Task Analysis, Temporal Analysis & \#4 \\ \hline 
10 & Customer requests for delivery, installation and return of products. & Lisa & 3/18/00\newline 5.00 
p.m\newline \newline 1.5 hrs. & Work process analysis, Task Analysis, Conceptual Analysis & \#4, \#5 \\ \hline 
\caption{ Knowledge Acquisition Plan}
\end{longtable}

\newpage

%% Knowledge Session Reports
%% Section Change

\chapter{ KA Session Reports}

\section{ KA Session \# 1}

\subsection{ Session Information}
Scenario Employed: Scenario \#1: Customer buys an in stock product.
Domain Expert: Jimmy Rice  
Title: Senior Designer 
Employer: Home Depot  
Background Information: Jimmy has been working at HD for the last three years and 
he is very experienced in the design center business. He helps the customer select 
product, processes the order for them and makes their visit to HD a pleasurable one. He is also responsible for all the number crunching that takes place at the design center. He has a couple of associates who assist him and who look up to him for guidance.
Knowledge Engineer: Haritha K Nandela
Session Time and Date: 5.00 p.m.  02/28/00
Session Location: Training room at Home Depot, South Austin
Session Purpose/Goal: To ascertain enough information for the work process Select 
Product, Process Order and Buy Product for an in stock product so that the different tasks and sub tasks in the process can be identified.

\subsection{Session Input} 
\begin{enumerate}
\item · ·The KE needs to read the textbook titled User-Centered requirements and 
class notes so as to prepare herself for the different type of questions to be asked. 
\item · ·Before the session the KE needs to determine the techniques she is going to be using for 
the coming session. 
\item · ·The KE also has to prepare a questionnaire that would help her in analyzing the work process and tasks. 
\item · ·The KE needs to fax a memo to the DE. Please refer Fig 2.7 on the next page.
\item · ·The KE also has to arrange for specific time and place to meet the Domain Expert.
\end{enumerate}

\subsection{Session Output}

\begin{enumerate}
\item · ·An overall understanding of the work process Select Product, Process Order 
and Buy Product and gain information regarding the resources and that data required 
for those processes.

\item · ·Identify different sub tasks and tasks for the work processes.
\end{enumerate}

KA Technique(s): Work Process Analysis, Task Analysis.

Type of Knowledge: Procedural knowledge

\subsection{Planned KA Questions}

\begin{enumerate}
\item Is scenario  \#1 a valid scenario?

\item Who are people involved in this process?

\item Would you like to change anything in process currently available?

\item For the given scenario what do you after customer walks in?

\item How do you extract the information from the customer to help him in choosing 
the product?

\item How do you approach a person trying to buy wallpaper?

\item Are there different grades of wallpaper? What are they?

\item If the customer knows the color of wallpaper that he likes how you direct 
them are the wallpaper books sorted by color?

\item What do you base the selection of wallpaper for a customer who has come 
to buy in stock?

\item How does the customer decide how much wallpaper he needs?

\item For an in stock product after the selection of the product what's next?

\item If the customer has windows requirements that you do not have in stock for example if he needs 15 and you have only ten of them what do you do?

\item Do you generate any quotes for in stock blinds?

\item Do you generate any invoices for in stock blinds?

\item For in stock product do you tell the customer how to install the blinds 
and do they need accessories?

\item Is there a difference in an individual homebuyer and a commercial buyer?
\end{enumerate}

Session Report Date: 03/11/00

\textit{To: Jimmy Rice Senior Designer Home Depot} 
\textit{From: Haritha K Nandela}

\textit{Hello Jimmy,
I was wondering if it would be okay with you to meet on 02/28/00 at 5.00 p.m. for may be an hour or so. I would like to discuss with you about the following process at home depot. I will be covering the ordering of a product that already available in stock. I will be doing it for blinds as well as wallpaper. If you have any questions please contact me at this number 512-723-8445.}

\textit{Thanks}
\textit{Haritha}

\begin{enumerate}

\item Is scenario  \#1 a valid scenario?
A:  The process is correct but for scenario \#1, you might need to change a couple 
things like the customer would buy aluminum blinds instead of wooden. The reason 
why the customer buys the in stock aluminum blind is because they want something 
for short term and something that easily replaceable. They might be landowner as 
opposed to homeowner. As a general rule of thumb homeowners who intend to stay for 
a long time would invest more time and money in buying blinds but someone looking 
for commercial reasons would go for something cheap and fast.

\item Who are people involved in this process?
A:  Definitely the homeowner, designer, vendor, installer and the cashier.

\item Would you like to change anything in process currently available?
A:  Something that I would like to change is that if it were possible to link 
the different Home Depots so that I could access the information regarding inventory at different locations. Also if I could access the data of different vendors so that I can verify that they will be able deliver the requests I make before I give the customer any kind of estimates etc.

\item For the given scenario what do you after the customer walks in?
A: When customer walks in I try and understand from the information what her goal 
of coming to home depot is. After she tells me that she is planning on installing 
blinds for her home which she is going to rent out to tenants, I realize that she 
is looking for something cheap and fast. I direct her to the display where we have 
different kinds of blinds on display. Since she would like something she could pick 
up and walk out, I'd suggest her to look at aluminum blinds because we have a lot 
of sizes and in that type and she might find all windows that she needs. Then she'd 
select the color and then depending on the window sizes I'd cut them the blinds if 
required.

\item How do you extract the information from the customer to help him in choosing 
the product?
A:     We call that process qualifying the customer. What it means is we need to 
know who, what, when, where and why. Who is the decision-maker, it's not good if 
your friend comes and tells that she want blinds for your house I'd prefer if it 
were you who comes in. When is the when are you going to buy the product is it going to be one year from now or is going to be next week. Where is where are you and where you would like to install the product. Do you want it in Montana so do I need to refer you to another Home Depot or can I help you. What is what are you trying to do are you? Are you remodeling the house ? or you want new windows for the whole house etc. Why is basically if know what your goals are for your home project. So this is the kind of information I try to gather from the customer and if can get the answers for all these questions then I'm in very good position to help the customer.

\item So what is the approach when someone come in to buy in stock blinds?
A: Well, first I'd like to know their purpose and then I'd direct them to the display from where they can decide what kind of blinds they want. For example horizontal, vertical etc.

\item How do you approach a person trying to buy wallpaper?
A:   Some of the customers do have an idea what kind of wallpaper they want if they 
do not I ask them to walk around and see the different types in stock and see if 
they like any of them. If they do not find anything interesting then I ask them to 
look into the wallpaper books. Where there are different sections for contemporary, 
kitchen, bathroom etc., which might aid them in choosing the product.

\item Is everything that is in stock on the shelf or is it somewhere in the warehouse?
A:  Everything is normally on the shelf other wise it might be in the overhead but everything is in the store.

\item If the customer knows the color of wallpaper that he likes how you direct 
them are the wallpaper books sorted by color?
A:  No the wallpaper books are not sorted by color but you find the color you want 
in every book. The first thing the customer needs to decide on is the style of wallpaper that are contemporary, traditional, kitchen and bath. Then the books are divided based on the section like for example if the customer wants cowboy wallpaper. I'd direct them to the country section and there they would find the cowboy wallpaper in beige color.

\item Are there different grades of wallpaper? What are they?
A:  Yes, there are different grades of wallpaper. They are Solid vinyl, Vinyl coated and Embossed. The solid vinyl is very durable and can be used in kitchen. The vinyl coated is has a thin layer of vinyl over the paper and the last one is the embossed wallpaper that is wallpaper with puffy impressions. The embossed in very expensive is not very durable. We suggest the solid vinyl because they are more durable.

\item For an in stock product after the selection of the product what's next?
A:  After the customer selects the product then I ask them if they have made a color selection. After they have decided I will go see if the list of window sizes they need are in stock. If the sizes do not match we can cut the blinds based on the measurements given by the customer. When done I load them on to the cart and send to them to the cash register.

\item If the customer has windows requirements that you do not have in stock 
for example if he needs 15 and you have only ten of them what do you do?
A: For in stock product we can cut the blinds in stock give them to the customer 
for the size he needs. We first can look at the overhead if its available or check 
if its available at other stores or we can tell him when they are going to in stock 
and ask him to come back again. They cannot pay ahead and come back and collect them

\item Can the associates help the customer with the in stock product?
A: Absolutely, the first thing the associates are taught is to cut the mini blinds 
and know the differences between the types. The in stock is what we sell most but 
it's not the place we make more money.

\item How do you cut the blinds? Do you have any equipment?
A: Yes we have a mini blind cutter that is given to us by Bali, we use it to cut 
the blinds. We cut I-inch aluminum and vinyl and we cut only Bali brand and no other.

\item Do you generate any quotes for in stock blinds?
A: Normally I do not. The customer himself or herself walk around look up the prices and the sizes they need and make the quote. For example for Bali blinds the there is list on the shelf that says 30x60 is 17 dollars etc. Sometimes if the customer needs some help we help them generate a quote or do it for them.

\item Is there an additional task if the product is wallpaper?
A: There is an additional task for wallpaper because the note on the wallpaper says 
that just dip it in the water and stick it. We recommend that they use the glue .We 
have a whole wall lined with different glues. We tell them how prime the wall, use 
the glue and hang the wallpaper. So I spend some time explaining to the customer 
the whole process.

\item How does the customer decide how much wallpaper he needs?
A:  Well the easiest way to decide is calculate the square footage for the walls 
in the room and deduct the doors and windows which would give an approximate square 
footage. Each roll has 56 square feet so from that I calculate and tell them how 
many rolls they need to buy.

\item Do you generate any invoices for in stock blinds?
A: No we do not do any invoice generation for in stock products

\item Can the customer take hand samples for the in stock blinds? What is the 
process for wallpaper?
A:  No we do not give sample books to customer to take home. But the wallpaper is 
on the stands and the customer can tear off a piece of the wallpaper and take it 
home to see it is a good selection.

\item What do you base the selection of wallpaper for a customer who has come 
to buy in stock?
A: The wallpaper for kitchen sells the fastest. But I'd ask them questions like the 
color of the cabinets, light or dark wallpaper, wall space etc. Then I'd direct them to the right section.

\item What are the types of blinds that you have in stock?
A:  In horizontal blinds I have 1-inch vinyl and aluminum and 2-inch vinyl and in 
vertical I have different styles in plastic.

\item For in stock product do you tell the customer how to install the blinds 
and do they need accessories?
A: Yes, I give them quick run down how to do it. The product has instructions on 
how to install it so that will help them. They do not need any accessories to install the blind they might need a drill but that's about it.

\item Is there a difference in an individual homebuyer and a commercial buyer?
A: Not really. The only thing I can think of is that the commercial customer might have more windows to cut and pull off the shelf which might take some time as opposed to a homeowner who might need only some windows.

\item What are different ways we can pay at home depot?
A: At home depot you can by cash, check or card. Any kind of credit card is okay.

\item How would you schedule the associates?
A: What I do is make sure that a fresh associate is not left all alone. And that 
there is some senior person with her all the time.

\end{enumerate}

\section{ KA Session \# 2}

\subsection{Session Information}
Scenario Employed: Scenario \#1: Customer buys an in stock product.
Domain Expert: Jimmy Rice  Title: Senior Designer 
Employer: Home Depot  
Background Information: Jimmy has been working at HD for the last three years and 
he is very experienced in the design center business. He helps the customer select 
product, processes the order for them and makes their visit to HD a pleasurable one. He is also responsible for all the number crunching that takes place at the design center. He has a couple of associates who assist him and who look up to him for guidance.
Knowledge Engineer: Haritha K Nandela
Session Time and Date: 11.00 am 03/04/00
Session Location: Training room at Home Depot, South Austin

\subsection{Session Purpose} To ascertain the temporal and conceptual information for the work process Select Product, Process Order, Buy Product for an in stock product as to derive time line diagrams and concept models.

\subsection{Session Input}

\begin{enumerate}
\item · ·The KE needs to read the textbook titled User-Centered requirements and 
class notes so as to prepare herself for the different type of questions to be asked. 

\item · ·Before the session the KE needs to determine the techniques she is going to be using for the coming session. 

\item · ·The KE also has to prepare a questionnaire that would help her in analyzing the work process and tasks. 

\item · ·The KE faxes a memo to the DE. Please refer to fig 2.8

\item · ·The KE also has to arrange for specific time and place to meet the Domain 
Expert.
\end{enumerate}

\subsection{Session Output}

\begin{enumerate}
\item · ·An overall understanding timing issues and concept related to the different work processes Select Product, Process Order and Buy Product.

\item · ·Identify for each task in the work processes the pre and post conditions, 
task frequency and task duration.
\end{enumerate}

KA Technique(s): Temporal Analysis, Conceptual Analysis.

Type of Knowledge: Semantic knowledge

Session Report Date: 03/12/00

\textit{To: Jimmy Rice Senior Designer Home Depot}
\textit{From: Haritha K Nandela}

\textit{Hi Jimmy,}
\textit{I was wondering if it would be okay with you to meet on 03/ 04/00 at 11.00 am for may be an hour. As before I will be covering the ordering of a product that already available in stock. But this time I will concentrate on the timing issues and concerns and also the any other terms that are used in this process. If you have any questions please contact me the number is 512-723-8445.}
\textit{Thanks}
\textit{Haritha}

\begin{enumerate}
\item How much time you spend in the selection process compared to the whole 
process of buying the product?
A:    The time depends if the customer wants a product that in stock or if he wants 
a custom order. For in stock product I'd just walk them to the display and show the 
product and let them select the product so I may be spend not more than one minute 
or so. If they want some help in color selection I might take some more time but 
otherwise I will spend a lot time. Normally a customer who wants an in stock product is not really interested in spending a lot of time at HD.  Compared to buying a special order product that you will be living in for the rest of your life to buying an in stock product the time spent is very less.

\item What is frequency of customers coming to buy?
A:   We have customers coming in to buy all the time the frequency is more during 
the weekend, evening. I normally have time at the end of the day to spend with the 
customer.

\item Of the customers coming how many really buy a product?
A: There are different customers coming into the design center about 50\% about of 
them have an idea of what they and just want you to confirm their decision and they 
buy the product and are on their way another 25\% want to you tell them how to install the product and the rest do not know anything about the product they are about buy and need to hand held all the way through.

\item If a customer does buy a product what is the most sold product?
A:  The in stock blinds are the most sold however it is not the most money making 
product the specialty order are the most money making sections.

\item Do you get any sort of reward for making sales?
A:  Not really. But at the end of the year all the number are drawn up for all the 
sales and the returned goods that kind of gives an idea how much money I made for 
the company. And gives an overall picture for the performance review.

\item Whom do you consult for questions regarding the product?
A: Normally I would consult the vendor for question I do not have answers.

\item What are different levels of expertise at design center?
A: Yes, Currently we have new associates who a week old. So we have different levels. The levels are beginner, moderate and expert.  A beginner with six months of experience can start to help the customer with a lot of efficiency

\item How long does it take to cut the blinds?
A: It generally depends on the number of blinds. Normally it takes about 4 minutes 
per blinds and that's what we tell the customer.

\item Can the associates do the cutting of blinds?
A: Yes they can. Normally they freeze up if the customer asks special order questions. But they can do other tasks like cutting the blinds.

\item What is frequency of cutting the blinds when a customer is buying an in 
stock product?
A:  For in stock blinds we have many sizes in stock for the popular brands. So may 
be half the time we might need to cut the blinds. So 50\% of the time the customer 
can just pick the product and walk to the cashier's desk. This is because for an 
in stock product you will not get a perfect fit as it is with a custom order blind.

\item Before you cut the blinds do you ask the customer to pay ahead of time?
A: No. After we cut the blinds they take the product to the front desk. Sometimes 
we get phone calls from a customer who has already been to the design center and 
they ask for the blinds to be cut and never come and pick them up.

\item How long does it take for a custom to pick, customize and buy an in stock 
product?
A:  It does not take more than twenty minutes for the customer to select, buy and 
take the product and walk away with the out. Since the time to cut each blind is 
four minutes it really depends on the number of blinds that need to be cut. The range might be 20 minutes to a maximum of two hours.

\item If someone wants to buy in stock blinds what is the most critical information 
that the person needs to have?
A:  The most popular type for in stock is the aluminum blinds so if they have the 
measurements then we can help them decide and buy.

\item What kind of measurements do you need to cut the blinds
A:  The instock blinds can never be as good fit as the custom order ones and that 
is why they are so cheap. But we can make adjustments with the in stock blinds so 
we may have a leeway of ¾ an inch. So your measurements need to be plus or minus 
¾ inch.

\item Where do you get data or information to contact the vendor?
A:  We have a price book that has the sku number and it also has the vendor contact 
information.

\item What are the different accessories you can buy for wallpaper? Do you give 
discounts?
A: For the wallpaper you can buy the wallpaper itself and the border and the fabric 
to match. Yes we do give discounts on the product.

\end{enumerate}

\section{ KA Session \# 3}

\subsection{Session Information}
Scenario Employed: Scenario \#2: Customer selects a custom order product.
Domain Expert: Julie   Title: Senior Designer 
Employer: Home Depot  
Background Information: She is working for the Home Depot Design center for the past 4 years. She helps the customer of Home Depot to make the right product selection and generates quotes for them to make a decision on the product.
Knowledge Engineer: Haritha K Nandela
Session Time and Date: 6.00 p.m. 03/06/00
Session Location: Home Depot Design Center, Round Rock, Austin.

\subsection{Session Purpose} To ascertain enough information for the work process Select Product for a custom order product so that the different tasks and sub tasks in the process can be identified.

\subsection{Session Input}

\begin{enumerate}
\item · ·The KE needs to read the textbook titled User-Centered requirements and 
class notes so as to prepare herself for the different type of questions to be asked. 

\item · ·Before the session the KE needs to determine the techniques she is going to be using for the coming session. 

\item · ·The KE also has to prepare a questionnaire that would help her in analyzing the work process and tasks. 

\item · ·The KE faxes a memo to the DE. Please refer to fig 2.9

\item · ·The KE also has to arrange for specific time and place to meet the Domain 
Expert.
\end{enumerate}

\subsection{Session Output}

\begin{enumerate}
\item · ·An overall understanding of the work process Select Product for a custom 
order and gain information regarding the resources and that data required for those 
processes.

\item · ·Identify different sub tasks and tasks for the work process.
\end{enumerate}

KA Technique(s): Work Process Analysis, Task Analysis.

Type of Knowledge: Procedural knowledge, Semantic knowledge

\subsection{Planned KA Questions}

\begin{enumerate}
\item Could you elaborate for me the different steps involved in selecting product 
like blinds?

\item If the customer walking in does not know anything about the blind u show 
him the different types but if has an idea about what he wants then how do approach?

\item So what are the different tasks you do when a customer comes in and says that I'd like blinds?

\item Which are critical factors that are important so as to be successful in 
the selection process?

\item What really helps the customer in deciding from the different types?

\item Are blinds specific to a particular room?

\item How do you make sure the customer choice is correct?

\item How long does it take to select a special order product?

\item When qualifying the customer what is the kind of information you are trying 
to get?

\item If the customer has a specific request what do you do?

\item When I make a selection what do? Do you contact the vendor for availability?

\item If the vendor has discontinued the product? What is your next step?

\item What kind of details do you give the vendor when you asking him for availability?

\item How do you contact the vendor? If you are not able to reach the vendor what do you do?

\item Does the vendor have any delivery date specified?

\item Do you suggest any accessories when the customer selects a blind?
\end{enumerate}

Session Report Date: 03/12/00

\textit{To:  Julie Senior Designer Home Depot} 
\textit{From: Haritha K Nandela}

\textit{Hi Jimmy,}
\textit{I was wondering if it would be okay with you to meet on 03/ /00 at 6.00 p.m. For this session I will be covering the selection process for buying a custom order product. I will concentrate on the requirements for the process and the different tasks involved in the process. If you have any questions please contact me the number is 512-723-8445.}

\textit{Thanks}
\textit{Haritha}

\begin{enumerate}
\item Could you elaborate for me the different steps involved in selecting product 
like blinds?
A:     The first thing I do when the customer walks in is take them to the display 
and show the different types available like the cellular shades, 2-inch and verticals. So I get them to first decide on the type of product and then show them the different brands available in that particular type. Normally I show them three brands because the customer gets confused if you show them more than three brands. Then I tell them the advantages, disadvantages, price, warranties and differences. Finally the decision is driven by the price and the color, whichever brand offers a color the customer wants at reasonable price that's the product they pick.

\item If the customer walking in does not know anything about the blind u show 
him the different types but if has an idea about what he wants then how do approach?
A:    In the beginning when deciding on the type of blinds price is the important factor that helps the customer decide. I pick out a standard size and give the cost for that size for the different types, which helps him decide.

\item What is the approach you take when the customer walks in design center?
A:    First thing we do is try and identify when the customer wants the window treatments or wallpaper so that we can assist him better.  Also understand their project whether it's a remodeling project or a new house that they trying to get window treatments for. Identify how serious the customer in buying the product.  If they are not sure I try and direct them to different books from where they can 
get some ideas. I suggest them some ideas, give them the options they have available. For example I had a customer yesterday who started building I realized that they would not need the window treatment until next year so I just showed them types and asked them to observe window treatments whenever they go anywhere.

\item So what are the different tasks you do when a customer comes in and says 
that I'd like blinds?
A:  I'd first take them and show them different types and give them the hand samples. So the first factor they need to decide is the type because I do not want to spend two hours and then find out that then customer would like another type. So qualifying the customer is the most important thing. Because I'd like the customer to decide what they want or take time to decide what they want. So after the customer decides the type and then I'd take them to the sample books show them the sample books in different brands like the levlor, Bali and vista. The sample books are actual pieces of the blind so that they know exactly what they are getting.

\item Which are critical factors that are important so as to be successful in 
the selection process?
A:   I'd say all the factors of qualifying the customer would be the most significant in the selection process because if I know how much the customer is willing to spend and if I know what he'd like the sale would be very successful.

\item What really helps the customer in deciding from the different types?
A:    Normally I'd give the customer the price and that really would help them in 
deciding. Like for example in Vijaya's case since she wanted arched vertical blinds. I kind of came to know that either she had lot money to spend or that she did not know what she was up against. So when she came in I told her that it would cost her 800 dollars a window and that made her decide that she did not want it So the price is one of the important factors. We really call that bracketing the customer that is nothing but getting a feel of how much the customer is willing to spend.

\item So if the type of product is known and then how do you proceed?
A:    So of someone calls me and says that they'd like arched verticals I call up 
the vendor and find out the price for standard window and give it to the customer 
so then the customer either gives his real measurements or decides the window treatment is too costly.

\item Are blinds specific to a particular room?
A:   Not really but when using a certain kind of blinds in particular it might not 
work out that good. Like if you put cellular in kitchen then cleaning them might 
be difficult they are better in bedroom. When qualifying the customer I'd tell them 
they better not use these in the kitchen. So there are advantages and disadvantages 
which I tell the customer during the selection process.

\item How do you make sure the customer choice is correct?
A:   I tell them that special order cannot be returned that helps in deciding for 
sure what they really want.

\item How long does it take to select a special order product?
A:  For special order product it might take me about half-hour to get all the information that I need to know and give them all the information that they need to know and complete the job.

\item When qualifying the customer what is the kind of information you are trying 
to get?
A: The price, type of product, measurements and the color.

\item If the customer has a specific request what do you do?
A: I'll have to call the vendor and ask them if they can deliver the requested feature and they can accommodate it.

\item When I make a selection what do? Do you contact the vendor for availability?
A:  Yes after I get the selection then I contact the vendor to check if they have it in stock and find how long it will take for them to get it. Normally it takes about 7 to ten weeks.

\item If the vendor has discontinued the product? What is your next step?
A:  If I call up the vendor and find that they have discontinued the product, then 
may be I'll try and see if the other stores have that product if not then I just 
have to ask the customer to select another product.

\item What kind of details do you give the vendor when you asking him for availability?
A: I have to give details regarding each blind. I have to give them the sku number for the product, the measurements and the quantity of the product.

\item How do you contact the vendor? If you are not able to reach the vendor 
what do you do?
A: I contact the vendor by looking up the number in the price book. During normal 
business hours I call the vendor and if I'm not able to contact I leave a number 
to contact me back I also give the information they need. Then I ask the customer 
to either contact me or I contact them.

\item Does the vendor have any delivery date specified?
A: Yes, Each vendor has a different lead time specifies for each product.

\item Do you suggest any accessories when the customer selects a blind?
A:   Everything that the customer needs for the blind comes complete in the order. 
No, I do not suggest any accessories to the customer.

\end{enumerate}

\section{ KA Session \# 4}

\subsection{Session Information}
Session \#4: Designer generates quotes for special order request
KA Report Description: When customer selects a product from catalog and approaches 
the designer, the designer confirms the selection and generates quotes.
Scenario Employed: Scenario \#2, Customer requests quotes on the product selected 
from catalog.
Domain Expert: Julie, Senior Designer 
She is working for the Home Depot Design center for the past 4 years. She helps the 
customer of Home Depot to make the right product selection and generates quotes for 
them to make a decision on the product.
Knowledge Engineer: Vijaya Balakrishna --Interview the designer
Session Time and Date: 5.00 p.m.  Feb 29th
Session Location: Home Depot Design Center, Round Rock, Austin.

\subsection{Session Purpose}
To obtain a detail knowledge about the work process involved 
after the customer selects a product from catalog and approaches the designer for 
quote, and what information the designer needs in order to generate a quote. And 
to understand the responsibilities and accomplishments of the designer to generate 
a quote.

\subsection{Session Input}
To fax the sample memo to domain expert, Julie so that she is prepared for the scheduled session. Refer to figure 2.11 for the memo sent to Julie on this session.  Confirming the session with the performer Julie.  Determining the KA technique to be used.  Preparing the questions on the subject matter to interview the domain expert.

\subsection{Session Output}  To understand the process flow of how the designer understands the customer requirements in order to generate a quote for them.

KA Technique: Work process Analysis

Type of Knowledge: Procedural knowledge. This is used to obtain a detailed knowledge of the task in a step by step method and how each of these steps are linked to accomplish the task.

Interview Plan: The total time 1hr allocated for the Interview is divided between 
the main tasks involved in this KA session. Refer to the figure-2.10 below for planning map.

               35 min            25 min

Figure-2.10-Planning map for KA session\#4

Memo sent to Julie for KA session\#4:
\textit{To: Julie, Senior Designer- Home Depot}
\textit{From: Vijaya Balakrishna, Software Engineer}

\textit{Hello,}
\textit{The Q\&A session which  I mentioned to earlier may take up to 1 hr, we are scheduled to meet you at 5.oo p.m on Thursday, the Feb 29th.  I plan to cover two topics  }
\textit{·	Customer approaches you after making a product selection}
\textit{·	You generating quote for the customer}
\textit{If there is any reference materials which I can see and use to understand the process, it would be very useful.  Thanks a lot for co-operating on this project.}
\textit{Vijaya}

KA session questions:

Task \#1: Designer qualifying the customer

\begin{enumerate}
\item When a customer selects the blinds from catalog and approaches you for quote, 
what do you do next?
A: I ask for the measurements of their windows to generate a quote

\item What if the customer does not have measurements?
A:  I tell them that in order to generate a quote I need measurements, but I give 
them rough estimate on an average window size, from that they can figure out what 
price range they are looking at. Whether it is for 1 window or 10 windows.  But it 
is important for me to know the exact measurements to generate a quote and actually 
we also go to the customers' house and measure their windows

\item Quotes can be generated on an average size window, why do you need measurements?
A:If it is for the entire house, the window size varies and of course the price for the different sizes and I have to know that to give them a quote.

\item What if the customer has measurements?
A: I ask them if they have measured their windows correctly, that is top, middle 
and bottom to 1/8th of an inch and if so give them three different quotes. Refer 
to DOC -1

\item What if the customer has wrong measurements?
A:  Most of the cases it is true, they often tend to measure length and width, but 
most of the windows are not perfect across its length and width and so I suggest 
them to re-measure.

\item Is the correct measurements required in order to generate a quote?
A: Actually not, rough measurements for all the windows is enough for me to generate a quote.

\item Do you tell the customer that Home Depot does measurements?
A: Yes, I tell them that we can measure their house but there is a measurement fee 
of \$25 and if they decide to buy the product then they get a credit for that amount.

\item Once you get the window measurements, do you go ahead and generate a quote?
A:I give them quotes on three different products.

\item What if the customer has made a wrong selection, do you let him/her be aware 
about it?
A: Yes, as a designer I have to understand what the customer really needs. I tell 
them the advantages and disadvantages of all the products and this is done before 
they make the decision on the product. But again the customer would select a product from the catalog which is in trend, like select wooden blinds for all the windows in their house, and wooden blind is not good next to the kitchen sink and in the bathroom next to the tub. So I have to make them be aware about it and give them an alternative, the faux wood which is similar to the wooden blind and is also water resistant.

\item What kind of product details that you confirm with a customer before generating the quote?
A:  It depends. If the blinds selected are 2'' horizontal, then I mention the controls that is associated with this blind -- tilt cord left, lift cord right. And there is an option to reverse the controls. If the blinds  is a 1'' horizontal, then there is wand towards the left and lift towards the right. Again there is a possibility of reversing the controls. Similarly, there are control options for cellular and verticals. I can note this down on a paper for you. Refer to DOC 4

\item Is the process different if the customer selects the wall paper from the catalog and requests a quote?
A: The process is much simpler, I just need to know how many double roles of wallpaper they require. But the rest of the process is same.

Task \#2: Designer generating quotes for the customer

\item Do you generate a quote on the product selected ?
A: Actually I generate quote on three different products.

\item Why do you give the customer three different quotes ?
A: First of all I give them only three different quotes because they are not over-whelmed to make a decision. And I may give them quotes on three different products so that they can differentiate between the price range on the type of product. I may give three different quotes on the same product -- but by three different vendors.

\item On what three different products do you generate the quotes ?
A: It is normally the Real Wood, Cellular, and the Faux Wood for blinds.

\item Why is that these three products are chosen among others ?
A: First of all these three products are the products that are mostly chosen by customers.  Thus they form a trend among buyers. Next, these products offer a good mark-up value to Home Depot.

\item Are there any specific situations where you generate more than three quotes 
?
A: Yes, by this time I know if the customer is really serious about making a purchase and if this also involves more profit for us, then I generate more than three quotes.

\item Do all the vendors make the same kind of product ?
A: Mostly Yes, but there may be some vendors who has limited choices in colors offered by vendors, design to install the blind and the controls, and other things like 1.  Limited choices in tapes for blinds 2.  Limited choices in blind-texture 3.  Limited choices in installing the cords for the blinds

\item Where do you get the necessary information to generate the quotes ?

A: I consult the price book to generate the quote. The price book lists all the vendors that we deal with and all the different products that they manufacture, and also the price range on each of the product on all window sizes. I use this information to generate a specific quote.

\item Is your quote generation process currently automated ?
A: No, I generate a quote on a sheet of paper and give it to the customer.

\item What kind of information does this quote contain ?
A: I list all the window measurements (width * length) and the different vendors, 
Then I list the different prices for each of these combinations.

\item Does the customer decide to buy the product right away?
A: No, in most of the cases, they take this information home to make a decision and 
then let us know if they have decided to go with us, or come back to us with more 
questions. They may also ask for a quote on a different product, by a different vendor altogether.

\item If the customer decides to buy the product, then what is the next step ?
A: I ensure that they have the right measurements before I generate an invoice. 

\end{enumerate}

\section{ KA Session \# 5}

\subsection{Session Information}
Designer generates quotes for special order request.  When customer selects a product from catalog and approaches the designer, the designer confirms the selection and generates quotes.  Scenario Employed: Scenario \#2, Customer requests quotes on the product selected from catalog.  Domain Expert: Julie, Senior Designer She is working for the Home Depot Design center for the past 4 years. She helps the 
customer of Home Depot to make the right product selection and generates quotes for 
them to make a decision on the product.  Knowledge Engineer: Vijaya Balakrishna --Interview the designer.  Session Time and Date: 10.00 a.m.  March 2nd  Session Location: Home Depot Design Center, Round Rock, Austin.

\subsection{Session Purpose} To understand the key strategies the designer uses to perform the task and generate task decomposition and to find out the resources involved in doing the task and sub tasks

\subsection{Session Input}

\begin{enumerate}
\item · ·To fax the sample memo to domain expert, Julie so that she is prepared for 
the scheduled session. Refer to figure 2.13  for the memo sent to Julie on this session  

\item · ·Confirming the session with the performer Julie

\item · ·Determining the KA technique to be used.

\item · ·Preparing the questions on the subject matter to interview the domain expert.
\end{enumerate}

\subsection{Session Output} 
To Identify the task and subtasks involved in designer generating quote for the customer.

KA Technique: Task Analysis

Type of Knowledge: Procedural knowledge. This is used to obtain a detailed knowledge of the task in a step by step method and how each of these steps are linked to accomplish the task.

Interview Plan: The total time 1hr allocated for the Interview is divided between 
the main tasks involved in this KA session. Refer to the figure-2.12 below for planning map.

 Memo sent to Julie for KA session\#5:

\textit{To: Julie, Senior Designer- Home Depot}
\textit{From: Vijaya Balakrishna, Software Engineer,}

\textit{Hello,}
\textit{This follow up Interview is to understand the }
1. Tasks and subtasks involved in qualifying the customer
2. Tasks and subtasks involved in generating the quote
\textit{I think this Interview would take approximately 1 hr to 1hr 15 minutes. We have agreed to meet on March 2nd Tuesday at 10.00a.m. Thanks a lot for your time.}
\textit{Vijaya}

KA session questions:

SubTask \#1: Designer qualifying the customer

\begin{enumerate}

\item What is your goal when customers approach you with their selections for a quote ?
A: I have to qualify the customer and generate a price quote which makes them buy 
the product.

\item What is the key accomplishment for you to perform this job ?
A: It is to qualify the customer. That is to understand the customer requirements 
so that I deliver what they really want.

\item Do you have customers approaching you for request all the time or is there any specific time of the day when they approach you?
A: We have customers walking into the design center all the time. But we are busy 
during the evenings and the weekends.

\item What is the prerequisite for you to generate a quote ?
A: I need the window measurements, type of product, and the color of the product, 
to generate a quote.

\item How much time does it take for you to generate a quote?
A: It may vary from 5 minutes to 30 minutes.

\item Why does the time vary ?
A: It depends, sometimes the customer may not be sure of the product that they have 
chosen. It also depends on the number of windows and the number of quotes that I 
am generating.

\item Does your task change if the customer has made a wrong selection ?
A: Yes, I explain to them the disadvantages of the product that they have chosen 
and they have to start the product selection process all over again.

\item Do you need to be experienced in order to handle all the tasks after a product is selected from the catalog?
A: Selecting a product from a catalog is a special order request and the designer 
has to know the details of the catalog products. If the designer is unable to handle all the tasks, then he/she would consult the other experienced designers in fulfilling this special order request.

\item Does the number of years of experience matter ?
A: On an average we need to have at least 6 months of experience. Every day we are 
asked new questions that  we have never been asked before.

\item Does your task change if the product they have selected involves one window 
or ten windows ?
A: No, it does not change. But it takes more time to generate generate the figures.

\item Does your task change if the customer does not have measurements ? 
A: Yes, I can give a rough estimate on one window size. I ask them to come back with precise measurements in order for me to generate a quote.

\item What is the average window size ?
A: It is 35' * 64' (width * height)

\item How do you generate a price based on this window size?
A: We have a price book which we refer to, in order to find the price listings for 
all different products, vendors and window sizes.

\item How do you present this quote to the customer ?
A: I write this information on a sheet of paper and give it to the customer.

\item If the customer has not measure their windows, what kind of assistance do you 
offer to the customer?
A: We tell them that we can measure the windows for them, but it would cost them 
\$25, which would be adjusted towards their payment if they decided to buy the product with us. Or I instruct them on how to measure the windows.

\item How do you instruct the customer to measure their windows?
A: I tell them to measure top, middle and bottom to 1/8th of an inch -- and as width * height

\item If the customer has measurements, does your task change?
A: Yes, in this case, I just have to generate a quote based on the product selected. I generate quotes for three different products. If they are firm on one particular product, I generate a quote on the same product -- but manufactured by three different vendors.

\item Does your task change if the customer buys a product to be installed in a house, apartment or business?
A: No, the process followed is the same in all the three cases.

\item Are there any situations where you approach your supervisor in order to do this task?
A: No, there are no such situations.

\item Are there any other persons involved it executing this task?
A: If the customer requests measurements to be done by us, then I have to contact 
the measurer to give the customers information.

\item Do you always have enough information to execute the task?
A: No, not always. The customer will select a product from a catalog and request 
for a special design. To fulfill this request, I have to contact the vendor of this 
product to find out if they carry the design.

\item What if you are not able to contact the vendor?
A: I record the customer information and later on contact the vendor. I then pass 
this information along to the customer at a later time.

\item Do you require customer information in order to generate a quote?
A: No, for generating a quote I do not need customer information. I need customer 
information only when decides to buy the product.

\item Does the amount of time spent on the customer depend on the time of the day?
A: I normally spend more time at the end of the day because I will be done with all 
the administrative tasks.

\end{enumerate}

\section{ KA Session \# 6}

\subsection{Session Information}
Designer generates quotes for special order request

KA Report Description: When customer selects a product from catalog and approaches 
the designer, the designer confirms the selection and generates quotes.

Scenario Employed: Scenario \#2, Customer requests quotes on the product selected 
from catalog.

Domain Expert: Julie, Senior Designer 

She is working for the Home Depot Design center for the past 4 years. She helps the 
customer of Home Depot to make the right product selection and generates quotes for 
them to make a decision on the product.

Knowledge Engineer: Vijaya Balakrishna --Interview the designer (contact Information)

Session Time and Date: 5.00 p.m. March 6th

Session Location: Home Depot Design Center, Round Rock, Austin.

\subsection{Session Purpose} To analyze the task frequency, the task duration, the conditions to execute the task, the sequence of tasks, and to find out the resources involved in doing the task and sub tasks

\subsection{Session Input}

\begin{enumerate}
\item · ·To fax the sample memo to domain figure 2.15  for the memo sent to Julie 
on this session expert, Julie so that she is prepared for the scheduled session. 
Refer to 

\item · ·Confirming the session with the performer Julie

\item · ·Determining the KA technique to be used.

\item · ·Preparing the questions on the subject matter to interview the domain expert.
\end{enumerate}

\subsection{Session Output} To identify the terms and resources used to do the task, timing constraints in doing the task and the duration of each task.

KA Technique: Temporal and Conceptual Analysis

Type of Knowledge: Semantic knowledge. To understand the major concepts, definition 
and the vocabulary used in the domain and the relationship between the tasks and 
the subtasks

Interview Plan: The total time 1hr allocated for the Interview is divided between 
temporal analysis and the Conceptual analysis of scenario\#1. Refer to 2.14 below 
for planning on KA session

                30 min            30 min

Figure-2.14-Planning map for KA session\#6

 Memo sent to Julie for KA session\#6:

Figure 2.15: Memo for KA session\#6

KA session questions:

Temporal Analysis
\begin{enumerate}

\item In how many cases do the customers come into the design center with the right 
selection?
A: In most of the cases the customers are not sure of the selection until they buy 
the product.

\item In how many instances do the customers come into the design center with both 
product and also has the measurements?
A: 80\% of the time they do not have the measurements.

\item How much time do you spend with a customer to make the right product selection?
A: It varies to either 5 minutes to 30 minutes or it may take few days.

\item What resources must be in place before you generate a quote?
A: I need to have the measurements of the windows, color of the product, type of 
product and the window treatment selected.

\item What is the output of this task?
A: Generating three different quotes for the customer based on the selections made.

\item How much time does it take to generate a quote?
A: It may take from 5 minutes to 30 minutes, based on the number of windows and different vendors and products.

\item What is the frequency of generating quote for a custom order product?
A: About half the time we encounter custom order product selections.

\item Is there any contingency in doing your task?
A: I cannot generate a quote if the customer does not have any measurements.

\end{enumerate}

Conceptual Analysis:

\begin{enumerate}

\item What do you mean by qualifying a customer?
A: To understand the requirements based on who, what, when, where, and why. 

\item How important is this to generate a quote?
A: On a scale of 1 to 10, it is a 10 because this makes me realize what the customer really wants and how I can help the customer make the right selection in order to buy the product.

\item What is the term used for selecting a product from a catalog?
A: It is called a special order product.

\item Will all the designers at the design center be able to help the customer involving a special order request?
A: The designers need to have atleast a minimum of 6 months experience in order to 
help the customer with special order products.

\item Can the quotes be obtained over the telephone or internet?
A: Currently we do not have the process of obtaining quotes over the internet and 
we do not encourage giving quotes over the telephone.

\item What is the measurement of the average window size?
A: It is 35' * 64' (width * height).

\item How do you instruct the customers to measure the windows?
A: I instruct the customer to measure top, middle, bottom 1/8th of an inch and also 
give the customers, vendor catalogs, which explains how to measure windows with pictures. There are catalogs by different vendors like: DesignView, Levolor. These catalogs have illustrations of the different products that they manufacture and how to take measurements for their windows with pictures.

\item What do you refer to in order to generate a quote for the customer?
A: I refer the price book which lists all the vendors, the different types of products that they manufacture, and window sizes, and the price for each window size.

\item Do all window sizes are entered in the price book?
A: The price for the window sizes are rounded to the next highest even integer.

\item How do you present the quote to the customer?
A: I write the quote on a sheet of paper and give it to the customer. Refer to DOC 
2

\item What kind of information does the quote contain?
A: The quote contains window measurements for the different products and their price range.

\item What are the components of the horizontal blinds?
A: The components are: 
The headrail is used to hold the slats. The horizontal blinds can come with or                                                                                                                 without valance. And the valance to cover the headrail.

\item What are the components used with Vertical Blinds?
A: The components are: 
The headrail is used to hold the vanes. The vertical blinds can come with or                                                                                                                 without valance. And the valance to cover the headrail.

\item What are the controls available with horizontal blinds?

A: It depends. If it is a 2'' blind, then the controls are tilt cord left, lift cord right. Or it may be the opposite. If it is a 1'' horizontal, it has wand on the left, and the lift cord on the right. Again the opposite is possible. If this is a cellular blind, there is a choice for the lift cord to be either left or right.

\item What are the controls associated with the vertical blinds?

A: There is a choice to stack the blinds towards left, right or split it to either 
side. Controls are always on the same side of the stack. The wand is used to tilt 
as well as traverse.

\item When do you realize that you need to generate an invoice?

A: If I am sure that customer is sure to buy the product and has accurate measurements, than I generate an invoice.

\item Why do you stress on customers getting accurate measurements ?

A: This concerns special order products, and the customers cannot return these special order products. The vendors will go ahead and deliver the blinds with whatever measurements that the customer supplies. If the measurements are wrong, the vendors will not accept the product back. Therefore, we stress that the customer measurements be accurate. We cast a shadow of doubt to the customer by making them sign a form which states that the measurements they have taken are accurate. Only after that we order the product. Refer to DOC 6

\item Will the customers never be able to return the product? 

A: Actually Home Depot has a policy to accept all returns. So we do accept return 
order products, but we do not let the customer know about it. This is because we 
cannot return the blinds to the vendor and we have to mark down the product, thereby incurring a loss.

\end{enumerate}

\section{ KA Session \# 7}

\subsection{Session Information}
Session \#7: Designer generating invoice on the product selected

KA Report Description: After the designer is sure that the customer has made the 
right selection and also has the right measurements and is also ready to buy the 
product, than she would go ahead and generate the invoice

Scenario Employed: Scenario \#3, Designer generates invoice based on customer selection.

Domain Expert: Julie, Senior Designer 

She is working for the Home Depot Design center for the past 4 years. She helps the 
customer of Home Depot to make the right product selection and generates quotes for 
them to make a decision on the product.

Knowledge Engineer: Vijaya Balakrishna --Interview the designer                                                 

Session Time and Date:10.00 a.m. March 14th

Session Location: Home Depot Design Center, Round Rock, Austin.

\subsection{Session Purpose} To analyze the process the designer follows to do the tasks, to decompose these tasks and also to note the resources and the equipment used to accomplish this task.

\subsection{Session Input}

\begin{enumerate}
\item · ·To fax the sample memo to domain expert, Julie so that she is prepared for 
the scheduled session. Refer to figure 2.17 for the memo sent to Julie on this session  

\item · ·Confirming 
the session with the performer Julie

\item · ·Determining the KA technique to be used.

\item · ·Preparing the questions on the subject matter to interview the domain expert.
\end{enumerate}

\subsection{Session Output} To identify the tasks and subtasks involved when designer generates quote for the customer

KA Technique: Task Analysis

Type of Knowledge: Procedural knowledge. This is used to obtain a detailed knowledge of the task in a step by step method and how each of these steps are linked to accomplish the task.

Interview Plan: The total time 1hr allocated for the Interview is divided between 
two topics, one the designer generating the invoice and the other customer paying 
the bill.

 Memo sent to Julie for KA session\#7:
\textit{To: Julie, Senior Designer- Home Depot}
\textit{From: Vijaya Balakrishna, Software Engineer, Siemens}

\textit{Hello,}
\textit{As discussed earlier, In our next Interview I would like to cover two topics 
1. Tasks you do to generate an invoice for the customer on a special order product
2.  and customer paying the bill}
\textit{I think this Interview would take approximately 1 hr 15 min . We have agreed to meet on march 14th Tuesday at 10.00a.m. Thanks again.
Vijaya}

KA session questions:

\begin{enumerate}

\item When do you know you need to generate the invoice for the customer?
A: When I am sure that customer has the right measurements for all the windows and 
has the right product selection, than I generate the invoice for them.

\item How do you generate the invoice?
A: I have an semi-automated system wherein I have to key in all the product selections, the prices for each selection and the measurements.

\item Do all designer have access to the system?
A: Yes, all designers have an access to the system, If we hire new designer the computer guy generates an user-id for him/her

\item When do you input the customer information?
A: This is one of the very first tasks that I do when I am sure that the customer 
is going to buy the product. I may do this even before I generate a quote, depending on how sure I am about the customer buying the product. But normally I do it after the customer decides to buy the product.

\item Can you explain in detail about the current invoice generation system?
A: I can use the system by entering my user--id and password, next it brings up a 
screen where in I can look up in to the existing orders or to create a new Invoice, 
when I select to create a new invoice, I need to enter the customer information, 
after I enter this information, it brings up a invoice form where I have to input 
measurements for each window, type of product, vendor information, the controls selected 
-- three times for every window one for the Receiving Dept, the installer, and the 
vendor. But if the customer has requested for measurements I'll have to type in twice 
-- one for the vendor and the other for the receiving department at Home Depot, so 
that the latter will have information about the product that we ordered to the vendor.

The invoice generated looks as if the information was automated, but I have manually typed all the information on every line.

\item How much time does this take?
A: This takes me at least one hour if I need to generate an invoice for the all th 
windows in the house. This process is time consuming and I really wish if this could be made much simpler. I always advise that the customer to come back after an hour to collect the invoice.

\item From where do you look up the information in order to enter it into the system?
A: I have the handwritten selections and measurements which I need ensure that it is entered correctly into the system.

\item Is there a possibility of entering wrong data?
A: Yes, there is always room for error when entering these kind of details about 
the product.  This may lead to ordering the wrong selection.

\item Will it help if the system gives you a choice to pick the product, vendor, etc.   when you generate an invoice
A:  Of course yes ! It will help us and save time.

\item Do you have any idea on how the current system could improve?
A: Kind off, I can write it for you. Refer to DOC 3

\item Can you explain this?
A: Yes, first of all I would like the screen to display the available vendors and 
than the type of the products they make and the type of fabric and I must be able 
to pick the window measurements width * length instead of typing it. After this the 
system must give an option to generate quote or stage for register. If its only quote, I don't have to enter customer information but if it is stage for register it must give an option to enter customer information and than must give an option to select the color and controls for the type of product and than the Information to where the customer wants the product to be delivered and than to print the invoice.

\item After you enter all the product information, what is the next thing you do? 
A: There is an option to generate quote, or stage for register. The stage for register is used to print the invoice with the total price on it.

\item After you generate the invoice, what happens?
A: First of all the customer has to pay for it, next the information typed to the 
vendor is faxed automatically by the system and also to the Home dept receiving dept, and also to the measurer if he is going to measure the windows.

\item What kind of details go to the vendor?
A: Most of the details such as the measurements, the controls and the type of product with the respective SKU number

\item What is the SKU number?
A: Every type of product such as horizontal, vinyl, cellular belonging to a particular vendor has a SKU\#. And SKU stands for Stock keeping Unit.

\item What details go to receiving dept?
A: This is the vague description of the product. They are not concerned about the 
controls of the product or the color, they just need to know what product, the quantity and the vendor information, so if they receive the product, they are sure that it is the one we ordered.

\item What information goes to the measurer?
A: All the information we send to the vendor like the size of the window, the controls, the color, etc.

\item What do you do after you generate the invoice?
A: yes I give it to him or her, there is a UPC code printed on the invoice. I walk 
with him to the cash register and the cashier at the cash register scans the UPC 
code and the customer pays for it.

\item Do you do any verification with the customer before they pay for the product?
A: Yes I verify the scheduled date for delivery and tell them that it may vary a little and than again confirm the price on the product before they pay for it.

\item If the customer has a discount coupon for a home depot product, can be taken 
care in the invoice?
A: Yes I can do it or the cashier at the cash registry can also do it. In the system there is an option to mark down the product, I have to enter this discount information in that place.

\item Do you give the discount on the total price?
A: It is not possible, because I can give discount only on the product and not on 
any services like measurement and installation and the total amount includes all 
charges, so I look up at the invoice and mark down, that is give a discount accordingly.

\item Are there any particular payments method that Home Depot follows?
A: We accept cash, cheque and all credit cards.

\item What are the critical factors for generating the quote?
A: The custom-order measurements, on the product selected -- the amount of light 
that has to be blocked, how often the blinds are raised or lowered, and the color 
match.

\item If you have to generate an invoice for custom-order wallpaper, is it the same 
process?
A: Yes, it is much simpler because I just have to enter a SKU number and the quantity. There are no other details involved in this case. It can vary from 1 double-rolled wall paper to any number of double-rolled wallpaper.

\item Once the bill is paid by the customer, do you still interact with the customer?
A: It depends. If the customer requests delivery to be made to their house, and if they are satisfied with the product, they will not call me, but the product is delivered to Home Depot, than I will call them once I get the product. But mormally they call me if there is a problem with the product.

\item If you need to access the customer information, how do you do it?
A: Our computer system provides an option when I login, to see the existing orders, 
I look up in to a particular order by the customer telephone number. That brings 
up their invoice, which I can use to access the necessary information.

\end{enumerate}

\section{ KA Session \# 8}

\subsection{Session Information}
Designer generating invoice on the product selected.  KA Report Description: After the designer is sure that the customer has made the right selection and also has the right measurements and is also ready to buy the product, than she would go ahead and generate the invoice. This session is focussed on the timing issues, the duration, the terms used and the pre-conditions and the post conditions to do the task.

Scenario Employed: Scenario \#3, Designer generates invoice based on customer selection.

Domain Expert: Julie, Senior Designer 
She is working for the Home Depot Design center for the past 4 years. She helps the 
customer of Home Depot to make the right product selection and generates quotes for 
them to make a decision on the product.

Knowledge Engineer: Vijaya Balakrishna --Interview the designer (contact Information)

Session Time and Date: 6.00 p.m. March  20th

Session Location: Home Depot Design Center, Round Rock, Austin.

\subsection{Session Purpose} To analyze the process the designer follows to do the tasks, to decompose these tasks and also to note the resources and the equipment used to accomplish this task.

\subsection{Session Input}

\begin{enumerate}
\item · ·To fax the sample memo to domain expert, Julie so that she is prepared for 
the scheduled session. Refer to figure 2.5.19  for the memo sent to Julie on this 
session  

\item · ·Confirming the session with the performer Julie

\item · ·Determining the KA technique to be used.

\item · ·Preparing the questions on the subject matter to interview the domain expert.
\end{enumerate}

\subsection{Session Output} To identify the terms and resources in the tasks and the pre and post conditions of the tasks and the frequency of each task.

KA Technique: Temporal and Conceptual Analysis

Type of Knowledge: Semantic knowledge. To understand the major concepts, definition 
and the vocabulary used in the domain and the relationship between the tasks and 
the subtasks

Interview Plan: The total time 1hr allocated for the Interview is divided between 
two topics, one the designer generating the invoice and the other customer paying 
the bill. Refer to figure 2.18 for the KA planning on session 8

 Memo sent to Julie for KA session\#8:

\textit{To: Julie, Senior Designer- Home Depot}
\textit{From: Vijaya Balakrishna, Software Engineer, Siemens}

\textit{Hi Julie,}
\textit{I would be covering the same topic as discussed last time, but now in this session I'll be covering the details such as frequency, the duration to do the task and the terms and resources used in doing this task. We have decided to meet on March 20th at 6.oo p.m. Thanks again for your time.
Thanks,
Vijaya}

\subsection{Temporal Analysis}

\begin{enumerate}

\item How long does it take you to generate an invoice for blinds?
A: based on the number of windows, I may take from 15 minutes to 1 hr. this is because I have to key in the same information three times for each window.

\item How long does it take you to generate an invoice for wall paper?
A: it may take up to 5 to 15 minutes because no details are involved

\item What is the precondition for you to generate a Invoice for the blinds?
A: The right measurements and the right product selection. The customer needs to 
be sure of his selections

\item What is the precondition for you to generate a Invoice for the wall papers?
A: The quantity of double roles they require and of course the selected product

\item What are the resources used to generate the invoice?
A: I use the computer system to type in the details of the product selected, and to 
generate the Invoice. I'll use the price book to verify the SKU\# and the price for 
the particular measurement, and I'll look in to the form where I have customer measurements.

\item Do you use the same resources for generating invoice for the wallpaper?
A: Mostly yes, but I don't need the measurements, I use the price book for SKU\# 
and the price list, and in the same screen in the system I'll enter the quantity 
and the SKU\#.

\item What are the different tasks that take place after you generate the invoice?
A: I hand over the invoice to customer to pay, after the customer pays the bill, a fax is automatically faxed to the vendor by the system and also to the receiving department and if the customer has requested for Installation, a fax is also sent to him automatically.

\item  Are there any problems you encounter while doing this task?
A: Yes sometimes, like in one case one of the designer gave a wrong quote for the 
customer, that is actually for a lesser price and once we realized this we called 
the customer, but customer wanted to the product on the earlier stated quote.

\item How do you handle these problems?
A: In this case we gave a discount of 20\% and did not charge the measurement fee.

\item Do you generate the invoice with the lesser price?
A: We generate the invoice on the actual price but we give a discount on certain 
products which is reflected on the invoice.

\item What if the manufacturer is not able to deliver the product on time?
A: This happens some times, that is the reason we give the delivery date between 
2 to 3 weeks.  But if it extends even this time frame, than the vendor will notify 
me and I'll call up the customer and let them know. The time frame for delivery is 
not in my hands, I always make this very clear to the customer and tell them again 
about it when they are about to pay the bill.
\end{enumerate}

\subsection{Conceptual Analysis}

\begin{enumerate}

\item What does SKU stand for?
A: It is the stock keeping unit, It can be for all products in the case of instock 
product or it on the type of product for a particular vendor for a special order 
product

\item Do all the designer generate an invoice for the special order product?
A: No they need to have at least 6 months of experience to do this on their own because of the details involved, but they can always do with assistance from the more experienced designer.

\item Are there any situations where is the less experienced designer does not have 
any assistance from the more experienced designer?
A: We always make sure that there are at least two designers at all times, one to 
be well experienced and the other person may be new in the store.

\item How do you schedule the delivery date?
A: if the customer has only requested the products but not installation, than we 
tell them 2 to 3 weeks, actually the system automatically generates the scheduled 
date for delivery based on the vendors and the product selected. But if the customer has requested for installation, than we give them a time frame between 3 to 4 weeks. This is because once the Installer gets the products, depending on his schedule he'll install the products.

\item What is the role of the Home depot receiving department?
A: The receiving department needs to know what products have been ordered from  all 
the departments at Home Depot, because if they receive the shipment they must be 
in a position to verify it.

\item What are the details that you enter in the system?
A: The vendor name (Like Gerber, Bali, Levolor), type of product that the vendor 
selected (like verticals, real wood 1'', cellular, faux wood) and their respective 
SKU numbers, type of fabric (like solitaire, midnight), measurements of the windows, customer color and control choices, and then I do Stage-4 register.

\item What is Stage-4 register?
A: This means I have entered all the selections and I am now generating the invoice. This will also print an invoice report which I give to the customer.

\item Do you have a sample invoice that I can document?
A: Yes, with customer permission. Refer to DOC 5

\end{enumerate}

\section{ KA Session \# 9}

\subsection{Session Information}
KA Session Title:  Discuss the process at design center, to entertain customer's 
request for measurement.
Scenario Employed: Scenario \#4
Domain Expert: Lisa   Title: Systems Manager  Employer: Home Depot Background Information: Lisa has been working at Home Depot for the last seven years and is experienced in handling retail problems and managing the staff working at the Design Center.  She is managerially responsible for all the tasks performed at the Design Center.   Lisa also decides on the policy and productivity matters.  Session Time and Date:  3/04/00   7:00 p.m.  Session Location: Training room at Home Depot, Round Rock

\subsection{Session Purpose} To find out details of activities performed at the Design Center to entertain and process the customer's request to have the site measured by measurer of Home Depot.

\subsection{Session Input} The Knowledge Engineer schedules an appointment with the domain expert and prepares for the session.  He reads and tries to acquire more information on the techniques that can be used in this session.  A list of questionnaire was prepared by the knowledge engineer and sent to the domain expert so that the domain expert is familiar with the process and the type of questions that will be posed to them during this session.  Arrangements were made to audio tape the session.

\subsection{Session Output} A detailed understanding of the work processes involved around processing customer's request for taking measurements of the site.   To outline any differences in the way selection of a product is processed if the customer comes without the measurements.

\subsection{KA Technique(s)}
Work Process Analysis, Task Analysis, and Temporal Analysis

Type of Knowledge: Procedural knowledge

Session Report Date:  03/04/00  5:00 p.m.

Report:

\begin{enumerate}
\item What does the designer do when they come across a customer who doesn't 
have measurements?

A:  The designer would either give a sample blind from the stock or ask the customer to get the measurements and come back again to the Design Center.  If the customer has some idea about how to take measurements the designer would hand over a sample blind.  In case the customer has no idea about the size of window, the designer will teach the customer how to take measurements of the windows and ask for an accuracy of 1/8th inch. Its important to have the exact measurements for the blinds before the designer cuts the blind.  The designer would also let customer know about the measurement and installation program that the design center offers on an extra charge.

\item What do you do if the customer doesn't want to go for measurements program?
A: The designer takes the customer's information like the name, address and the phone number.  The designer gives the customer a measurment form that has to be filled and signed by the customer.  The customer comes back with the measurements and the designer processes the order further.  The designer usually doesn't qualify the customer if they don't have accurate measurement.  Opens a request that is faxed to the measurer. The customer goes to the front desk and pays for the service.  The measurer contacts the customer and fixes an appointment and turns up at the site and takes the measurements of the necessary, windows etc.

\item What happens after the customer decides to go for the measurements and installation program?
A:  After the customer decides for the service the design center employee takes customer's personal information in the system and the SKU of the item they intend to buy. If the record of their information pre- exists in the system, the designer verifies the information. The designer prints the information and an invoice is generated which is handed to the customer.  The customer takes the invoice goes to the cashier at the front desk and pays.  As soon as the customer's payment is received a request is opened for the customer and fax is sent to the measurer.  The measurer calls the customer within two business days and sets up a time that is convenient for both.  After the measurer measures, he faxes the measurements back to the designer. The designer generates the quote for the product and installation.   The quote has all the details about the product and the exact sizes.

\item Does Design Center have its own measurer?
A:  Yes, the measurer and the installer work only for the design centers of Home 
Depot.  He covers all the Home Depots in Austin and has to entertain service requests from all locations.

\item How does the designer convey special requests of timing that comes from 
customer to the measurer?

A:  The designer enters the special timing request from the customer in the system 
that is faxed to the measurer.  For instance, the customer could be out of town for 
two days etc. or the measurer might be busy on that day.  The designer does not have any control over the measurer's schedule and it is between the customer and the measurer to agree on a suitable date and time.  The measurer has an agreement with the Home Depot to contact the customer within two days.

\item In case of custom order, how does the designer send the measurements to 
the vendor?

A:  The designer faxes the measurements along with the product type and details to 
the vendor and the product is delivered to the installer who goes and installs the 
product at the site.

\item What if customer would like to go only for installation and not for measurements or vice versa?
A:  The customer can choose either for `measurements and installation' or `measurements only'.  As a policy the installer doesn't install products, which was not measured by him.  This is because he can't guarantee the installation if the measurements have not been taken correctly.  The home depot follows this policy.

\item Where are measurements recorded after it is received from the measurer?
A:  After designer receives the measurements form measurer it is keyreced into the system.  All the measurements along the customer information are recorded in the system.
\end{enumerate}

\section{ KA Report  \#10}
\subsection{Session Information}
Session Title: To discuss the process of installation, delivery and return of 
products.  Scenario Employed: Scenarios \#4 and \#5  Domain Expert: Lisa Title: Systems Manager Employer: Home Depot.  Background Information: Lisa has been working at Home Depot for the last seven years and is experienced in handling retail problems and managing the staff working at the Design Center.  She is responsible for all the tasks performed at the Design 
Center.   Lisa also decides on the policy and productivity matters, which concerns 
customer's satisfaction.  Session Time and Date: 3/18/00  5:00 p.m.  Session Location: Training room at Home Depot, Round Rock

\subsection{Session Purpose} To find out the details about the way customer's request for installation is processed.  

\subsection{Session Input}
The KE schedules an appointment with the domain expert and prepares 
for the session.  He reads and tries to acquire more information on the techniques 
that can be used in this session.  A list of questionnaire was prepared and sent 
to the domain expert to keep the domain expert familiar with the process and the 
type of questions which will be posed to them during this session.

\subsection{Session Output}
A detailed understanding of the work processes involved around processing customer's request for taking measurements of the site.  To outline any differences in the way selection of a product is processed if the customer comes without the measurements.  

KA Technique(s): Work Process Analysis, Task Analysis, and Temporal Analysis

Type of Knowledge: Procedural knowledge

Session Report Date: 03/19/00

Report:

\begin{enumerate}
\item How does the customer come to know about the installation program?

A:  There are banners in the design center that inform the customer about it.  It 
says \textbf{Professional installation available}.  For the details of the program the customer will have to approach the designer.  The designer explains about the program to the customer and lets him know the rates charged by the design center.

\item How is the information about the delivery of the product passed to the 
delivery team?  

A:  For a custom order the options are indicated to the vendor.  The customer has 
three delivery options of either accepting the delivery at home or to the installer.  The product can also be delivered at the Home Depot.  The front desk usually handles this.  Sometimes they buy other items from Home Depot too and ask for combined delivery.  The delivery is made by a special department which takes care of all this, I am not sure about the detailed working of this department.

\item What happens after the vendor has delivered the product. (In case of custom 
order).
A:  The installer installs the product only if the measurements were taken by Home 
Depot.  This is because the installer can't guarantee the installation if the measurements were not taken correctly.  The measurer approaches the customer to fix up a suitable time and goes and installs the product.

\item What is done to product returned by the customer?
A:  The customer returns the product and receives the money back at the service desk.  At the end of each day the designer of the design center picks up the returned product from the front desk.  The designer inspects each product.  If the product is an in stock product and if its condition is good, it's placed back on the shelf.  If the product is an ex-stock product, an RTV (Return to Vendor) is generated.  Then the designer marks down each product.   The returned items are marked down to zero as opposed to a certain percentage of the value in case of discounts. The instock items which are not in condition to be placed back on shelf is trashed.
\footnote{ The return of product is not part of the domain and therefore not modeled.  Refer to the domain mental model.}

\item What do you mean by marking down of the product?

A:  Mark down means lowering the value of the product.  Markdowns are done in two 
cases.  First, to give the customer some discount at the time of purchase.  This 
is nothing but discount.  Second, to account for the lost, damaged or returned product.  In the second case, the goods are marked down to zero, meaning the product has been lost completely in value.  There is a register wherein the details of the product SKU/UPC Number and item descriptions are recorded.  The reason for damage is also recorded.  (A sample form is attached with the report).
\footnote{ Same as above.}

\item Can customer pay for the product before the measurements are taken, to 
avoid coming again to Home Depot?

A:  Yes. The customer can pay by his credit card in that case because the exact measurements are not taken and therefore the exact amount can't be calculated.  This happens quite often, the customer gives his credit card number and the expiration date of the card to the designer and after the designer gets the measurements from the installer he/she charges the credit card of the customer.

\item Does Home Depot deliver in stock product?

A:  Yes. Home Depot delivers the product to customers.  The customer goes to the 
cashier and pays for the delivery along with the payment for the product and the 
goods are taken out in the trucks to the customer's location.

\item How does the delivery work in case of custom order?

A:  In case of special order, at the end of order the designer has to select the 
destination of delivery.  There are three options of shipping the product delivered.  The destination could either be Home Depot or Installer or to the customer.  By this time designer knows where the product has to be shipped and picks the correct response.   The customer could have either selected of a self-installation or for installation by the Home Depot installer.  If the customer requested for delivery to their home address the designer takes down the instructions from the customer.  This includes the day and time they want the items to be delivered.  The designer also asks if the customer is comfortable with the items being left at the doorstep in case they are not present at the time of delivery.  In case they are not, the goods are delivered to Home Depot and the customer comes and picks up the product from Home Depot.  The Home Depot staff informs the customer when the items arrive.  As the third option, the items can be delivered to the installer and the installer schedules with the customer for installation.

\item How does Home Depot track the orders delivered to Home Depot for its customers?  And how do they inform the customer?

A:  The Inventory department receives the order and informs the service desk and 
design center about the receipt.  Service desk calls the customer and informs them 
that the order has been received.  The customer comes to the service desk with the 
receipt of payment made to pick the items.  The system is updated when the customer 
picks up the order.

\item Does the process of selection change if the customer doesn't have measurements?  

A:  The customer is advised by the designer to get the measurements before designer can help them.  The designer tells the customer how to take measurements so that they can get the right measurements.  Often customers return with the measurements.  The measurements are important in product selection.  The designer gets an idea what the customer's site is like and is able to help sell an appropriate product to the customers needs.  This is one of the ways to qualify the customer.

\item What are the changes that can be made in the system to improve it further?

A:  Every invoice has an estimated date of arrival at the end of it.  The estimated 
arrival date could vary depending on the type of product and the specialty requirements, which goes along with it.  For e.g. the arches are takes time to cut, pack and deliver.  The estimated date of arrival often confuses the customer and wrongly sets the customers expectation.  Another problem that occurs is the breakdown of communication channel between the design center and the vendor.  The fax is either lost on the way or the receptions are not clear.  The designer doesn't come to know about the loss of information during the communication.  It will be good to have a better communication where the design center employees can check the status of any order.  The designers would like to know if the customer received the product and also find out if the vendor had received the communication and is processing the order at their end.  The design center employees would like to keep track of certain orders and the system should be able to provide them with some kind of automation to achieve this, so that the designer can act whenever there is a communication breakdown.

\end{enumerate}

\newpage
%% Knowledge Models
%% Section Change
\chapter{ Knowledge Models }

\section{ Overview}

The knowledge models are representations of the information obtained in the KA sessions regarding the problem domain. An atom of information in the domain would focus on a concept in the domain, it could be a task, resource, state, event, or a relationship. The definition of an atom would include the name, synonyms, description etc similar to a domain dictionary. These atoms can be tasks, resources, states, events, or relationship. A collection on the other hand would provide the structure and relationships between atoms of information. Examples of collections are task templates, task hierarchies, sequence diagrams, state charts, etc.  Following are the different kinds of models that we will be deriving from our KA sessions.  Here is the brief description of these models.

\begin{enumerate}
 
\item \textbf{Event Trace: }An event trace will represent the interaction between resources along a timeline this interaction may be with regards to data, events or tasks.

\item \textbf{Task Decomposition: }This model represents the task breakdown in the work process and identifies all the subtasks for each super task. This diagram represents all the tasks that take place in the problem domain. It also shows the relation between two tasks, like what are a task's super task or its sub tasks.

\item \textbf{Task Hierarchies: }The task hierarchy diagrams will specify the different ways to perform work process or tasks.

\item \textbf{Task Templates: }The task templates will have all the information regarding a task. It consists of details such as the frequency, duration, resources, input-output, pre and post conditions and its relation to other tasks.

\item \textbf{Concept Hierarchy: }The concept hierarchy models represent the 'kind of' relations between the resources.

\item \textbf{Concept Maps: }These models are also known as the entity relation diagrams. These models are used to characterize attributes of resources and relationships between data elements.

\item \textbf{Domain Dictionary: }The domain dictionary is a definition of a concept in the domain. 

\item \textbf{Document Analysis: }The document analysis has the document resources that were used for modeling.

\end{enumerate}

\newpage
\section{ Models}

\subsection{ Model: Designer\_CA\_01}

Knowledge Model Number: Designer\_CA\_01

Model Description: Conceptual Hierarchy of Resources

KA Source: KA Session Report \#1

Model Perspective: Designer

Date: 04/03/00

\begin{figure}[h]
\centering
\fbox{\includegraphics[width=5in,height=4in]{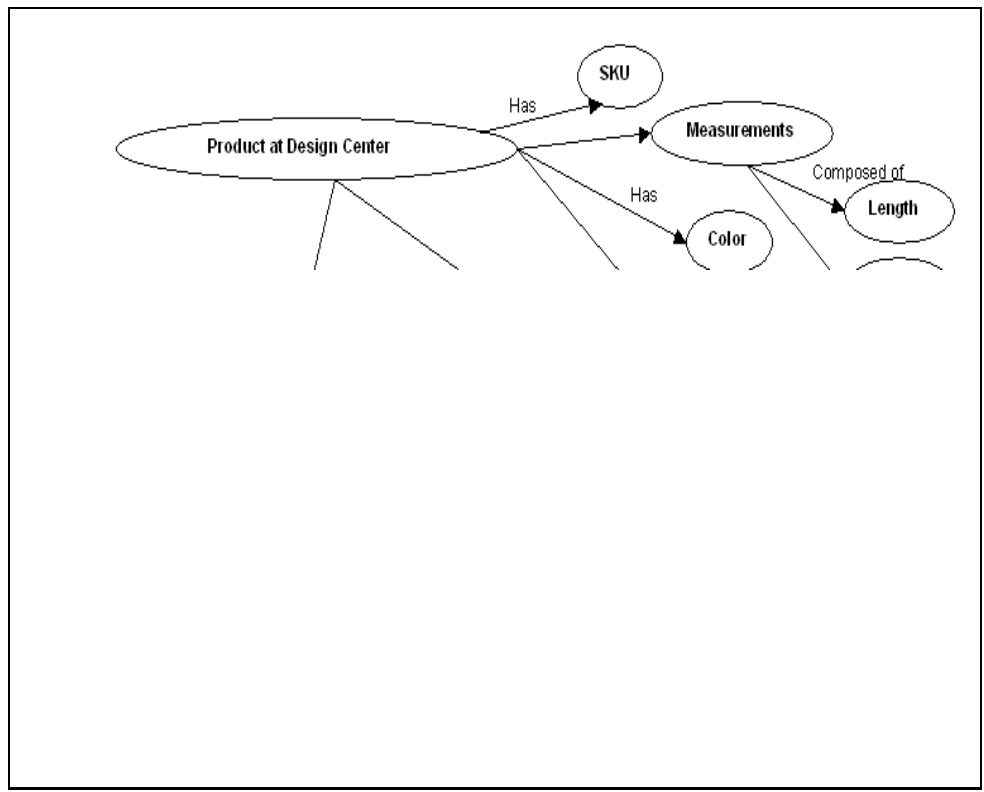}}
\caption{Conceptual Hierarchy of Resources }
\end{figure}

\newpage
\subsection{ Model: Designer\_CA\_33}

Knowledge Model Number: Designer\_CA\_33

Model Description: Concept Map

KA Source: Introduction, KA Session Report \#1

Model Perspective: Designer

Date: 04/13/00

\begin{figure}[h]
\centering
\fbox{\includegraphics[width=5in,height=4in]{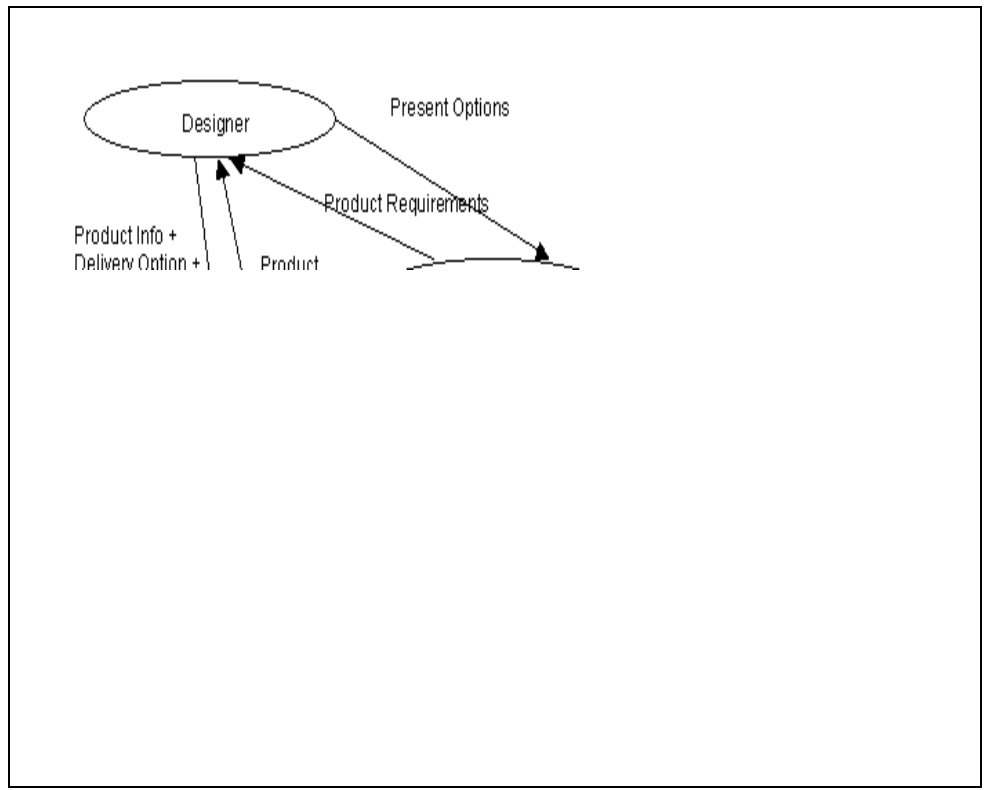}}
\caption{Concept Map }
\end{figure}

\newpage
\subsection{ Model: Designer\_CA\_08}

Knowledge Model Number: Designer\_CA\_09

Model Description: Entity relationship between data concepts of the Quote Sheet 

KA Source: KA Session \#6 

Model Perspective: Designer

Date: 04/03/00

\begin{figure}[h]
\centering
\fbox{\includegraphics[width=5in,height=4in]{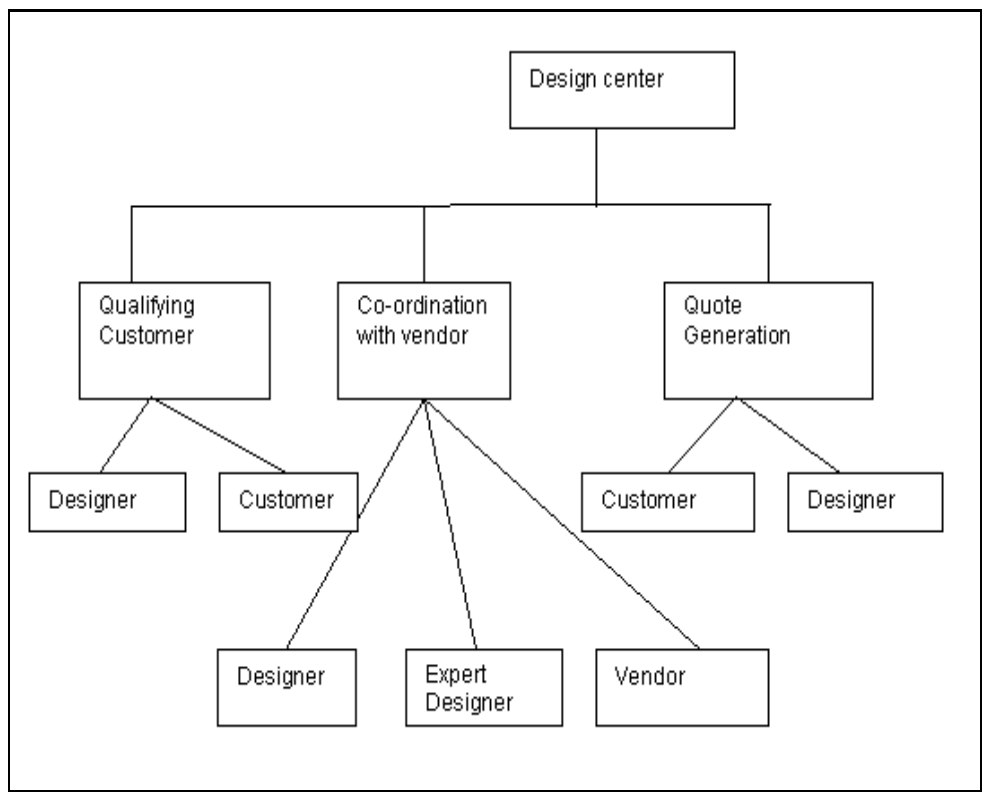}}
\caption{Entity relationship between data concepts of the Quote Sheet }
\end{figure}

\newpage
\subsubsection{ Model: Designer\_CA\_09}

Knowledge Model Number: Designer\_CA\_09

Model Description: Entity relationship between data concepts of the Quote Sheet 

KA Source: KA Session \#6 

Model Perspective: Designer

Date: 04/03/00

\begin{figure}[h]
\centering
\fbox{\includegraphics[width=5in,height=4in]{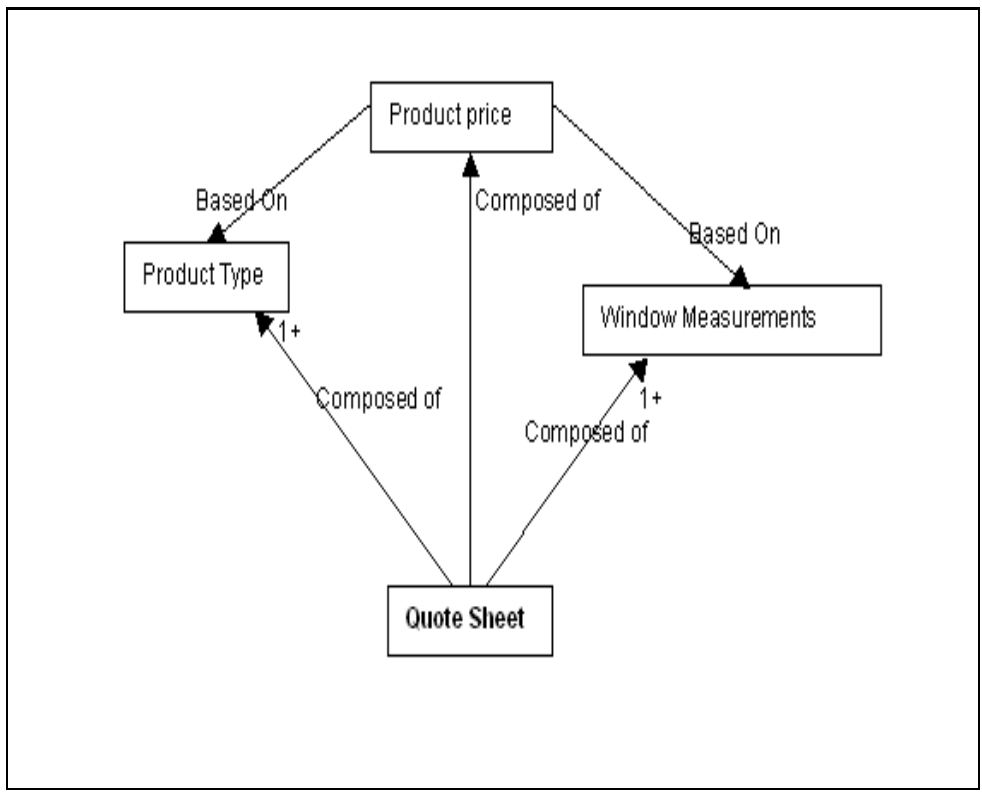}}
\caption{Entity relationship between data concepts of the Quote Sheet }
\end{figure}

\newpage
\subsection{ Model: Designer\_CA\_10}

Knowledge Model Number: Designer\_CA\_10

Model Description: Entity relationship between data concepts of Accurate Measurement form

KA Source: KA Session \#6 

Model Perspective: Designer

Date: 04/03/00

\begin{figure}[h]
\centering
\fbox{\includegraphics[width=5in,height=4in]{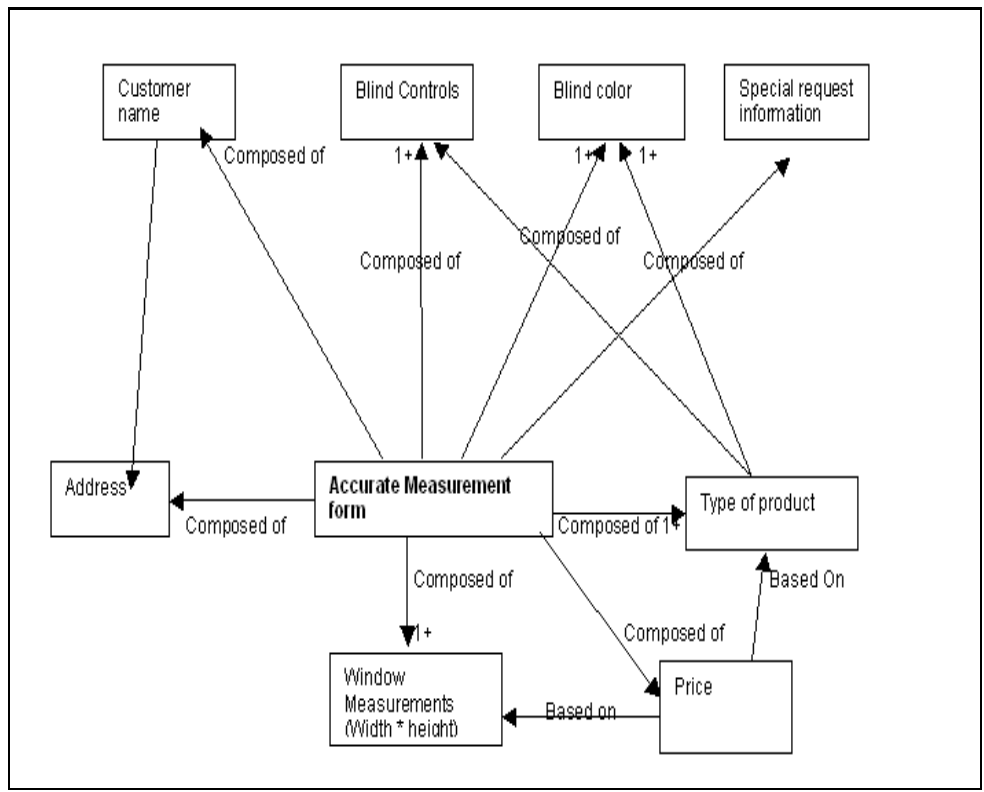}}
\caption{Entity relationship between data concepts of Accurate Measurement form }
\end{figure}

\newpage
\subsection{ Model: Designer\_CA\_18}

Knowledge Model Number: Designer\_CA\_18

Model Description: Entity relationship between data concepts of Invoice

KA Source: KA Session \#8 

Model Perspective: Designer

Date: 04/03/00

\begin{figure}[h]
\centering
\fbox{\includegraphics[width=5in,height=4in]{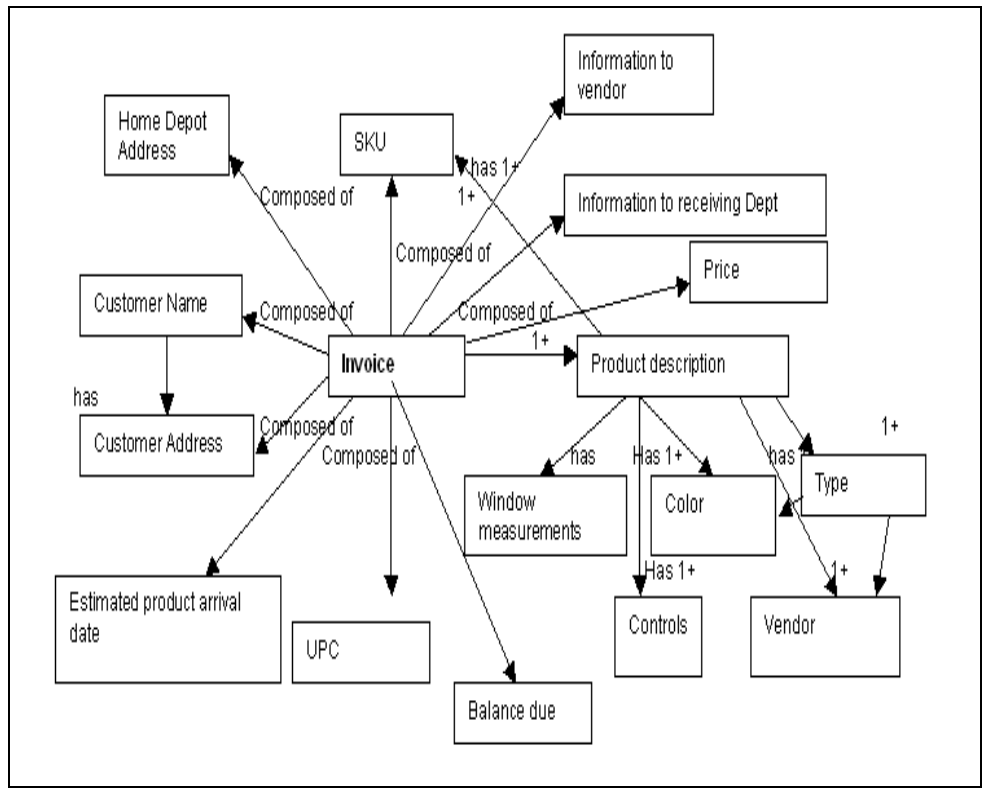}}
\caption{Entity relationship between data concepts of Accurate Measurement form }
\end{figure}

\newpage
\subsection{ Model: Designer\_CA\_19}

Knowledge Model Number: Designer\_CA\_19

Model Description: Concept hierarchy for performer Role

KA Source: Introduction, KA Session \#8

Model Perspective: Designer

Date: 04/03/00

\begin{figure}[h]
\centering
\fbox{\includegraphics[width=5in,height=3in]{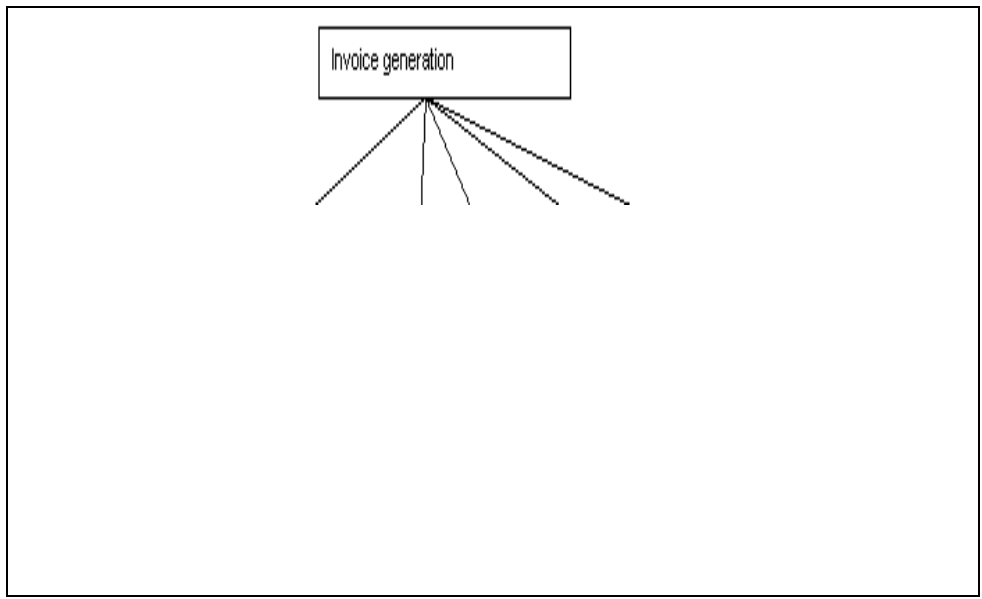}}
\caption{ Concept hierarchy for performer Role}
\end{figure}

\newpage
\subsection{ Model: Designer\_CA\_22}

Knowledge Model Number: Designer\_CA\_22

Model Description: Concept Map representing performer roles.

KA Source: KA Session Report \#9 Section 2.5.9

Model Perspective: Designer

Date: 03/04/00

\begin{figure}[h]
\centering
\fbox{\includegraphics[width=4in,height=3in]{mdesca22.eps2}}
\caption{ Concept Map representing performer roles}
\end{figure}

\newpage
\subsection{ Model: Designer\_CA\_24}

Knowledge Model Number: Designer\_CA\_24

Model Description: Concept Map representing performer roles and information/entity flow

KA Source: KA Session \#10 Section 2.5.10

Model Perspective: Designer

Date: 04/09/00

\begin{figure}[h]
\centering
\fbox{\includegraphics[width=4in,height=3in]{mdesca24.eps2}}
\caption{ Concept Map representing performer roles and information/entity flow}
\end{figure}

\newpage
\subsection{ Model: Designer\_CA\_25}

Model Description: Entity relationship diagram

KA Source: KA Session \#10 Section 2.5.10

Model Perspective: Designer

Date: 04/09/00

\begin{figure}[h]
\centering
\fbox{\includegraphics[width=4in,height=3in]{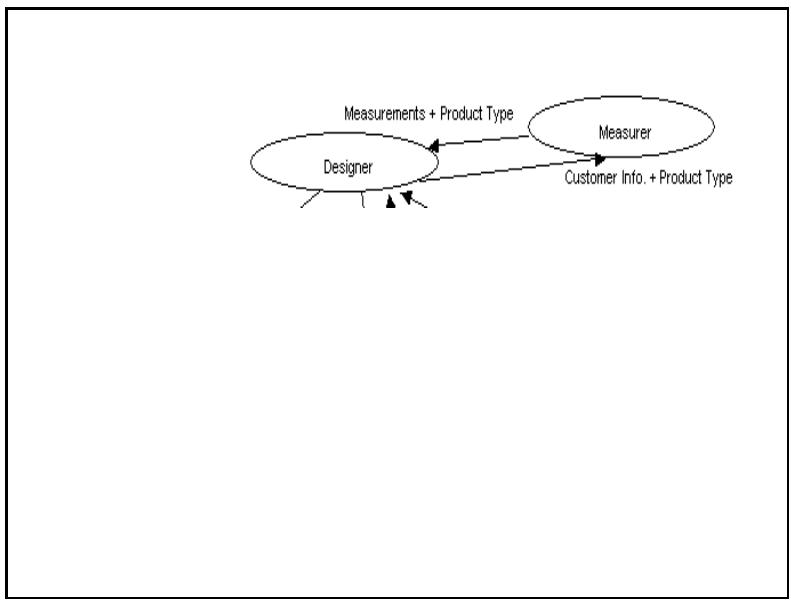}}
\caption{ Concept Map representing performer roles and information/entity flow}
\end{figure}

\newpage
\subsection{ Model: Designer\_TA\_02}

Knowledge Model Number: Designer\_TA\_02

Model Description: Task Decomposition for In stock Product

KA Source:  KA Session Report \#1

Model Perspective: Designer

Date: 04/01/00

\begin{figure}[h]
\centering
\fbox{\includegraphics[width=5in,height=4in]{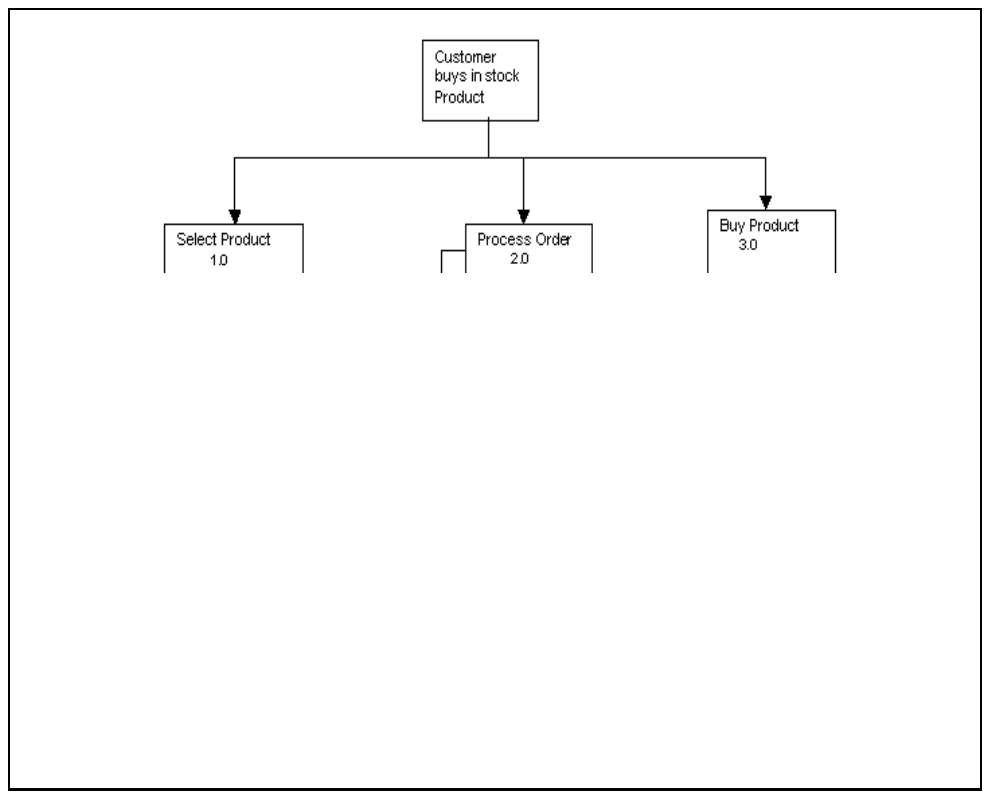}}
\caption{ Task Decomposition for In stock Product}
\end{figure}

\newpage
\subsection{ Model: Designer\_TA\_03}

Knowledge Model Number: Designer\_TA\_03

Model Description: Task Hierarchy for: 

Task 2.1 - Check Inventory
Task 2.2 - Contact Vendor
Task 3.1 - Pay cashier

KA Source:  KA Session Report \#1 

Model Perspective: Designer

\begin{figure}[h]
\centering
\fbox{\includegraphics[width=5in,height=4in]{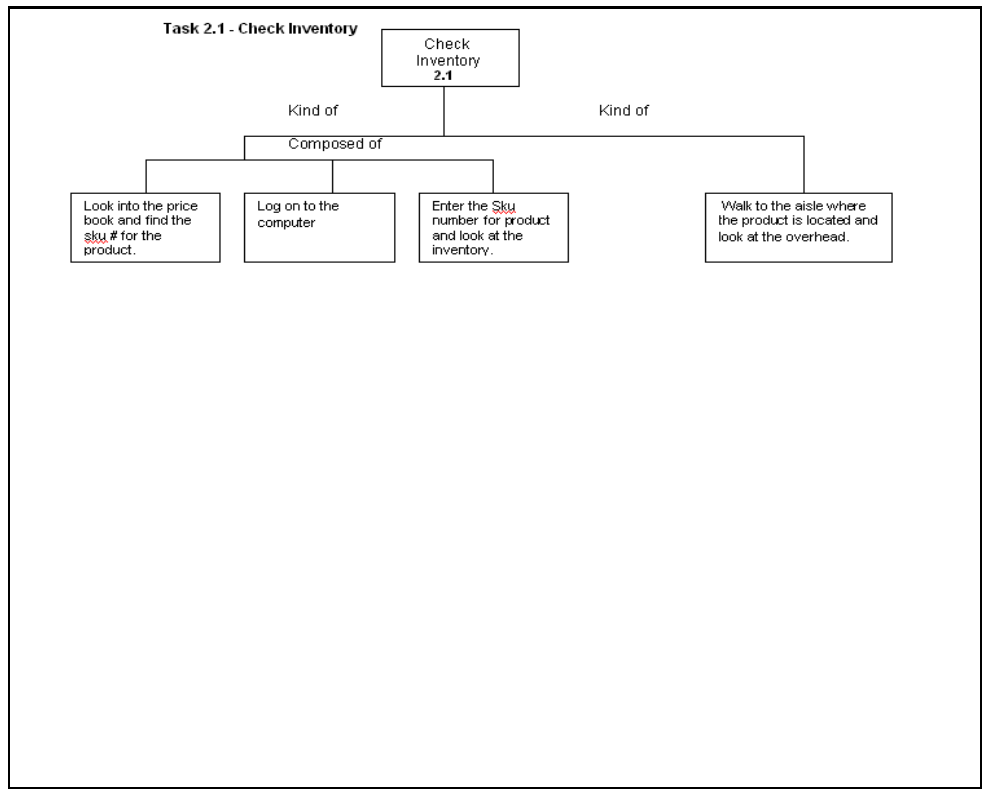}}
\caption{ Task Hierarchy}
\end{figure}

\newpage
\subsection{ Model: Designer\_TA\_04}

Knowledge Model Number: Designer\_TA\_04

Model Description: Task template for task \# 1.1.1 from Designer\_TA\_02.

KA Source:  KA Session Report \#1 \& KA Session Report \#2

Model Perspective: Designer

Date: 04/01/00

\newpage
\subsection{ Model: Designer\_TA\_05}

Knowledge Model Number: Designer\_TA\_05

Model Description: Task template for task \# 1.1.2 from Designer\_TA\_02.

KA Source: KA Session Report \#1 \& KA Session Report \#2

Model Perspective: Designer

Date: 04/01/00

Task Name: Obtain measurements and price range

Task Description:

 Execution Duration: 1-5 minutes

 Execution Frequency: Discrete

 Pre-condition \& Criticality Constraint pairs: Homeowner --- High

Pre-condition Statement: The person coming in to buy the product should be the none 
other than the homeowner because only the homeowner can make the decision. Also the 
customer should get the measurements then only the designer can help him. Having 
the price range what the customer plans to stay also will be helpful.

Post-condition Statement \& Criticality Constraint Pairs: Information Gained -- High

Post-condition 
Statement: When the designer knows the price range the selection process becomes 
faster having information regarding what the customer wants then he can assist the 
customer with correct choices.

 Input

  Input Name: measurements and price range obtained

  Input Type: Event

  Criticality:   High

  Sender of Input: Designer

 Output

  Output Name: Measurements

  Output Type: Data

  Criticality:  High

  Sender of Output: Customer

 Output

  Output Name: Price range

  Output Type: Data

  Criticality:  Medium

  Sender of Output: Customer

 Location Task is performed: This task may be performed on the phone or at the Design Center

 Resources:

  Personnel:  Any Design Center employee

  Equipment: None

Sub Tasks: None

\newpage
\subsection{ Model: Designer\_TA\_18}

Model Description: Scenario Elicitation outcome matrix

KA Source: KA Session \#4, section 2.5.4

Model Prespective: Designer

Date: 04/06/00

Date Revised:

\begin{longtable}{|p{0.9in}|p{0.6in}|p{0.7in}|p{0.7in}|p{0.9in}|p{0.9in}|} \hline 
Work Process & Performer & Resources Used & Interactions & Result & Problems Executing \\ \hline 
Qualify Customer to generate a Quote on the Blinds or wallpaper selected & Designer & Price book & No & Qualifies the customer by verifying\newline \newline 1. .Customer      has 
made the right product selection\newline 2. .Customer has measured their windows 
and have the right measurements\newline  & 1. .Customer does not have measurements\newline 2. .Customer 
not selected the right product\newline 3. .Customer does not have the right measurements \\ \hline 
Qualifying 
Customer -- Helps the customer to make the right  Blind or wallpaper selection by 
giving all the advantages and details based on customer needs & Designer & Catalogs & Yes-sometimes 
expert designer & Customer requests a quote from the designer &  \\ \hline 
Generate rough estimation on an average window size & Designer & Price book & No & Helps 
the Customer on the price range of the product they have selected &  \\ \hline 
Qualifying the customer --\newline Requests the customer for Windows measurements 
and provide details for measuring windows and also give an option of Home Depot measuring 
customer's window\newline \newline  & Designer & Nil & Yes- If the customer requests 
Home Depot for measuring their windows, than the Designer co-ordinates with Installer & 1. 
.Expects the customer to return with measurements\newline 2. .Co-ordinates with the 
Installer for measuring the customer windows\newline  &  \\ \hline 
Verifies the customer has right measurements \newline  & Designer & Nil & No & Decides 
to Generate Quotes for the product selected & Product selected is discontinued \\ \hline 
Generating quotes for the customer on a sheet of paper & Designer & Price book & Yes-sometimes the expert designer on special order products & Generates quotes  on three different products for the customer & As the process is manual, it takes a long time to generate the quote   \\ \hline 
\end{longtable}

\newpage
\subsection{ Model: Designer\_TA\_20}

Model Description: Task Decomposition

KA Source: KA Session Report\#4 \& KA Session Report \#5(2.5.4 \&2.5.5)

Model Perspective: Designer

Date: 04/06/00

\begin{figure}[h]
\centering
\fbox{\includegraphics[width=5in,height=4in]{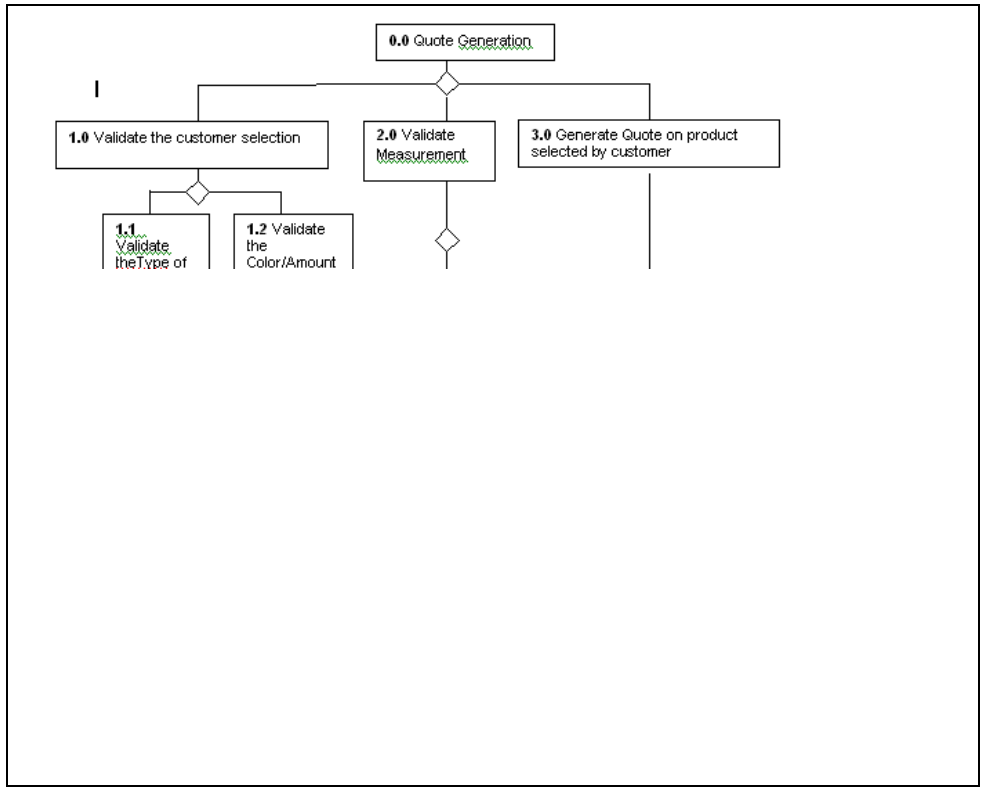}}
\caption{ Task Decomposition}
\end{figure}

\newpage
\subsection{ Model: Designer\_TA\_26}

Model Description: Task Template for Task 3.1 of Knowledge model number Designer\_TA\_20

KA Source: KA Session Report\#5 (Section 2.5.5)
  
Model Perspective: Designer

Date: 04/06/00

Task Name: Co-ordinate with the vendor on product availability

Task Description: If the customer is interested to buy the special order product 
and requests a quote from the designer, the designer checks with the vendor on product availability

Execution Duration: 5 minutes 

Execution Frequency: Discrete

Pre-condition \& Criticality Constraint pairs: Customer requests quote- Customer 
would like to buy the product-High  

Pre-condition Statement: Customer requests a quote on special order product. And 
is interested to but the product

Post-condition Statement \& Criticality Constraint Pairs: - Designer checks for the 
product availability -High

Post-condition Statement: The designer verifies with the vendor if the product selected by the customer is available for delivery. Based on that information, the designer generates quote

\textbf{Input}

Input Name: Designer checks for product availability, Selected products

Input Type: data

Criticality:   Medium

Sender of Input: designer verifies with the vendor

\textbf{Output}

Output Name: Product available for delivery in the lead time, could not contact the 
vendor, product not in stock or discontinued

Output Type: Event 

Criticality:  Medium

Sender of Output: Designer passes the information to customer

Location Task is performed: This task is performed at the Design Center, Home Depot.

\textbf{Resources}

\textbf{Personnel}  Designer, Vendor

\textbf{Equipment} Price book

\newpage
\subsection{ Model: Designer\_TA\_45}

Model Description: Task template for task \#3.0

KA Source: KA Session Report \#9 Section 2.5.9

Model Perspective: Designer

Date: 03/04/00

Task Name: Send Request to Measurer

Task Description:

Execution Duration: 1-10 minutes

Execution Frequency: Discrete

Pre-condition \& Criticality Constraint pairs: Task \#2

Pre-condition Statement: The customer should have paid at the payment center.

Post-condition Statement \& Criticality Constraint Pairs: The designer transmits 
the request to the measurer.

Post-condition Statement: The measurer receives the request along with customer information.

Input Name: Customer Information

Input Type: Entity

Criticality: High

Sender of Input: Designer

Output Name: Request Transmitted

Output Type: Event

Criticality:  High

Sender of Output: Designer

Location Task is performed: The task is performed at the Design Center.

\textbf{Resources}

Personnel:  Designer

Equipment:  Communication System 

\footnote{ Communication System: A resource in the domain.  Refer to domain dictionary for definition}

\newpage
\subsection{ Model: Designer\_DOC\_01}

Model Description: Document Framework Form - Instructions to measure windows

KA Source: KA session \#4, Section 2.5.4, Attachment DOC\#1  

Model Perspective: Designer

Date: 04/06/00

Topic: Instructions to measure windows

Key Concepts:

\begin{enumerate}
\item To measure the window (width*Length) to the nearest 1/8th of an inch.

\item To measure top, middle and bottom of the window
\end{enumerate}

\textbf{SubTopic\#1: Instructions for horizontal or cellular blinds}

Key Concepts for SubTopic\#1: 
\begin{enumerate}
\item To measure the smallest width and the Longest length

\item To measure windows with or without head rail
\end{enumerate}

\textbf{SubTopic\#2: Instructions for Vertical Blinds}

Key Concepts for SubTopic\#2:
\begin{enumerate}
\item To measure the top width and the smallest length of their window

\item To measure windows with or without the head rail
\end{enumerate}

Reference: Doc\#1

Doc\#1 User: Customer uses the above information provided by the designer to measure their windows

\newpage
\subsection{ Model: Designer\_ DOC\_02}

Knowledge Model Number: Designer\_DOC\_02

Model Description: Document Framework Form -Controls available with window Blinds

KA Source: KA session \#4, Section 2.5.4, Attachment Doc\#2  
 
Model Perspective: Designer

Date: 04/06/00

Topic: The different controls available with Window Blinds

Key Concepts:

\begin{enumerate}
\item Standard controls 

\item Reverse Standard controls
\end{enumerate}

SubTopic\#1: 2'' Horizontal blinds

Key Concepts: Tilt cord and Lift cord on right  side or left side of the window

SubTopic\#2: 1'' Horizontal blinds

Key concepts: Tilt wand and Lift wand on right side or left side of the window

SubTopic\#3: Cellular blinds

Key Concepts: Lift cord on right or left side of the window

SubTopic\#4: Vertical Blinds

Key concepts: Wands on same side as stack or on either side of the window

Reference: Doc\#2

Doc\#2 User: Customer uses this information provided by the designer to decide on 
what controls they need for their windows

\newpage
\subsection{ Model: Designer\_CA\_02}

Knowledge Model Number: Designer\_CA\_02

Model Description: Domain Dictionary

KA Source: KA Session Report \#1
 
Model Perspective: Designer

Date: 04/03/00

Domain Term Name: Wallpaper books

Related Knowledge Models: Designer\_CA\_01

Domain Term Definition: The wallpaper books contain sample wallpaper and manufacturer information. The customer looks at the book to decide the type of product he needs.

Sample Usage: The wallpaper books are not sorted by color but you will find the color you want in every book. The first thing the customer needs to decide on is the style of wallpaper that are contemporary, traditional, kitchen and bath.

\newpage
\subsection{ Model: Designer\_CA\_03}

Knowledge Model Number: Designer\_CA\_03

Model Description: Domain Dictionary

KA Source: KA Session Report \#1
 
Model Perspective: Designer

Date: 04/03/00

Domain Term Name: Mini Blind- Cutter

Related Knowledge Models: None

Domain Term Definition: The mini blind cutter is used at the design center to cut 
in stock blinds according to customer requirement

Sample Usage: Jane wanted to buy in stock blinds but the size she wanted was not 
in stock. So Jimmy suggested that she could pick a larger size in the brand she liked and that he could cut it to size she needed using the mini blind cutter.

\newpage
\subsection{ Model: Designer\_CA\_04}

Knowledge Model Number: Designer\_CA\_04

Model Description: Domain Dictionary

KA Source: KA Session Report \#2
 
Model Perspective: Designer

Date: 04/03/00

Domain Term Name: Price Book

Related Knowledge Models: None

Domain Term Definition: The price book is the book with all the product sku's the 
price and the vendor contact information.

Sample Usage: When Jane wanted the Bali 1 --inch blinds in 40 by 60 and the size 
was not in stock, Jimmy looked at the Price book and found the vendor number and 
called him up to ask for the delivery date for that particular type.

\newpage
\subsection{ Model: Designer\_CA\_05}

Knowledge Model Number: Designer\_CA\_05

Model Description: Domain Dictionary

KA Source: Introduction, KA Session Report \#3
 
Model Perspective: Designer

Date: 04/03/00

Domain Term Name: Catalog

Related Knowledge Models: None

Domain Term Definition: The catalogs are the product manufacturer published magazines that have all the different models manufactured by the company for the customer to select.

Sample Usage: When Cynthia wanted to decide on the different types of the blinds 
then Julie directed her towards the catalogs to see the different types catalogs 
stacked up on the shelf.

\newpage
\subsection{ Model: Designer\_CA\_06}

Knowledge Model Number: Designer\_CA\_06

Model Description: Domain Dictionary

KA Source: Introduction, KA Session Report \#1
 
Model Perspective: Designer

Date: 04/03/00

Domain Term Name: Mobile cart

Context: Introduction

Domain Term Definition: The mobile cart is a like a wireless computer on the cart 
which is used by the design center employees to place order for inventory and check 
inventory etc.

Sample Usage: When Jane wanted to buy a bali I-inch blinds in a quantity of 10 then 
jimmy logged to the mobile cart and saw that they had only 8 in stock.

\newpage
\subsection{ Model: Designer\_CA\_26}

Knowledge Model Number: Designer\_CA\_26

Model Description: Domain Dictionary

KA Source: Introduction, KA Session Report \#1
 
Model Perspective: Designer

Date: 04/13/00

Domain Term Name: Measurements

Context: Select Product

Domain Term Definition: Each blind has measurements, which are its length and the 
breadth they are in inches. These measurements are used determine fits customer's 
window. So when a customer wants to buy the product he needs to get the window size.

Sample 
Usage: Jane measured her window found that it was 30 X 60 so with these measurements she went to the design center to reinstall her blind.

\newpage
\subsection{ Model: Designer\_CA\_27}

Knowledge Model Number: Designer\_CA\_27

Model Description: Domain Dictionary

KA Source: Introduction, KA Session Report \#1
 
Model Perspective: Designer

Date: 04/13/00

Domain Term Name: Verticals

Context: Select Product

Domain Term Definition: This is a type of blind that is sold at the design center. 
They have the individual pieces of the blinds arranged in a vertical pattern.

Sample Usage: Jane looked at the large window facing the sea and decided that vertical blinds there would give her easy access to the patio.

\newpage
\subsection{ Model: Designer\_CA\_28}

Knowledge Model Number: Designer\_CA\_28

Model Description: Domain Dictionary

KA Source: Introduction, KA Session Report \#1
 
Model Perspective: Designer

Date: 04/13/00

Domain Term Name: Horizontal

Context: Select Product

Domain Term Definition: This is a type of blind that is sold at the design center. 
They have the individual pieces of the blinds arranged in a horizontal pattern.

Sample Usage: Jane looked at her kitchen window and felt that horizontal blinds there would be easy to maintain.

\newpage
\subsection{ Model: Designer\_CA\_29}

Knowledge Model Number: Designer\_CA\_29

Model Description: Domain Dictionary

KA Source: Introduction, KA Session Report \#1
 
Model Perspective: Designer

Date: 04/13/00

Domain Term Name: Cellular

Context: Select Product

Domain Term Definition: This is a type of blind that is sold at the design center. 
They have the individual pieces of the blinds arranged in a horizontal pattern but 
they give more privacy than the horizontal.

Sample Usage: Jane looked at her bedroom window and felt that cellular blinds there 
would give her enough privacy.

\newpage
\subsection{ Model: Designer\_CA\_30}

Knowledge Model Number: Designer\_CA\_30

Model Description: Domain Dictionary

KA Source: Introduction, KA Session Report \#1
 
Model Perspective: Designer

Date: 04/13/00

Domain Term Name: Embossed

Context: Select Product

Domain Term Definition: This is a type of wallpaper that is sold at the design center. They have the puffy impressions.

Sample Usage: Jane felt that embossed wallpaper would give her more elegance to her 
dining room

\newpage
\subsection{ Model: Designer\_CA\_31}

Knowledge Model Number: Designer\_CA\_31

Model Description: Domain Dictionary

KA Source: Introduction, KA Session Report \#1
 
Model Perspective: Designer

Date: 04/13/00

Domain Term Name: Solid Vinyls

Context: Select Product

Domain Term Definition: This is a type of wallpaper that is sold at the design center. Its a thick layer of vinyl.

Sample Usage: Jane felt that solid vinyl wallpaper would be easier to maintain in 
her kitchen.

\newpage
\subsection{ Model: Designer\_CA\_32}

Knowledge Model Number: Designer\_CA\_32

Model Description: Domain Dictionary

KA Source: Introduction, KA Session Report \#1
 
Model Perspective: Designer

Date: 04/13/00

Domain Term Name: Vinyl Coated

Context: Select Product

Domain Term Definition: This is a type of wallpaper that is sold at the design center. 
It has a layer of vinyl over the wallpaper.

Sample Usage: Jane felt that vinyl coated wallpaper would be easier to maintain in 
her family room.

\newpage
\subsection{ Model: Designer\_CA\_07}

Knowledge Model Number: Designer\_CA\_07

Model Description: Domain Dictionary

KA Source: Introduction, KA Session Report \#3
 
Model Perspective: Designer

Date: 04/04/00

Domain Term Name: lead-time

Related Knowledge Models: Designer\_TA\_18

Domain Term Definition: Every manufacturer has different lead-times for different 
product. The lead-time is nothing but the normal delivery time for a product.

Sample Usage:  The lead time for the 2-inch wooden blinds that Cynthia ordered was 
4 weeks but still she decided to buy the product.

\newpage
\subsection{ Model: Designer\_CA\_11}

Knowledge Model Number: Designer\_CA\_11

Model Description: Domain Dictionary

KA Source: Introduction, KA Session \#6
 
Model Perspective: Designer

Date: 04/03/00

Domain Term Name: Price Book

Context: Generate Quote or invoice

Domain Term Definition: Price book consists of price list of all the blinds the design 
center carries. It has the information of all vendors, special products they make 
and measurements for the different window sizes

Sample Usage: Designer refers to the price book to generate quote and to generate 
invoice

\newpage
\subsection{ Model: Designer\_CA\_12}

Knowledge Model Number: Designer\_CA\_12

Model Description: Domain Dictionary

KA Source: Introduction, KA Session \#6
 
Model Perspective: Designer

Date: 04/03/00

Domain Term Name: Headrail

Context: Blinds

Domain Term Definition: Headrail is used to support blinds (Horizontal or vertical)

Sample Usage: when customer decides to buy the blinds, the designer ask the customer if they would like the headrail to be mounted inside the window or outside

\newpage
\subsection{ Model: Designer\_CA\_13}

Knowledge Model Number: Designer\_CA\_13

Model Description: Domain Dictionary

KA Source: Introduction, KA Session \#6
 
Model Perspective: Designer

Date: 04/03/00

Domain Term Name: Slats

Context: Blinds

Domain Term Definition: the individual blinds is called slats when installed vertically

Sample Usage: Designer provides information to the customer on the size of the slat, precisely the 1'' or 2'' depending on the manufacturer

\newpage
\subsection{ Model: Designer\_CA\_14}

Knowledge Model Number: Designer\_CA\_14

Model Description: Domain Dictionary

KA Source: Introduction, KA Session \#6
 
Model Perspective: Designer

Date: 04/03/00

Domain Term Name: vanes

Context: Blinds

Domain Term Definition: The individual blind when the blinds are installed horizontally

Sample Usage: Designer asks the size of the vane the customer is interested in, it can 1'' or 2'' and it depends on the manufacturer

\newpage
\subsection{ Model: Designer\_CA\_15}

Knowledge Model Number: Designer\_CA\_15

Model Description: Domain Dictionary

KA Source: Introduction, KA Session \#6
 
Model Perspective: Designer

Date: 04/03/00

Domain Term Name: Valence

Context: Blinds

Domain Term Definition: Valence is used to cover the headrail

Sample Usage: Designer asks the customer if they would like to buy the valence along with the headrail and also tells them that some manufacturers provide the valence along with the headrail.

\newpage
\subsection{ Model: Designer\_CA\_16}

Knowledge Model Number: Designer\_CA\_16

Model Description: Domain Dictionary

KA Source: Introduction, KA Session \#6
 
Model Perspective: Designer

Date: 04/03/00

Domain Term Name: Cords

Context: Blinds

Domain Term Definition: Controls provides with blinds to lower or raise the blinds 
or to tilt  the blinds

Sample Usage: Designer asks the customers what kind of controls they are interested 
in, if they want the cords towards the right or left for tilting and to lower or 
raise the blinds.

\newpage
\subsection{ Model: Designer\_CA\_17}

Knowledge Model Number: Designer\_CA\_17

Model Description: Domain Dictionary

KA Source: Introduction, KA Session \#6
 
Model Perspective: Designer

Date: 04/03/00

Domain Term Name: Wands

Context: Blinds

Domain Term Definition: Controls provides with blinds to tilt the blinds

Sample Usage: Designer asks the customers on what side they  would like the wand 
to be.

\newpage
\subsection{ Model: Designer\_CA\_20}

Knowledge Model Number: Designer\_CA\_20

Model Description: Domain Dictionary

KA Source: Introduction, KA Session \#8
 
Model Perspective: Designer

Date: 04/03/00

Domain Term Name: SKU --Stock keeping unit

Context: Invoice

Domain Term Definition: A number to identify all the products in case of Instock 
product, and a number to identify a type of product manufactured by a particular 
vendor

Sample Usage: Designer needs to enter the SKU number for each product in the invoice

\newpage
\subsection{ Model: Designer\_CA\_21}

Knowledge Model Number: Designer\_CA\_21

Model Description: Domain Dictionary

KA Source: Introduction, KA Session \#8
 
Model Perspective: Designer

Date: 04/03/00

\newpage
\subsection{ Model: Designer\_CA\_23}

Knowledge Model Number: Designer\_CA\_23

Model Description: Domain Dictionary

KA Source: KA Session \#9 and \#10 Section 2.5.9 and 2.5.10
 
Model Perspective: Designer

Date: 04/09/00

Domain Dictionary

Domain Term Name: Payment Desk

Related Knowledge Models: Customer\_TA\_37

Domain Term Definition:

The front desk is a man driven counter that accepts payment from the customer.  It 
has system connected to the inventory that updates the inventory after the items 
have been checked out.  It also has systems that trigger the dispatch or finalize 
the customer's request for measurement/installation and delivery after the payment 
has been made.

Sample Usage: The customer pays for the measurements at the payment desk.

Domain Term Name: Recording System

Related Knowledge Models: Designer\_TA\_34

Domain Term Definition:

Recording system is an electronic or manual form of filing information.  The system 
saves and stores different types of information and often has the feature of searching through the stored information.  

Sample Usage: The designer records the measurements on the recording system.

Domain Term Name: Sample Blinds

Related Knowledge Models: 

Domain Term Definition:

Blinds which are pre-measured and are used by the designer to be given away to help 
the customer in measuring the windows and doors.  These blinds give an idea to the 
customer how to take measurements.

Sample Usage: The designer gives the sample blind to the customer for measurements.

Sample Usage: The designer gives a sample blind to the customer to use it to take measurements.

Domain Term Name: Payment Center

Related Knowledge Models: Customer\_TA\_04

Domain Term Definition:

Place where payment is accepted from the customer and a receipt generated by the 
system. The payment system usually has infrastructure for accepting and authenticating the credit card and accepting checks.

\newpage 
%% Unified Knowledge Models
%% Section Change

\chapter{ Unified Knowledge Models}

\section{ Model: UKM\_CA\_01}

\begin{figure}[h]
\centering
\fbox{\includegraphics{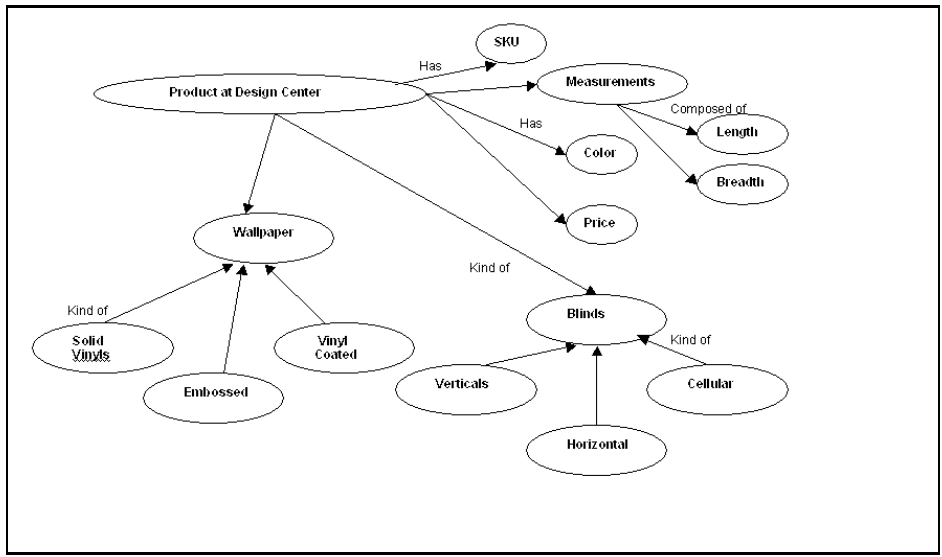}}
\caption{ Unified Knowledge Model UKM\_CA\_01}
\end{figure}

\newpage
\section{ Model: UKM\_CA\_02}

\begin{figure}[h]
\centering
\fbox{\includegraphics{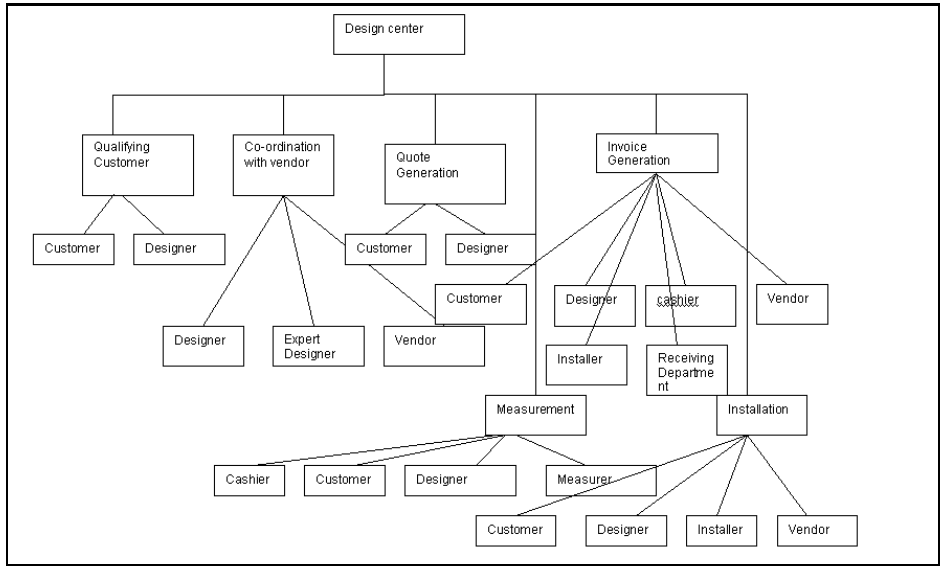}}
\caption{ Unified Knowledge Model UKM\_CA\_02}
\end{figure}

\newpage
\section{ Model: UKM\_CA\_03}

\subsection{ Instructions to measure windows}
Key Concepts:
\begin{enumerate}
\item To measure the window (width*Length) to the nearest 1/8th of an inch.
\item To measure top, middle and bottom of the window
\end{enumerate}

\subsection{ Instructions for horizontal or cellular blinds}
Key Concepts: 
\begin{enumerate}
\item To measure the smallest width and the Longest length
\item To measure windows with or without headrail
\end{enumerate}

\subsection{ Instructions for Vertical Blinds}
Key Concepts:
\begin{enumerate}
\item To measure the top width and the smallest length of their window
\item To measure windows with or without the headrail
\end{enumerate}
User: Customer uses the above information provided by the designer to measure the windows

\subsection{ Possible controls on Blinds}
Key Concepts:
\begin{enumerate}
\item Standard controls 
\item Reverse Standard controls
\end{enumerate}

\subsection{ 2'' Horizontal blinds}
Key Concepts: Tilt cord and Lift cord on right  side or left side of the window

\subsection{ 1'' Horizontal blinds}
Key concepts: Tilt wand and Lift wand on right side or left side of the window

\subsection{ Cellular blinds}
Key Concepts: Lift cord on right or left side of the window

\subsection{ Vertical Blinds}
Key concepts: Wands on same side as stack or on either side of the window
User: Customer uses this information provided by the designer to decide on 
what controls they need for the windows

\newpage
\section{ Model: UKM\_CA\_04}
\begin{figure}[h]
\centering
\fbox{\includegraphics{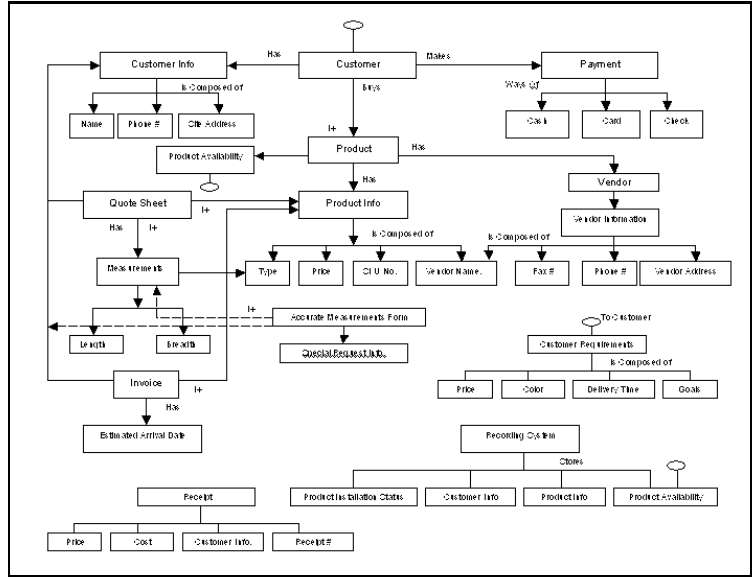}}
\caption{ Unified Knowledge Model UKM\_CA\_04}
\end{figure}

\newpage
\section{ Model: UKM\_CA\_05}

\textbf{Domain Dictionary}

\subsection{ Recording system}

Domain Term Definition: Recording system is an electronic or manual form of filing 
information.  The system saves and stores different types of information and often 
has the feature of searching through the stored information.  In the design center 
it is used to store customer information and product details.

\subsection{ Payment Acceptance System}

Domain Term Definition: The payment desk is a man driven counter that accepts payment from the customer.  It has system connected to the inventory that updates the inventory after the items have been checked out.  It also has systems that trigger the dispatch or finalize the customer's request for measurement or installation and delivery after the payment has been made. The payment center generates the receipt after payment has been accepted. The payment system usually has infrastructure for accepting and authenticating the credit card and accepting checks.

\subsection{ Wallpaper books}

Domain Term Definition: The wallpaper books contain sample wallpaper and manufacturer information. The customer looks at the book to decide the type of product he needs.

Sample Usage: The wallpaper books are not sorted by color but you will find the color you want in every book. The first thing the customer needs to decide on is the style of wallpaper that are contemporary, traditional, kitchen and bath.

\subsection{ Mini Blind- Cutter}

Related Knowledge Models: None

Domain Term Definition: The mini blind cutter is used at the design center to cut 
in stock blinds according to customer requirement

Sample Usage: Jane wanted to buy in stock blinds but the size she wanted was not 
in stock. So Jimmy suggested that she could pick a larger size in the brand she liked and that he could cut it to size she needed using the mini blind cutter.

\subsection{ Price Book}

Related Knowledge Models: None

Domain Term Definition: The price book is the book with all the product sku's the 
price and the vendor contact information.

Sample Usage: When Jane wanted the Bali 1 --inch blinds in 40 by 60 and the size 
was not in stock, Jimmy looked at the Price book and found the vendor number and 
called him up to ask for the delivery date for that particular type.

\subsection{ Catalog}

Related Knowledge Models: None

Domain Term Definition: The catalogs are the product manufacturer published magazines that have all the different models manufactured by the company for the customer to select.

Sample Usage: When Cynthia wanted to decide on the different types of the blinds 
then Julie directed her towards the catalogs to see the different types catalogs 
stacked up on the shelf.

\subsection{ Lead time}

Domain Term Definition: Every manufacturer has different lead-times for different 
product. The lead-time is nothing but the normal delivery time for a product.

Sample Usage:  The lead time for the 2-inch wooden blinds that Cynthia ordered was 
4 weeks but still she decided to buy the product.

\subsection{ Headrail}

Context: Blinds

Domain Term Definition: Headrail is used to support blinds (Horizontal or vertical)

Sample Usage: when customer decides to buy the blinds, the designer ask the customer if they would like the headrail to be mounted inside the window or outside

\subsection{ Slats}

Context: Blinds

Domain Term Definition: the individual blinds is called slats when installed vertically

Sample Usage: Designer provides information to the customer on the size of the slat, precisely the 1'' or 2'' depending on the manufacturer

\subsection{ Vanes}

Context: Blinds

Domain Term Definition: The individual blind when the blinds are installed horizontally

Sample Usage: Designer asks the size of the vane the customer is interested in, it can 1'' or 2'' and it depends on the manufacturer

\subsection{ Valence}

Context: Blinds

Domain Term Definition: Valence is used to cover the headrail

Sample Usage: Designer asks the customer if they would like to buy the valence along with the headrail and also tells them that some manufacturers provide the valence along with the headrail.

\subsection{ Cords}

Context: Blinds

Domain Term Definition: Controls provides with blinds to lower or raise the blinds 
or to tilt the blinds

Sample Usage: Designer asks the customers what kind of controls they are interested 
in, if they want the cords towards the right or left for tilting and to lower or 
raise the blinds.

\subsection{ Wands}

Context: Blinds

Domain Term Definition: Controls provides with blinds to tilt the blinds

Sample Usage: Designer asks the customers on what side they would like the wand to 
be.

\subsection{ SKU --Stock Keeping Unit}

Context: Invoice

Domain Term Definition: A number to identify all the products in case of Instock 
product, and a number to identify a type of product manufactured by a particular 
vendor

Sample Usage: Designer needs to enter the SKU number for each product in the invoice

\subsection{ Stage for register}

Context: Invoice generation

Domain Term Definition: After all the product information is entered by the designer in to the system, Stage for register is the command used in the Invoice generation software to generate the Invoice. 

Sample Usage: Designer generates the invoice for the customer

\subsection{ Measurements}

Context: Select Product

Domain Term Definition: Each blind has measurements, which are its length and the 
breadth they are in inches. These measurements are used determine fits customer's 
window. So when a customer wants to buy the product he needs to get the window size.

Sample Usage: Jane measured her window found that it was 30 X 60 so with these measurements she went to the design center to reinstall her blind.

\subsection{ Verticals}

Context: Select Product

Domain Term Definition: This is a type of blind that is sold at the design center. 
They have the individual pieces of the blinds arranged in a vertical pattern.

Sample Usage: Jane looked at the large window facing the sea and decided that vertical blinds there would give her easy access to the patio.

\subsection{ Horizontal}

Context: Select Product

Domain Term Definition: This is a type of blind that is sold at the design center. 
They have the individual pieces of the blinds arranged in a horizontal pattern.

Sample Usage: Jane looked at her kitchen window and felt that horizontal blinds there would be easy to maintain.

\subsection{ Cellular}

Context: Select Product

Domain Term Definition: This is a type of blind that is sold at the design center. 
They have the individual pieces of the blinds arranged in a horizontal pattern but 
they give more privacy than the horizontal.

Sample Usage: Jane looked at her bedroom window and felt that cellular blinds there 
would give her enough privacy.

\subsection{ Embossed}

Context: Select Product

Domain Term Definition: This is a type of wallpaper that is sold at the design center. They have the puffy impressions.

Sample Usage: Jane felt that embossed wallpaper would give her more elegance to her 
dining room

\subsection{ Solid Vinyls}

Context: Select Product

Domain Term Definition: This is a type of wallpaper that is sold at the design center. Its a thick layer of vinyl 

Sample Usage: Jane felt that solid vinyl wallpaper would be easier to maintain in 
her kitchen.

\subsection{ Vinyl Coated}

Context: Select Product

Domain Term Definition: This is a type of wallpaper that is sold at the design center. It has a layer of vinyl over the wallpaper.

Sample Usage: Jane felt that vinyl coated wallpaper would be easier to maintain in 
her family room.

\newpage 
%% Reference Architecture
%% Section Change

\chapter{ Reference Architecture}

\section{ Architecture Overview}

The Domain Reference Architecture has four classes. - Customer, Measurer, Installer 
and Designer.  The architecture has role based DRAC Responsibility assignment.  Primary goal of this architecture is to reduce coupling, roundtrip and dependencies between the DRAC.   The Derivation Process was also aimed at maximizing Reusability.  Assigning the services to the DRAC that needs it, reduced coupling.   Services, which are shared between the DRACs, are the services, which are needed by both the DRAC.  Assigning data/event to the DRAC that needs it most also reduced the roundtrip.  The architecture is independent of any technology classes.  This ensures reusability of the architecture.

\newpage
\section{ Component Organization}

\begin{figure}[h]
\centering
\fbox{\includegraphics[width=5in,height=4in]{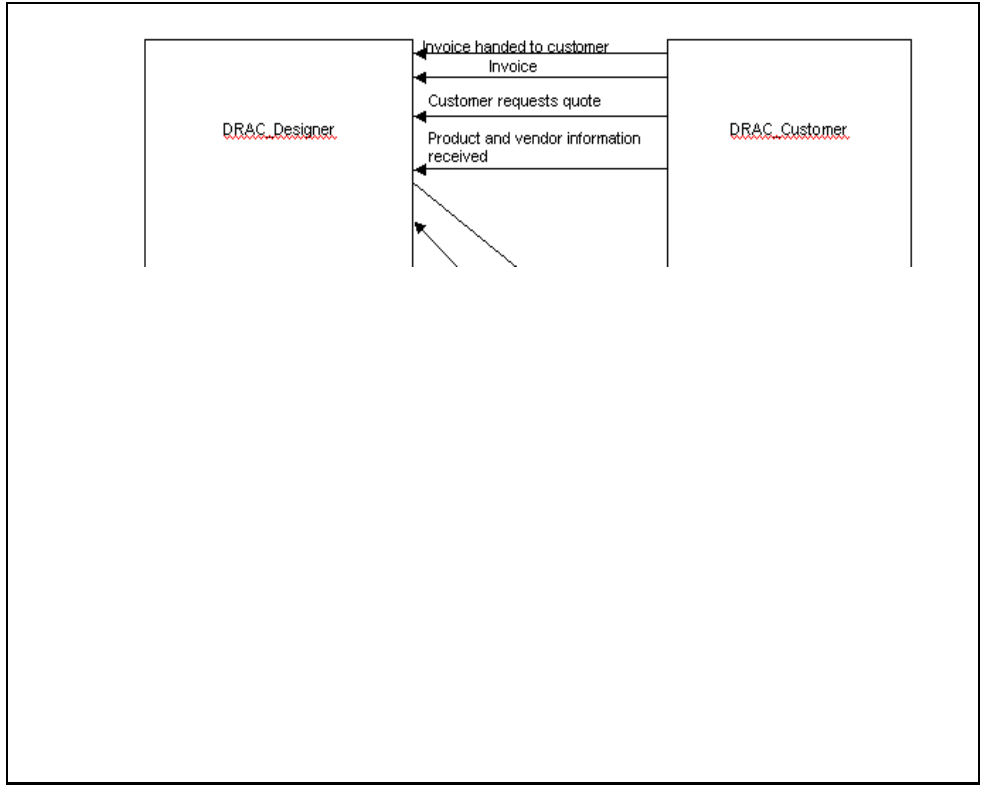}}
\caption{Component Organization.}
\end{figure}

\newpage
\section{ Primitive Components}

\subsection{ DRAC : Designer}

Declarative Model:

DRAC Name: Designer

Created by:  Haritha Nandela

Domain Reference: Designer \_UKM\_TA\_13, Designer \_UKM\_TA\_04, Designer \_UKM
\_TA\_15, Designer \_UKM\_TA\_14, Designer \_UKM\_TA\_24, Designer \_UKM\_TA\_06, 
Designer \_UKM\_TA\_05, Designer \_UKM\_TA\_07, Designer \_UKM\_TA\_12, Designer 
\_UKM\_TA\_08, Designer \_UKM\_TA\_16, Designer \_UKM\_TA\_18, Designer \_UKM\_TA
\_17, Designer \_UKM\_TA\_19, Designer \_UKM\_TA\_25, Designer \_UKM\_TA\_26, Designer 
\_UKM\_TA\_27, Designer \_UKM\_TA\_28, Designer \_UKM\_TA\_29,

Description: This DRAC based on the performer DESIGNER.

Rationale:

DRAC Attributes:

 Name: Measurements

Description: Each blind has measurements, which are its length and the breadth they 
are in inches. These measurements are used determine fits customer's window. So when 
a customer wants to buy the product he needs to get the window size.

 Cardinality: 1

 Domain Model Reference Number:UKM\_CA\_04, UKM\_CA\_05

 Name: Customer Information

Description: Name, Delivery site address, Installation site address, Phone number

Cardinality: 1

 Domain Model Reference Number: UKM\_CA\_04

 Name: Product Information

Description: Product Type, Price, SKU, Vendor Name

Cardinality: 1

 Domain Model Reference Number: UKM\_CA\_04

 Name: Vendor Information

Description: Name, Address, Phone number, Fax Number

Cardinality: 1

 Domain Model Reference Number: UKM\_CA\_04

 Name: Product Availability

Description: Yes or no indicating product availability

Cardinality: 1

 Domain Model Reference Number: UKM\_CA\_04

 Name: Customer Requirements

Description: Price, Range, color, who-the decision maker, when- when customer requires, where- where customer wants the product to be installed, why- what the customer is trying to achieve

Cardinality: 1

 Domain Model Reference Number: UKM\_CA\_04

 Name: Quote

Description: Measurements , Product Information

Cardinality: 1

 Domain Model Reference Number: UKM\_CA\_04

 Name: Installation Status Updated

Description: None

Cardinality: Event

 Domain Model Reference Number: UKM\_TA\_01, UKM\_CA\_04

 Name: Quote Generated

Description: None

Cardinality: Event

 Domain Model Reference Number: UKM\_TA\_01, UKM\_CA\_04

 Name: Customer information Stored

Description: None

Cardinality: Event

 Domain Model Reference Number:UKM\_TA\_01, UKM\_CA\_04

 Name: Measurements recorded

Description: None

Cardinality: Event

 Domain Model Reference Number:UKM\_TA\_01, UKM\_CA\_04

 Name: Designer note price from price list into system

Description: None

Cardinality: Event

 Domain Model Reference Number:UKM\_TA\_01, UKM\_CA\_04

 Name: Designer has keyed in the information

Description: None

Cardinality: Event

 Domain Model Reference Number:UKM\_TA\_01, UKM\_CA\_04

 Name: Stage for Register

Description: After all the product information is entered by the designer in to the 
system, Stage for register is the command used in the Invoice generation software 
to generate the Invoice. 

Cardinality: Event

 Domain Model Reference Number:UKM\_TA\_01, UKM\_CA\_05

Name: Product information recorded

Description: None

Cardinality: Event

 Domain Model Reference Number:UKM\_TA\_01, UKM\_CA\_04

Name: Generated invoice correct

Description: None

Cardinality: Event

 Domain Model Reference Number: UKM\_CA\_04, UKM\_TA\_01

Name: Product and vendor information sent

Description: None

Cardinality: Event

 Domain Model Reference Number:UKM\_TA\_01, UKM\_CA\_04

Name: Product and Measurements sent

Description: None

Cardinality: Event

 Domain Model Reference Number:UKM\_TA\_01, UKM\_CA\_04

Name: Blinds cut

Description: None

Cardinality: Event

 Domain Model Reference Number:UKM\_TA\_01, UKM\_CA\_04

DRAC Services:

Service Name: Coordinate with vendor on product availability

 Domain Reference: Designer\_UKM\_TA\_13

 Service Duration: 10- 15 Minutes

 Execution Frequency: Discrete

  Input Data---

  Input Data Name: Vendor Information

  Received from Service: Present Brands

  Received from Attribute: Vendor Information

  Received from DRAC: Designer

  Input Data Name: Product Information

  Received from Service: Present Brands

  Received from Attribute: Product Information

  Received from DRAC: Designer

  Input Events---

  Input Event Name: Customer requests quote

  Received from Service: None

  Received from Attribute: None

  Received from DRAC: Customer

 Output Data---

  Output Data Name: Product Availability

Sent to Service: Generate quote for 3 brands Generate quote on average window size, 
Generate quote on three different product types

  Sent to DRAC: Designer

  Pre Condition---

  Name: Product Information

  Pre-condition: Present Brands should be completed before this task

Criticality: High

  Name: Vendor Information

  Pre-condition: Present Brands should be completed before this task

Criticality: High

  Post Condition---

  Name :Product Availability

Post Condition: Product Availability is required before Generate quote for 3 brands 
Generate quote on average window size, Generate quote on three different product 
types

Service Name: Gather customer requirements

 Domain Reference: Designer\_UKM\_TA\_04

 Service Duration: 1- 2 hours

 Execution Frequency: Discrete

  Input Events---

  Input Event Name: Customer approached

  Received from Service: None

  Received from Attribute: None

  Received from DRAC: Designer

  Output Data---

  Output Data Name: Customer requirements

Sent to Service: Present Types

  Sent to DRAC: Designer

Service Name: Generate quote on three different brands

  Domain Reference: Designer\_UKM\_TA\_15

 Service Duration: 5- 30 Minutes

  Execution Frequency: Discrete

  Input Data---

   Input Data Name: Product Information

  Received from Service: Present Brands

   Received from Attribute: Product Information

   Received from DRAC: Designer

  Input Data Name: Measurements

   Received from Service: Record Measurements

   Received from Attribute: Measurements recorded

  Received from DRAC: Designer

   Input Data Name: Product Availability

   Received from Service: Coordinate with vendor on product availability

  Received from Attribute: Product Availability

   Received from DRAC: Designer

   Output Data---

  Output Data Name: Quote

Sent to DRAC: Customer

  Output Event---

   Output Event Name: Quote generated

Sent to DRAC: Customer

  Pre Condition---

  Name: Product Information

  Pre-condition: Present Brands service should be completed before this task

Criticality: High

  Name: Measurements

  Pre-condition: Present Brands service should be completed before this task

Criticality: High

  Name :Product Availability

Post Condition: Coordinate with vendor on product availability should be completed 
before this task

Service Name: Generate quote on average window size

  Domain Reference: Designer\_UKM\_TA\_14

 Service Duration: 5- 10 Minutes

  Execution Frequency: Discrete

  Input Data---

  Input Data Name: Product Information

  Received from Service: Present Brands

  Received from Attribute: Product Information

  Received from DRAC: Designer

  Input Data Name: Product Availability

  Received from Service: Coordinate with vendor on product availability

  Received from Attribute: Product Availability

  Received from DRAC: Designer

  Input Event---

  Input Event Name: Customer requests a quote

  Received from DRAC: Customer

  Output Data---

  Output Data Name: Quote

Sent to DRAC: Customer

  Output Event---

  Output Event Name: Quote generated

Sent to DRAC: Customer

  Pre Condition---

  Name: Product Information

  Pre-condition: Present Brands service should be completed before this task

Criticality: High

  Name :Product Availability

Pre Condition: Coordinate with vendor on product availability should be completed 
before this task

Service Name: Generate quote on three different product types

 Domain Reference: Designer\_UKM\_TA\_24

  Service Duration: 10- 15 Minutes

 Execution Frequency: Discrete

  Input Data---

  Input Data Name: Product Information

  Received from Service: Present Brands

  Received from Attribute: Product Information

  Received from DRAC: Designer

  Input Data Name: Measurements

  Received from Service: Record Measurements

  Received from Attribute: Measurements recorded

  Received from DRAC: Designer

  Input Data Name: Product Availability

  Received from Service: Coordinate with vendor on product availability

  Received from Attribute: Product Availability

  Received from DRAC: Designer

  Output Data---

  Output Data Name: Quote

Sent to DRAC: Customer

  Output Event---

  Output Event Name: Quote generated

Sent to DRAC: Customer

  Pre Condition---

  Name: Product Information

  Pre-condition: Present Brands service should be completed before this task

Criticality: High

  Name: Measurements

  Pre-condition: Present Brands service should be completed before this task

Criticality: High

  Name :Product Availability

Post Condition: Coordinate with vendor on product availability should be completed 
before this task

Service Name: Present Brands

 Domain Reference: Designer\_UKM\_TA\_06

 Service Duration: 5- 10 Minutes

 Execution Frequency: Discrete

  Input Data---

  Input Data Name: Product Type

  Received from Service: Present Types

  Received from Attribute: Product Type

  Received from DRAC: Designer

  Output Data---

  Output Data Name: Product Information

Sent to Service: Generate quote for 3 brands, Coordinate with vendor on product Availability, Generate quote on three different product types, Notes price list of price list product, enter product information. Send Production information to vendor, Send product Information to Installer

  Sent to DRAC: Designer, Installer

  Output Data Name: Product Information

Sent to Service: Coordinate with vendor on product Availability, Send Production 
information and vendor to vendor

  Sent to DRAC: Designer, Installer

  Pre Condition---

  Name: Product Types

  Pre-condition: Present Types should be completed before this task

Criticality: High

  Post Condition---

  Name :Product Information

Post Condition: Product  Information  is obtained when this task is completed.

Service Name: Present Types

 Domain Reference: Designer\_UKM\_TA\_05

 Service Duration: 5 minutes- 1 hours

 Execution Frequency: Discrete

  Input Data---

  Input Data Name: Customer Requirements

  Received from Service: Gather Customer requirements 

  Received from Attribute: Customer Requirements

  Received from DRAC: Designer

  Output Data---

  Output Data Name: Product Type

Sent to Service:  Generate quote for 3 product types, Send request measurer,

  Sent to DRAC: Designer

  Pre Condition---

  Name: Customer Requirements

  Pre-condition: Gather customer requirements should be completed before this task

Criticality: 
High

  Post Condition---

  Name :Product Type

Post Condition: Product Type  is obtained when this task is completed.

Service Name: Record Customer information

 Domain Reference: Designer\_UKM\_TA\_07

 Service Duration: 5- 15 Minutes

 Execution Frequency: Discrete

  Input Data---

  Input Data Name: Customer Information

  Received from Service: None

  Received from Attribute: None

  Received from DRAC: Designer

  Output Events---

  Output Event Name: Customer Information recorded

  Sent to Service: Send request to measurer, send request to installer

  Sent to DRAC:  Measurer, Installer

  Pre Condition---

  Name: Customer Information

  Pre-condition: Customer information needed to complete this service.

Criticality: High

  Post Condition---

  Name :Product Information

Post Condition: Customer Information  is recorded when this task is completed.

Service Name: Record Measurements

 Domain Reference: Designer\_UKM\_TA\_12

 Service Duration: 5- 10 Minutes

 Execution Frequency: Discrete

  Input Data---

  Input Data Name: Measurements

  Received from Service: Send measurements to designer

  Received from Attribute: Measurements and customer information received by designer

  Received from DRAC: Designer

  Input Data Name: Customer Information

  Received from Service: Send measurements to designer

  Received from Attribute: Measurements and customer information received by designer

  Received from DRAC: Designer

  Output Events---

  Output Event Name: Measurements recorded

  Sent to Service: Send request to measurer, send request to installer

  Sent to DRAC: Measurer, Installer

  Pre Condition---

  Name: Customer Information

  Pre-condition: Customer information needed to complete this service.

Criticality: High

  Name: Measurements 

  Pre-condition: Measurements needed to complete this service.

Criticality: High

  Post Condition---

  Name: Measurements 

Post Condition: Measurements is recorded when this task is completed.

Service Name: Send request to measurer

 Domain Reference: Designer\_UKM\_TA\_08

 Service Duration: 1- 10 Minutes

 Execution Frequency: Discrete

  Input Data---

  Input Data Name: Customer Information

  Received from Service: Record customer information

  Received from Attribute: Customer information stored

  Received from DRAC: Designer

  Input Data Name: Product types

  Received from Service: Present types

  Received from Attribute: Product type

  Received from DRAC: Designer

  Output Events---

  Output Data Name: Measurer receives customer information and product type information

Sent 
to Service: Check Availability with customer

  Sent to DRAC: Measurer

  Pre Condition---

  Name: Customer Information

  Pre-condition: Customer information should be completed before this task

Criticality: High

  Name: product Type

  Pre-condition: Present types should be completed before this task

Criticality: High

Service Name: Designer notes price list of selected product

 Domain Reference: Designer\_UKM\_TA\_16

 Service Duration: 5- 15 Minutes

 Execution Frequency: Discrete

  Input Data---

  Input Data Name: Measurements

  Received from Service: Record Measurements

  Received from Attribute: Measurements recorded

  Received from DRAC: Designer

  Input Data Name: Product Information

  Received from Service: Present Brands

  Received from Attribute: Product Information

  Received from DRAC: Designer

  Output Event---

  Output Event Name: Designer notes price from price list into system

Sent to Service: Enter price information

  Sent to DRAC: Designer

  Pre Condition---

  Name: Product Information

  Pre-condition: Present Brands should be completed before this task

Criticality: High

  Name: Measurements

  Pre-condition: Measurements should be completed before this task

Criticality: High

Service Name:  Enter price information

 Domain Reference: Designer\_UKM\_TA\_18

 Service Duration: 10- 25 Minutes

 Execution Frequency: Discrete

  Input Events---

  Input Event Name: Entering price list for the selected product

  Received from Service: Designer notes price list of the selected product

  Received from Attribute: Designer notes price from price list into system

  Received from DRAC: Designer

  Output Event---

  Output Event Name: Stage for register

Sent to Service: Generate invoice and hand it to customer

  Sent to DRAC: Designer

  Pre Condition---

  Name: Notes price list for the selected product

  Pre-condition: Designer notes price list of selected product should be completed 
before this task

Criticality: High

  Post Condition---

  Name : Stage for register

Post Condition:  Stage for register is completed at the end of this service.

Service Name: Enter product information

 Domain Reference: Designer\_UKM\_TA\_17

 Service Duration: 10- 25 Minutes

 Execution Frequency: Discrete

  Input Data---

  Input Data Name: Measurements

  Received from Service: Record measurements

  Received from Attribute: Measurements recorded

  Received from DRAC: Designer

  Input Data Name: Product Information

  Received from Service: Present Brands

  Received from Attribute: Product Information

  Received from DRAC: Designer

  Output Event---

  Output Event Name: Product information recorded

Sent to Service: Sent product and measurements to vendor, send request to installer

  Sent to DRAC: Designer

  Pre Condition---

  Name: Product Information

  Pre-condition: Present Brands should be completed before this task

Criticality: High

  Name: Measurements

  Pre-condition: Measurements should be completed before this task

Criticality: High

  Post Condition---

  Name: Product Information

  Pre-condition: Present Brands should be completed before this task

Criticality: High

Service Name: Generate invoice and hand it to customer

 Domain Reference: Designer\_UKM\_TA\_19

 Service Duration: 5 Minutes

 Execution Frequency: Discrete

  Input Events---

  Input Event Name: Stage for register

  Received from Service: Enter price information

  Received from Attribute:  Stage for register

  Received from DRAC:  Designer

  Output Data---

  Output Data Name:  Invoice

Sent to Service: Pay cashier

  Sent to DRAC: Customer

  Pre Condition---

  Name: Stage for register

  Pre-condition: Stage for register should be completed before this task

Criticality: High

  Post Condition---

  Name : Invoice

Post Condition: Invoice is required before service Pay cashier

Service Name: Send product info and measurements to receiving dept

 Domain Reference: Designer\_UKM\_TA\_25

 Service Duration: 10- 15 Minutes

 Execution Frequency: Discrete

  Input Data---

  Input Data Name: Vendor Information

  Received from Service: Present Brands

  Received from Attribute: Vendor Information

  Received from DRAC: Designer

  Input Data Name: Product Information

  Received from Service: Present Brands

  Received from Attribute: Product Information

  Received from DRAC: Designer

  Output Event---

  Output Event Name: Product info and vendor info sent

Sent to Service: None

  Sent to DRAC: External

  Pre Condition---

  Name: Product Information

  Pre-condition: Present Brands should be completed before this task

Criticality: High

  Name: Vendor Information

  Pre-condition: Present Brands should be completed before this task

Criticality: High

Service Name: Send product information and measurements to vendor

 Domain Reference: Designer\_UKM\_TA\_26

 Service Duration: 10- 15 Minutes

 Execution Frequency: Discrete

  Input Data---

  Input Data Name: Measurements

  Received from Service: Record measurements

  Received from Attribute: Measurements recorded

  Received from DRAC: Designer

  Input Data Name: Product Information

  Received from Service: Present Brands

  Received from Attribute: Product Information

  Received from DRAC: Designer

  Output Events---

  Output Event Name: Product info and measurements send to vendor

  Sent to Service: None

  Sent to DRAC: External

  Pre Condition---

  Name: Product Information

  Pre-condition: Present Brands should be completed before this task

Criticality: High

  Name: Measurements

  Pre-condition: record measurements should be completed before this task

Criticality: High

Service Name: Send request to installer

 Domain Reference: Designer\_UKM\_TA\_27

 Service Duration: 10- 15 Minutes

 Execution Frequency: Discrete

  Input Data---

  Input Data Name: Measurements  Received from Service: Record Measurments

  Received from Attribute: Measurements recorded

  Received from DRAC: Designer

  Input Data Name: Product Information

  Received from Service: Present Brands

  Received from Attribute: Product Information

  Received from DRAC: Designer

  Input Data Name: Vendor Information

  Received from Service: Present Brands

  Received from Attribute: Vendor Information

  Received from DRAC: Designer

  Output Events---

  Output Event Name: Product and vendor information received

Sent to Service: Check availability with customer

  Sent to DRAC: Measurer

  Pre Condition---

  Name: Product Information

  Pre-condition: Present Brands should be completed before this task

Criticality: High

  Name: Measurements

  Pre-condition: record measurements should be completed before this task

Criticality: High

Service Name: Check Inventory

 Domain Reference: Designer\_UKM\_TA\_28

 Service Duration: 10- 15 Minutes

 Execution Frequency: Discrete

  Input Data---

  Input Data Name: Measurements  Received from Service: Record Measurments

  Received from Attribute: Measurements recorded

  Received from DRAC: Designer

  Input Data Name: Product Information

  Received from Service: Present Brands

  Received from Attribute: Product Information

  Received from DRAC: Designer

  Output Data---

  Output Data Name: Product Availability

Sent to Service: Cut Blinds

  Sent to DRAC: Designer

  Pre Condition---

  Name: Product Information

  Pre-condition: Present Brands should be completed before this task

Criticality: High

  Post Condition---

  Name :Product Availability

Post Condition: Product Availability is required before we can cut the blinds

Service Name: Cut Blinds

 Domain Reference: Designer\_UKM\_TA\_29

 Service Duration: 10- 15 Minutes

 Execution Frequency: Discrete

  Input Data---

  Input Data Name: Measurements

  Received from Service: Record Measurements

  Received from Attribute: Measurements recorded

  Received from DRAC: Designer

  Input Data Name: Product Information

  Received from Service: Present Brands

  Received from Attribute: Product Information

  Received from DRAC: Designer

  Input Data Name: Product Availability

  Received from Service: Check Inventory

  Received from Attribute: Product Availability

  Received from DRAC: Designer

  Output Events---

  Output Data Name: Blinds cut

Sent to Service:  Pay cashier

  Sent to DRAC: Customer

  Pre Condition---

  Name: Product Information

  Pre-condition: Present Brands should be completed before this task

Criticality: High

  Name: Measurements

  Pre-condition: Record Measurements should be completed before this task

Criticality: High

  Name:  Product Availability

  Pre-condition: Check Inventory should be completed before this task

Criticality: High

  Post Condition---

  Name : Blinds Cut

Post Condition: Blinds need to be cut  before the cashier can be paid.

Integration Model:

Dependencies on other DRAC's:

Services Required: None

Attributes Required:

  Attribute Name: Invoice

  DRAC that owns the attribute: Customer

  Attribute Name: Customer requests quote

  DRAC that owns the attribute: Customer

  Attribute Name: Customer approached

  DRAC that owns the attribute: Customer

  Attribute Name: Invoice handed over to customer

  DRAC that owns the attribute: Customer

  Attribute Name: Measurements and customer information received

  DRAC that owns the attribute: Measurer

  Attribute Name: Measurer receives customer information and product type information

  DRAC 
that owns the attribute: Measurer

  Attribute Name: Product information and vendor information received

  DRAC that owns the attribute: Installer

\subsection{ DRAC : Measurer }

Declarative Model:

DRAC Name: Measurer

Created by: Anshuman Sinha

Domain Reference: Measurer \_UKM\_TA\_09, Measurer \_UKM\_TA\_22, Measurer \_UKM
\_TA\_21, Measurer \_UKM\_TA\_23,

Description: This DRAC based on the performer MEASURER.

Rationale:

DRAC Attributes:

Name: Measurements and customer information received.

Description: The event is triggered when the measurements and customer information 
is recorded into the recording system.

Cardinality: Event

Domain Model Reference Number: UKM\_TA\_01

Name: Availability confirmed.

Description: The event is triggered when the availability of the customer is confirmed.  The customer will be available for when the measurer will visit the site.

Cardinality: Event

Domain Model Reference Number: UKM\_TA\_01

Name: Measurer receives customer and product type information.

Description: This event is marked when the measurer receives the customer and product type information.

Cardinality: Event

Domain Model Reference Number: UKM\_TA\_01

Name: Measurements received.

Description: This event is marked when the designer receives the measurements.

Cardinality: Event

Domain Model Reference Number: UKM\_TA\_01

DRAC Services

Service Name: Arrive at site and take measurements.

 Domain Reference: Designer\_UKM\_TA\_11

 Service Duration: 15-30 Minutes

 Execution Frequency: Discrete

  Input Data---

  Input Data Name: Customer Information

  Received from Service: Record customer information.

  Received from Attribute: Event: Customer information stored.

  Received from DRAC: Designer

  Input Event Name:  Availability confirmed

  Received from Service: Check availability with customer.

  Received from Attribute: Availability confirmed

  Received from DRAC: Measurer.

  Input Data Name: Product Type

  Received from Service: Present Types

  Received from Attribute: Product Type.

  Received from DRAC: Designer.

  Output Data---

  Output Data Name: Measurements.

Sent to Service: None

  Sent to DRAC: Measurer.

  Pre Condition---

  Name: Customer Information (Data)

  Pre-condition: Customer information should be available for this service.

Criticality: High

  Name: Customer Availability.

  Pre-condition: Customer has to be available for the measurements to be taken.

Criticality: High

  Name: Product Type.

  Pre-condition: Measurer has to know the product type before he can take measurements.

Criticality: High

  Post Condition---

  Name: Measurements

Post Condition: Measurements are taken and this is the output of the service.

Service Name: Send measurements to designer.

 Domain Reference: Designer\_UKM\_TA\_10

 Service Duration: 10-15 Minutes

 Execution Frequency: Discrete

  Input Data---

  Input Data Name: Measurements.

  Received from Service: Arrive at site and take measurements.

  Received from Attribute: Data: Measurements.

  Received from DRAC: Measurer.

  Output Event---

  Output Data Name: Measurements and customer information received.

Sent to Service: Record Measurements.

  Sent to DRAC: Designer.

  Pre Condition---

  Name: Measurements (Data)

  Pre-condition: Measurements should be available for this service.

Criticality: High

  Post Condition---

  Name: Measurements and customer information received.

Post Condition: Measurements and customer information is received by the designer 
at the end of the service.

Service Name: Check availability with customer.

 Domain Reference: Designer\_UKM\_TA\_9

 Service Duration: 10-15 Minutes

 Execution Frequency: Discrete

  Input Data---

  Input Data Name: Customer Information

  Received from Service: Record customer information.

  Received from Attribute: Event: Customer information stored.

  Received from DRAC: Designer

  Output Event---

  Output Data Name: Availability confirmed.

Sent to Service: Arrive at site and take measurements.

  Sent to DRAC: Measurer.

  Pre Condition---

  Name: Customer Information (Data)

  Pre-condition: Customer information should be available for this service.

Criticality: High

  Post Condition---

  Name: Availability confirmed

Post Condition: The availability of the customer is confirmed at the end of the service.

Integration Model:

Dependencies on other DRAC's:

Services Required: None

Attributes Required:

  Attribute Name: Measurements

  DRAC that owns the attribute: Designer

  Attribute Name: Customer Information

  DRAC that owns the attribute: Designer

  Attribute Name: Product Information

  DRAC that owns the attribute: Designer

\subsection{ DRAC : Installer }

Declarative Model:

DRAC Name: Installer

Created by: Anshuman Sinha

Domain Reference: Installer \_UKM\_TA\_09, Installer \_UKM\_TA\_10, Installer \_UKM
\_TA\_11

Description: This DRAC is based on the performer INSTALLER.

Rationale:

DRAC Attributes:

Name: Products delivered.

Description: The event is marked when the products are delivered by the vendor.  This is a precondition to product installation.

Cardinality: Event

Domain Model Reference Number: UKM\_TA\_01

Name: Products Installed.

Description: The event is marked when the products are installed by the installer 
to customer's satisfaction.

Cardinality: Event

Domain Model Reference Number: UKM\_TA\_01

Name: Product and vendor information received.

Description: This event is marked when the designer receives the product and vendor 
information about the product.

Cardinality: Event

Domain Model Reference Number: UKM\_TA\_01

DRAC Services

Service Name: Arrive at site and install products.

 Domain Reference: Designer\_UKM\_TA\_22

 Service Duration: 30-50 Minutes

 Execution Frequency: Discrete

  Input Data---

  Input Data Name: Products Delivered.

  Received from Service: Check delivery of product.

  Received from Attribute: Event: Product delivered.

  Received from DRAC: Installer

  Input Event Name: Availability confirmed

  Received from Service: Check availability with customer.

  Received from Attribute: Availability confirmed

  Received from DRAC: Measurer.

  Input Data Name: Measurements

  Received from Service: Arrive at site and take measurments.

  Received from Attribute: Data: Measurements.

  Received from DRAC: Measurer.

  Output Event---

  Output Data Name: Product delivered.

Sent to Service: Arrive at site and install product.

  Sent to DRAC: Installer.

  Pre Condition---

  Name: Products Delivered (Event)

  Pre-condition: Products should be delivered before this service.

Criticality: High

  Name: Availability confirmed. (Event).

  Pre-condition: Customer availability should be verified before the service .

Criticality: High

  Name: Measurements. (Data).

  Pre-condition: Measurements should be available before the service.

Criticality: High

  Post Condition---

  Name: Product Installed. (Event)

Post Condition: Product is installed after this service.

Service Name: Check delivery of product.

 Domain Reference: Designer\_UKM\_TA\_21

 Service Duration: 10-15 Minutes

 Execution Frequency: Discrete

  Input Data---

  Input Data Name: Products Information.

  Received from Service: Present brands.

  Received from Attribute: Data: Product information.

  Received from DRAC: Designer

  Input Data Name: Vendor Information.

  Received from Service: Present brands.

  Received from Attribute: Data: Vendor Information.

  Received from DRAC: Designer.

  Output Event---

  Output Data Name: Products delivered.

Sent to Service: Arrive at site and install product.

  Sent to DRAC: Installer.

  Pre Condition---

  Name: Product Information (Data)

  Pre-condition: Products Information must be received before the service begins..

Criticality: High

  Name: Vendor Information. (Data).

  Pre-condition: Products Information must be received before the service begins..

Criticality: High

  Post Condition---

  Name: Product Delivered. (Event)

Post Condition: Product is delivered at the end of the service.

Service Name: Record Status

 Domain Reference: Designer\_UKM\_TA\_23

 Service Duration: 10-15 Minutes

 Execution Frequency: Discrete

  Input Event---

  Input Data Name: Product Installed.

  Received from Service: None.

  Received from Attribute: None.

  Received from DRAC: Installer

  Output Event---

  Output Data Name: Installation status updated.

Sent to Service: The installation status is updated in the recording system.

  Sent to DRAC: Installer.

  Pre Condition---

  Name: Product Installed (Event)

  Pre-condition: Products is installed before the service starts.

Criticality: High

  Post Condition---

  Name: Installation status updated. (Event)

Post Condition: The status is recorded in the system.

Integration Model:

Dependencies on other DRAC's:

Services Required: 

  Name of event required: Availability confirmed

  Name of Service: Check availability with Customer

  Name of the DRAC that owns the service: Measurer

Attributes Required:

  Attribute Name: Measurements

  DRAC that owns the attribute: Designer

  Attribute Name: Customer Information

  DRAC that owns the attribute: Designer

  Attribute Name: Product Information

  DRAC that owns the attribute: Designer

  Attribute Name: Vendor Information

  DRAC that owns the attribute: Designer

  Attribute Name: Availability Confirmed

  DRAC that owns the attribute: Designer

\subsection{ DRAC : Customer }

Declarative Model:

DRAC Name: Customer

Created by: Anshuman Sinha

Domain Reference: Measurer \_UKM\_TA\_09, Measurer \_UKM\_TA\_10, Measurer \_UKM
\_TA\_11

Description: This DRAC is based on the performer CUSTOMER.

Rationale:

DRAC Attributes:

 Name: Invoice

Description: The invoice has the details of the order and the product information.  
It also has the customer information and the measurements along with the price and 
the total amount to be received from the customer.

 Cardinality: 1

 Domain Model Reference Number: UKM\_CA\_04

 Name: Payment for Invoice made.

Description: The event is marked when payment for Invoice is made by the customer.  

 Cardinality: Event

 Domain Model Reference Number: UKM\_TA\_01

 Name: Customer Approaches cashier

Description: The event is marked when the customer goes upto the payment desk to 
make a payment.

 Cardinality: Event

 Domain Model Reference Number: UKM\_TA\_01

 Name: Customer Requests Quote

Description: If the customer wants to buy the product he needs to look at the quote 
to decide if the product suit the budget.

 Cardinality: Event

 Domain Model Reference Number: UKM\_TA\_01

 Name: Customer Approached

Description: The customer is approached by the designer to gather requirements.

 Cardinality: Event

 Domain Model Reference Number: UKM\_TA\_01

 Name: Invoice handed over to customer.

Description: The Invoice is handed by designer the design center.

 Cardinality: Event

 Domain Model Reference Number: UKM\_TA\_01

DRAC Services

Service Name: Pay Cashier

 Domain Reference: Designer\_UKM\_TA\_20

 Service Duration: 10-15 Minutes

 Execution Frequency: Discrete

  Input Data---

  Input Data Name: Invoice

  Received from Service: Generate Invoice and hand it to customer.

  Received from Attribute: Data: Invoice;  Event: Invoice handed to customer.

  Received from DRAC: Designer

  Input Data Name: Customer Approaches cashier.

  Received from Service: None

  Received from Attribute: None

  Received from DRAC: Customer

  Input Events---

  Input Event Name: Customer Approaches cashier.

  Received from Service: None

  Received from Attribute: None

  Received from DRAC: Customer

  Output Events---

  Output Data Name: Payment for Invoice made.

Sent to Service: None

  Sent to DRAC: Customer

  Output Data Name: Receipt handed to customer.

Sent to Service: None

  Sent to DRAC: Customer

  Pre Condition---

  Name: Invoice (Data)

  Pre-condition: Invoice should be generated before this service.

Criticality: High

  Name: Customer approaches cashier.

  Pre-condition: Customer has to approach the cashier.

Criticality: High

  Post Condition---

  Name: Payment for Invoice made.

Post Condition: This marks the end of the service.

  Name: Receipt handed to customer.

Post Condition: The customer is handed the receipt after the payment is made to the 
cashier.

\newpage 
%% Unified Conclusion
%% Section Change

\chapter{ Conclusion}

Home Depot is the largest network of home-product retail shops and is made up of 
several departments. Each department needs to keep track of the product availability, product choices and order backlogs and generate invoice based on the selected product. The project that we have analyzed, is to help the customer associates keep track of the products, know the availability of the customer choices, generate invoice, co-ordinate with vendor, measurer and Installer. Incorporating this Information technology product will translate into better customer satisfaction and more revenues for Home Depot.

In retrospect, there were several things we could of have done differently for this 
exercise.  The interview could have been structured a bit more efficiently.  We notice that many of our questions were not useful for our final modeling.  However, we did cover this deficiency by conducting more interviews and as many questions. In the end we did get most of the information we need for our Domain Reference Architecture.  Our initial modeling also needed more organization because it did took us some effort to refine our model before they can be synthesized into our Domain Reference Architecture.  Overall, we did overcome most of the set backs we encountered, thus in our next software engineering project we are confidant that we can do an even better job.

Overall, in doing this exercise we learned a great deal of what automation can and 
can not accomplish.  We also gained experience in gathering the information we need 
from experts in different discipline.  The experience of converting the gathered 
knowledge in to a framework for a possibly real application is very exciting and 
rewarding.

\chapter{ Acknowledgements}
We would like to thank our professor Dr. Suzanne Barber of University of Texas at Austin for the great learning at her lecture class on Domain Specific Software Architecture.  We are also greatly indebted to all employees of the store especially Julie who had numerous lengthy discussions on the specifics of the working of ordering system and its maintenance.  She displayed lots of patience in knowledge acquisitions sessions.  Our special acknowledgement to teaching assistant for the course who helped us at various stages of report submission, correction and its pickup.

\chapter{ References}

\begin{enumerate}

	\item Domain Specific Software Architecture Lecture Notes by Dr. Suzanne Barber, University of Texas at Austin

	\item Domain Specific Software Architecture Program, Software Engineering Institute at Carnegie Mellon University - Technical Report by LTC Erik Mettala and Marc H. Graham

	\item A formal approach to software architecture by Robert J. Allen

	\item A domain specific software architecture for adaptive intelligent systems by Barbara Hayes-Roth, Karl Pfleger, and Marko Balabanovic 

	\item A domain specific software architecture engineering process outline by Will Tracz, Lou Coglianese and Patrick Young

	\item User-Centered Requirements: The scenario-based engineering process by Karen McGraw and Karan Harbison

	\item The scenario based engineering process(SEP): A user centered approach for the development of health care systems by K. Harbinson, L. Burnell, J. Kelly and G. Haddock

	\item Knowledge Acquisition for expert systems by A. Hart

	\item Thinkertoys by M. Michalko

	\item Knowledge Acquisition in the development of a large expert system by D. Prerau

\end{enumerate}

\end{document}